\documentclass[12pt]{article}
\usepackage{amsmath,amssymb,amscd,amsthm,amsfonts}
\usepackage{graphicx}
\usepackage{hyperref}
\usepackage{listings}
\usepackage[svgnames]{xcolor}
\usepackage[most]{tcolorbox}
\usepackage{graphicx}
\usepackage{bm}
\usepackage{pdflscape}
\usepackage{euscript}
\usepackage{amsthm}
\usepackage{mathrsfs}
\allowdisplaybreaks
\usepackage{float}
\usepackage{caption}
\usepackage{subcaption}
\usepackage{multirow}
\usepackage{tikz}
\usetikzlibrary{shapes,arrows}
\usepackage{multicol}
\usepackage{natbib}
\usepackage{mathabx}
\usepackage{cleveref}
\usepackage{enumitem,kantlipsum}
\usepackage{bibunits}
\usepackage{makecell}
\usepackage{longtable}
\usepackage{xpatch}
\xapptocmd{\proof}{\mbox{}\par\nobreak}{}{}
\usepackage[toc,page,header]{appendix}
\usepackage{titletoc}
\usepackage{minitoc}

\mtcsetdepth{parttoc}{4}
\makeatletter
\renewcommand{\paragraph}{\@startsection{paragraph}{4}{0ex}%
   {-3.25ex plus -1ex minus -0.2ex}%
   {1.5ex plus 0.2ex}%
   {\normalfont\normalsize\bfseries}}
\makeatother
\stepcounter{secnumdepth}
\stepcounter{tocdepth}

\tcbset{
    frame code={}
    center title,
    left=0pt,
    right=0pt,
    top=0pt,
    bottom=0pt,
    colback=blue!3,
    colframe=white,
    width=\dimexpr\textwidth\relax,
    enlarge left by=0mm,
    boxsep=5pt,
    arc=0pt,outer arc=0pt,
    }

\hypersetup{
  colorlinks,
  citecolor=Blue,
  linkcolor=Red,
  urlcolor=Blue}

\usepackage{setspace}
\usepackage[margin=1in]{geometry}

\newtheorem{definition}{Definition}
\newtheorem{theorem}{Theorem}
\newtheorem{proposition}{Proposition}

\newtheorem{lemma}{Lemma}

\newtheorem{remark}{Remark}
\newtheorem{assumptionInner}{}

\newenvironment{assumption}
{\pushQED{\qed}\assumptionInner}{\popQED\endassumptionInner}
\newcounter{assumptionInnerB}

\newenvironment{assumptionB}
{\pushQED{\qed}%
 \stepcounter{assumptionInnerB}
 \begingroup%
\assumptionInner}
{\popQED\endassumptionInner\endgroup}
\newcounter{assumptionInnerC}

\newcounter{assumptionInnerD}

\newenvironment{assumptionD}
{\pushQED{\qed}%
 \stepcounter{assumptionInnerD}
 \begingroup%
\assumptionInner}
{\popQED\endassumptionInner\endgroup}

\newcommand{\equalInLaw}{\stackrel{{\cal{L}}}{=}}
\newcommand{\convergeInLaw}{\stackrel{{\cal{D}}}{\longrightarrow}}
\newcommand{\convergeInP}{\stackrel{i.p.}{\longrightarrow}}

\newcommand{\T}{^{\top}}
\newcommand{\inv}{^{-1}}
\newcommand{\pseudoinv}{^{+}}

\DeclareMathOperator{\Var}{Var}

\DeclareMathOperator{\Tr}{Tr} 
\DeclareMathOperator{\argmin}{arg\,min}
\newcommand{\probP}{\text{I\kern-0.15em P}}

\renewcommand\Re{{\operatorname{Re}}}
\renewcommand\Im{{\operatorname{Im}}}
\newcommand{\bome}{\bm{\omega}}
\newcommand{\bomeast}{{\bome^{\ast}}}
\newcommand{\hatbomeast}{\widehat{{\bome}^{\ast}}}
\newcommand{\hatbomeastlam}{\widehat{{\bome}^{\ast}_{\lambda}}}
\newcommand{\hatbomeastpseudo}{\widehat{{\bome}^{\ast}_{\pseudoinv}}}

\newcommand{\bmu}{{\bm{\mu}}}
\newcommand{\hatbmu}{{\widehat{\bmu}}}

\newcommand{\bnu}{{\bm{\nu}}}
\newcommand{\htheta}{{\widehat{\theta}}}
\newcommand{\ttheta}{{\Tilde{\theta}}}
\newcommand{\sigmae}{\sigma_{\epsilon}}
\newcommand{\tsigma}{\tilde{\sigma}}
\newcommand{\bSig}{{\bm{\Sigma}}}

\newcommand{\hatbSig}{{\widehat{\bm{\Sigma}}}}
\newcommand{\hatSig}{{\widehat{\Sigma}}}
\newcommand{\hatbSiglam}{{\widehat{\bm{\Sigma}}_{\lambda}}}

\newcommand{\talpha}{{\tilde{\alpha}}}
\newcommand{\ttalpha}{{\tilde{\tilde{\alpha}}}}
\newcommand{\ualpha}{{\underline{\alpha}}}
\newcommand{\utalpha}{{\tilde{\underline{\alpha}}}}
\newcommand{\uttalpha}{{\tilde{\tilde{\underline{\alpha}}}}}
\newcommand{\tbeta}{{\tilde{\beta}}}

\newcommand{\ubeta}{{\underline{\beta}}}
\newcommand{\utbeta}{{\tilde{\underline{\beta}}}}

\newcommand{\ugamma}{{\underline{\gamma}}}
\newcommand{\uzeta}{{\underline{\zeta}}}
\newcommand{\utzeta}{{\underline{\tilde{\zeta}}}}
\newcommand{\uxi}{{\underline{\xi}}}
\newcommand{\bepsilon}{\bm{\epsilon}}
\newcommand{\bupsilon}{\bm{\upsilon}}
\newcommand{\hatbupsilon}{\widehat{\bm{\upsilon}}}

\newcommand{\hphi}{\widehat{\phi}}
\newcommand{\tphi}{\tilde{\phi}}
\newcommand{\hattau}{\widehat{\tau}}
\newcommand{\bA}{{\bm{A}}}

\newcommand{\bB}{{\bm{B}}}

\newcommand{\bb}{{\bm{b}}}

\newcommand{\bC}{{\bm{C}}}

\newcommand{\be}{\bm{e}}

\newcommand{\hatG}{{\widehat{G}}}
\newcommand{\bH}{\bm{H}}
\newcommand{\hatH}{\widehat{H}}
\newcommand{\bh}{\bm{h}}
\newcommand{\bI}{\bm{I}}
\newcommand{\bk}{\bm{k}}

\newcommand{\bsm}{{\bm{m}}}
\newcommand{\hatbsm}{{\widehat{\bm{m}}}}
\newcommand{\bM}{{\bm{M}}}
\newcommand{\bn}{{\bm{n}}}
\newcommand{\NormDis}{\mathcal{N}}
\newcommand{\bO}{{\bm{O}}}
\newcommand{\bp}{{\bm{p}}}
\newcommand{\bq}{{\bm{q}}}
\newcommand{\bR}{{\bm{R}}}

\newcommand{\rsvt}{{\mathcal{R}}}
\newcommand{\br}{\bm{r}}

\newcommand{\calS}{{\bm{\mathcal{S}}}}
\newcommand{\bs}{{\bm{s}}}

\newcommand{\barbs}{{\bar{\bs}}}

\newcommand{\bu}{{\bm{u}}}
\newcommand{\bv}{{\bm{v}}}

\newcommand{\bX}{\bm{X}}
\newcommand{\bx}{{\bm{x}}}

\newcommand{\bY}{\bm{Y}}
\newcommand{\by}{{\bm{y}}}
\newcommand{\bty}{{\tilde{\bm{y}}}}

\newcommand{\bz}{{\bm{z}}}

\begin{document}
\doparttoc 
\faketableofcontents
\title{\Large ``Double Descent” in Portfolio Optimization: Dance between Theoretical Sharpe Ratio and Estimation Accuracy}
\author{Yonghe Lu\thanks{College of Business and Economics, Australian National University. Email: \href{mailto: yonghe.lu@anu.edu.au}{yonghe.lu@anu.edu.au}} \and Yanrong Yang\thanks{College of Business and Economics, Australian National University. Email: \href{mailto: yanrong.yang@anu.edu.au}{yanrong.yang@anu.edu.au}} \and Terry Zhang\thanks{Corresponding author. College of Business and Economics, Australian National University.Email: \href{mailto: terry.zhang@anu.edu.au}{terry.zhang@anu.edu.au}}}

\date{\today}
\maketitle

\begin{abstract}
\begin{spacing}{1}
We study the relationship between model complexity and out-of-sample performance in the context of mean-variance portfolio optimization. Representing model complexity by the number of assets, we find that the performance of low-dimensional models initially improves with complexity but then declines due to overfitting. As model complexity becomes sufficiently high, the performance improves with complexity again, resulting in a ``double ascent'' Sharpe ratio curve similar to the ``double descent'' phenomenon observed in artificial intelligence. The underlying mechanisms involve an intricate interaction between the theoretical Sharpe ratio and estimation accuracy. In high-dimensional models, the theoretical Sharpe ratio approaches its upper limit, and the overfitting problem is reduced because there are more parameters than data restrictions, which allows us to choose well-behaved parameters based on inductive bias.  
\end{spacing}

\end{abstract}

{\bf Keywords:} Asset Pricing, Double Descent, High-Dimensional Data, Machine Learning, Mean-Variance Portfolio Optimization, Random Matrix Theory

\newpage
\section{Introduction}

Neural networks have recently achieved remarkable success in various fields, including natural language processing (e.g., ChatGPT), computer vision (e.g., face recognition, Tesla’s Autopilot), medical diagnosis (e.g., cancer classification), and stock return predictions (\cite{gu2020empirical}). The phenomenon of ``double descent'' is an important explanation for the success of large neural network models (\cite{nakkiran2021deep}). The term ``double descent'', first introduced by \cite{belkin2019reconciling}, refers to the relationship between the prediction error of a model and the model's complexity, where model complexity can be considered as the number of parameters describing the functional form of the model.\footnote{\cite{belkin2019reconciling} also use the term model capacity and model complexity interchangeably.} 

In standard textbooks, the relationship between the prediction error and model complexity is U-shaped. If a model is too simple, it does not capture the important features in the data, making poor predictions both in-sample and out-of-sample. As the model complexity increases, its performance begins to improve, generating the first descent in prediction loss. When a model becomes too complex, its prediction error starts to rise with complexity, because the model begins to overfit spurious patterns in the training data, resulting in good in-sample performance but poor out-of-sample prediction. 

Recent studies in machine learning and statistics have discovered that after a model becomes complex enough to perfectly fit all training data, a further increase in the model's complexity reduces its prediction error, generating the second descent in prediction loss. The second descent can be stronger than the first such that the optimal complex model that perfectly fits in-sample data outperforms all parsimonious models. This double descent phenomenon has been shown to exist in very general settings (\cite{hastie2022surprises}) and fundamentally challenges the traditional view of the bias-variance trade-off in model selection. Understanding its ubiquitousness and the underlying mechanism is the subject of ongoing research (\cite{muthukumar2020harmless,bartlett2020benign,mei2022generalization,li2023benign,ullah2024effect}).

This paper studies the ``double descent'' phenomenon in a classic finance problem, namely the mean-variance portfolio optimization. Portfolio optimization is different from a prediction problem, so we develop new tools that are more suitable for our setting.\footnote{The mathematical foundation of these tools are based on the random matrix theory (\cite{marchenko1967distribution,silverstein1995strong,bai2010spectral}).} In portfolio optimization, model complexity can be described by the complexity of the asset space, which we model as the number of assets. The parameters of the model are the optimal weights invested in each asset. Increasing the number of assets increases the complexity of the model.\footnote{A different approach is to model the optimal weights directly as a function of the asset's characteristics (\cite{brandt2009parametric}). Under this approach, the complexity of the model can be expanded by increasing the number of characteristics in the weight function. However, the model needs more information under this approach and is different from the classic \cite{Markowitz1952} problem.}

The portfolio optimization literature has long recognized the challenge of estimating the parameters for a portfolio optimization model, especially when the model is complex (\cite{chopra1993aathe,kacperczyk2004asset,demiguel2009optimal}). \cite{kan2007optimal} show that the out-of-sample performance of the estimated optimal portfolio decreases with the number of assets due to uncertainties about model parameters. The focus of the literature has been largely on how to improve the estimation procedure, holding the model complexity unchanged.\footnote{Various ways to improve the estimation have been proposed in the literature, for example, by using a Bayesian approach to place a prior on the parameters (\cite{garlappi2007portfolio,kan2007optimal,kan2022optimal} or applying shrinkage to parameters (\cite{ledoit2017nonlinear,ao2019approaching}). See \cite{markowitz2010portfolio,xidonas2020robust,gunjan2023brief} for reviews of the portfolio optimization literature. } 
Our focus is entirely different. We investigate how the portfolio's performance changes with a model's complexity, holding the estimation procedure fixed.\footnote{Ultimately, the performance of the optimal portfolio depends on the model, which specifies the number of assets and their mean-variance structure, and how the model is estimated. Many estimation procedures have regularization features that explicitly or implicitly reduce the complexity of a model. Different estimation procedures are designed for different models in different data settings. The performances of two models with different estimation procedures are not directly comparable. Therefore, to study how performance changes with model complexity, we need to fix a particular estimation procedure.} We use a well-known estimator, the generalized inverse estimator, to train our portfolio optimization models. 

In contrast to \cite{kan2007optimal}, we find that, in the high-dimensional regime, the performance of the estimated optimal portfolio can improve with the number of assets. 
We prove that under certain conditions, the asymptotic behavior of the out-of-sample Sharpe ratio displays a pattern of ``double ascent'': the performance of the portfolio first increases with the number of assets in the low-dimensional regime, then it declines with the number of assets, and in the high-dimensional regime, it increases again with the number of assets again. This pattern is analogous to the ``double descent'' phenomenon documented in various artificial intelligence problems.


To understand this phenomenon, our theory produces an elegant equation of the out-of-sample performance of complex portfolio models. It allows us to analytically answer several important questions regarding the relationship between portfolio performance and model complexity. For example, what mechanisms explain each ascent in performance? How complex a model needs to be to generate a good performance? Based on the empirical performance, what can we learn about the nature of asset returns?

Our theory highlights two mechanisms that jointly explain why model complexity improves portfolio performance. The first one is an economic mechanism and the second one is a statistical mechanism. The economic mechanism relates to a theoretical quantity, which we call the clairvoyant Sharpe ratio. This is the Sharpe ratio of the optimal portfolio based on the true parameters, which are not observable. Although the theoretical or clairvoyant Sharpe ratio is not observable, we can apply economic principles to imagine how it changes with model complexity. For example, the clairvoyant Sharpe ratio should be non-decreasing in the number of assets. It should also display decreasing returns to scale with respect to the number of assets.\footnote{For example, suppose there are $N$ assets with i.i.d returns, the clairvoyant Sharpe ratio grows at a rate of $\sqrt{N}$.} If there is no arbitrage opportunity in the economy, the clairvoyant Sharpe ratio should have an upper bound.\footnote{The upper bound of the clairvoyant Sharpe ratio has an important meaning in asset pricing models. For example, in the CAPM economy, it is the Sharpe ratio of the tangency portfolio, the market portfolio.}

Holding constant the clairvoyant Sharpe ratio, the out-of-sample Sharpe ratio (of the estimated optimal portfolio) also depends on the model complexity. We characterize this relationship based on the principles of statistics. In the low-dimensional regime, the out-of-sample Sharpe ratio always decreases with the number of assets, regardless of the factor structure of asset returns. However, in the high-dimensional regime, it can increase with the number of assets, depending on the factor strength of the underlying assets. The stronger the factor structure, the better the performance. 

Both mechanisms jointly determine the performance curve. The first ascent in performance is driven by the economic mechanism, because in the low dimensional regime, the clairvoyant Sharpe ratio increases rapidly with the number of assets despite the rising estimation error. The second ascent in performance is driven by the statistical mechanism, because in the high-dimensional regime, model complexity can improve estimation accuracy. 

Why does complexity improve estimation accuracy in the high-dimensional regime? The reason is that in the high-dimensional regime, there are more parameters than what the data can identify. In other words, the model can perfectly ``fit'' the data with different sets of parameters, which gives the modeler freedom to choose the best parameters.\footnote{This situation is analogous to an under-identified model in which the modeler calibrates parameters based on prior knowledge or intuition.} The criteria for choosing parameters, known as the inductive bias in artificial intelligence, play a key role in the success of complex models. The criteria that we apply through the generalized inverse estimator are closely related to the equal-weighted portfolio, which has been shown to work well in practice (\cite{demiguel2009optimal}). More specifically, we show that in complex models, optimal portfolio weights do not need to be overly stretched to fit the training data.\footnote{This argument is made more precisely through eigen-analysis.} The generalized inverse estimator chooses the portfolio with the smallest $\ell_2$ norm from all feasible portfolios, resulting in an optimal portfolio closer to the equal-weighted portfolio.

Our theory provides practical insights on the optimal design of complex portfolio models. To take advantage the benefit of model complexity, the number of assets should exceed the number of observations in the training sample (even by many times the sample size). For example, if a portfolio model is trained based on 10 years of monthly data, the number of assets should be greater than 120. The empirical exercises in the portfolio optimization literature rarely use more than 100 assets to test model performance, making it unlikely to observe any gain associated with complex models.\footnote{For example, \cite{ao2019approaching} test portfolio performance with up to 100 assets; \cite{anderson2022portfolio} use up to 26 assets to test portfolio performance; \cite{barroso2022lest} use 50 assets to test portfolio performance; \cite{kan2022optimal} use up to 100 assets, and \cite{kan2024optimal} use up to 48 assets.}

We test our theory empirically using actual US equity returns. We construct a large cross-section of test assets by grouping the largest 1500 stocks into portfolios based on various characteristics. These characteristics-sorted portfolios serve as the building blocks for the portfolio optimization models that we test.\footnote{We only require information on stock characteristics to create test assets. Our portfolio models do not need information on the asset's characteristics. Grouping stocks with similar characteristics into portfolios helps to avoid regime shifts in the mean and variance of an asset, since it is well-known that a company's risk and return change over its life cycle (\cite{carlson2004corporate,zhang2005value}).} Each time, we choose $N$ assets from these building blocks and measure the out-of-sample Sharpe ratio of the estimate optimal portfolio. We gradually increase $N$ from 2 to 700. We use ten years of rolling monthly data to train these models. The empirical performance improves with model complexity in both the low-dimensional and high-dimensional regimes. In the low-dimensional regime ($N<120$), the optimal performance is achieved by the model that includes 39 different assets, and their out-of-sample annualized Sharpe ratio is around 0.8. In the high-dimensional regime ($N>120$), the optimal performance is achieved by models that include more than 450 assets, and the out-of-sample Sharpe ratio is around 1.4, significantly outperforming the best low-dimensional model. 

Our theory also allows us to estimate the clairvoyant Sharpe ratio, which is an important theoretical quantity in asset pricing. Based on the empirical performance curve, we compute the clairvoyant Sharpe ratio as a function of the number of assets. Our calibration shows that the upper limit of the clairvoyant Sharpe ratio is 1.44 annualized as the number of assets $N\rightarrow\infty$. When $N=39$, the clairvoyant Sharpe ratio is 1.11, about three-quarter of its upper bound. When $N=121$, entering the high-dimensional regime, the clairvoyant Sharpe ratio is 1.42, reaching 98\% of its upper bound. This shows that the benefit of complex models in the high-dimensional regime (when $N>120$) is largely due to the improvement in estimation accuracy, since additional gain in the clairvoyant Sharpe ratio is small. 

Overall, we develop a theoretical framework to understand the ``double descent'' phenomenon in portfolio optimization. Our framework attributes the success of large models to an economic mechanism and a statistics mechanism. We also show that the benefit of highly complex models depends on the underlying factor strength of asset returns. Further investigating the benefit of complexity under different estimators is an interesting area of future research.

\subsection{Contribution and Literature Review}
We contribute to the large literature that studies mean-variance portfolio optimization. Earlier works highlight the challenge of parameter estimation (\cite{kan2007optimal,demiguel2009optimal}) and propose various ways to mitigate the estimation error, for example, adding constraints to portfolio weights (\cite{jagannathan2003risk,demiguel2009generalized}), adopting the Bayesian approach in estimating the optimal portfolio (\cite{kan2007optimal,tu2011markowitz,bogle2010keynes,kan2022optimal}), applying shrinkage to the parameters (\cite{ledoit2017nonlinear,ao2019approaching}, and learning from past estimation errors (\cite{barroso2022lest}). Most of the proposed methods in this literature are designed for low-dimensional models in which the number of assets is smaller than the number of observations. Our paper highlights the potential gain of expanding model complexity to the high-dimensional regime. In particular, we show that the superior performance of the equal-weighted portfolio (\cite{demiguel2009optimal}) manifests in the high-dimensional regime through the use of $\ell_2$-norm based inductive bias. It is an interesting area of future research to study whether the estimation methods proposed in the literature can deliver further gain when combined with high-dimensional models. 

Our paper is also closely related to the rapidly growing literature that studies asset pricing in the age of artificial intelligence. Several papers have demonstrated the empirical success of machine learning models in cross-sectional stock return predictions (\cite{gu2020empirical,cong2021alphaportfolio,cong2023asset,bryzgalova2019forest,avramov2023machine,chen2024deep}), in predicting market returns (\cite{dong2022anomalies,liao2023economic,kelly2024virtue}), and in building efficient portfolios (\cite{jensen2024machine,kelly2024universal}).\footnote{For a broader review of this literature, see \cite{giglio2022factor,kelly2023financial}} A few recent papers delve deeper into the benefit of complex models. In particular, \cite{kelly2024virtue} show that expanding the complexity of the predictor space significantly improves the out-of-sample prediction of the market return, which they refer to as ``benign complexity''. \cite{liao2023economic} finds that adding a large number of noise variables into the prediction model can improve the out-of-sample prediction performance due to diversifying away the overall variance. \cite{didisheim2024apt} demonstrate ``double ascent'' in the performance of optimal factor portfolios and highlight the superiority of high-dimensional factor models in asset pricing. We contribute to this literature by developing a theory of complex portfolio based on the modeling of individual assets and derive the relationship between the theoretical Sharpe ratio and the observed Sharpe ratio.


Lastly, our paper builds on the ongoing research in machine learning theory. The ``double descent" risk curve was first proposed by \cite{belkin2019reconciling}. 
\cite{muthukumar2020harmless} provided a lower bound on the mean squared error and showed that this bound approaches zero as the number of features goes to infinity. \cite{hastie2022surprises} analyzed the asymptotic generalization behavior across a range of setups, including isotropic and correlated features. 
\cite{bartlett2020benign} studied non-asymptotic upper and lower bounds of the generalization error for  minimum norm interpolating estimator. 
\cite{belkin2020two} considered a misspecified setting for Gaussian and Fourier features and recovered the double descent phenomenon. 
\cite{mitra2019understanding} conducted an asymptotic analysis with a specific focus on the magnitude of the peak at the ``interpolation threshold ($N=T$)” for both $\ell_2$ and $\ell_1$ minimizing estimators. 
\cite{shi2022} relaxed $\ell_2$ optimization problems to
tackle forecast combination with many forecasts or minimum variance portfolio with many assets.
\cite{mei2022generalization} considered interpolating random feature regression and obtained the asymptotic behavior, thereby connecting regression models to neural networks. \cite{liang2020just} derived the learning risk of the interpolating estimator in kernel ridgeless regression. 
\cite{li2023benign}  expanded on the framework established by \cite{bartlett2020benign}, showing that adding noise to features effectively acts as a form of implicit regularization, which can induce the double descent phenomenon in the model. The review paper \cite{bartlett2021deep} considered two-layer networks and provided an exact asymptotic analysis of the impact of over-parameterization.

\subsection{Organization}
The rest of this paper is organized as follows. Section \ref{Sec:Problem} introduces the mean-variance portfolio problem and the generalized inverse estimation approach. Section \ref{Sec:Main_Results} studies the statistical properties of Sharpe ratio and mean squared error along the number of assets, the sample size, the clairvoyant Sharpe ratio and the factor strength. Section \ref{Section:induction} discusses the intuition behind the double ascent phenomenon in Sharpe ratio. Section \ref{Sec:empirical_rolling} reports empirical results on real data analysis. Section \ref{Section:Conclusion} concludes this paper and discusses potential future works related to this topic. All theoretical results are proved in the supplementary material, as well as some additional empirical analysis.

\section{Problem Setup}\label{Sec:Problem}
This section introduces our portfolio optimization models, our estimation strategy, and the performance metrics we use to evaluate different models. 

\subsection{Portfolio optimization model}
We consider a capital market comprising $N$ assets. We use the term asset in a general sense, which could be a single stock, a portfolio of stocks, or any trading strategy. The random excess returns of these assets are represented by the vector $\br =
\left(r_{1},r_{2},...,r_{N}\right)\T$, with $\bmu=\left(\mu_{1},\mu_{2},...,\mu_{N}\right)\T$ denoting their population mean. We denote $\bSig$ as the population covariance matrix of asset returns. The objective of the investor is to develop a portfolio strategy to maximize reward, for a given risk constraint $\sigma$
\begin{eqnarray}
\underset{\bome}{\arg\max}\text{ }\mathbb{E}\left(\bome\T\br\right)=\bome\T\bmu \qquad s.t.\quad \Var{\left({\bome}\T{\br}\right)}={\bome}\T\bSig{\bome}\leq {\sigma}^2,
\nonumber
\end{eqnarray}
where $\bome$ denotes an $N\times 1$ vector of portfolio weights. This optimization problem omits the sum-to-one constraint on portfolio weights, as well as leverage and short sale considerations, in order to emphasize on the core of the trade-off between reward and risk.\footnote{The optimizaton problem specified above is equivalent to maximizing a mean-variance utility function with a corresponding risk aversion coefficient that matches with the risk constraint $\sigma$.} 

We denote $\bomeast$ as the optimal portfolio. It admits the following explicit expression:
\begin{eqnarray}
\label{omega_star}
\bomeast=\dfrac{\sigma}{\sqrt{\theta}}\bSig\inv\bmu,
\end{eqnarray}
where $\theta=\bmu\T\bSig\inv\bmu$ denotes the square of the theoretical Sharpe ratio of the optimal portfolio. 

Important information on the design of a portfolio optimization model is encoded in Equation (\ref{omega_star}). In particular, the complexity of the model is determined by the number of assets, which determines the dimensionality of $\bSig$ and $\bmu$. Just as a prediction modeler should carefully decide how many variables to include in a regression model, portfolio modelers should choose how many assets to include in an optimizer with care. However, the literature has largely overlooked this decision, and the number of assets ranges from as low as 2 to a maximum of 100 in many papers.

\subsection{\texorpdfstring{Clairvoyant Sharpe ratio $\sqrt{\theta}$}%
{Clairvoyant squared Sharpe ratio theta}}

The parameter $\theta=\bmu\T\bSig\inv\bmu$ is an important parameter. It not only directly enters the optimal portfolio weight, but also plays a key role in model comparison. The square root of $\theta$ represents the highest attainable Sharp ratio by an investor if the true parameters are known to her. This is the theoretical Sharpe ratio that is not observable. Hence, we also refer to it as the clairvoyant Sharpe ratio. Two portfolio models with different levels of $\theta$ are likely to have different performance. Let us make this point more concrete with two examples. In the first example, there are $N$ assets with i.i.d. excess returns. Denote their mean and standard deviation as $u$ and $s$ respectively. The clairvoyant Sharpe ratio in this example depends on $N$
\begin{equation}
    \sqrt{\theta} =\frac{u}{s}\sqrt{N}.
\end{equation}
As $N$ approaches to infinity, $\theta$ also increases to infinity. Consider the second example, which is the CAPM world, the market portfolio has the highest Sharpe ratio, and all other assets have zero CAPM alpha and i.i.d. idiosyncratic risk. In this example, the clairvoyant Sharpe ratio is capped by the Sharpe ratio of the market portfolio and does not depend on the number of assets $N$ as long as the market index is part of the portfolio assets. 

The benefits of complexity in these two examples are entirely different. In the first example, adding more assets is more likely to generate better outcomes, whereas in the second example, adding more assets is likely to harm the performance. Later, we will prove that the out-of-sample performance depends on both $\theta$ and model complexity. When we compare the performance of different models, we take this effect into account. Even though $\theta$ is unobservable and its relationship with model complexity is ex ante unclear, our theory allows us to draw inferences about $\theta$ based on the observable empirical performance of models with different complexity.

\subsection{Estimation strategy}
For any model to work empirically, it needs to be estimated based on the data. Different estimation approaches affect the performance of a model. We focus on the pseudoinverse estimator to estimate Equation (\ref{omega_star}). Let $\hatbmu$ and $\hatbSig$ represent the sample mean and sample covariance matrix. Then, 
\begin{align}
\hatbmu=\dfrac{1}{T}\sum^{T}_{t=1}\br_t,\quad\text{and}\quad \hatbSig = \dfrac{1}{T-1}\sum^{T}_{t=1}\left(\br_t-\hatbmu\right)\left(\br_t-\hatbmu\right)\T.
\nonumber
\end{align}
Note that our application of sample mean and sample covariance matrix adheres to their standard, traditional definitions. We define our estimator as following:
\begin{definition}[Pseudoinverse Estimator]
Given the sample mean vector $\hatbmu$ and the sample covariance matrix $\hatbSig$ from the observed data, along with a specified risk constraint $\sigma$, we define the following estimator of the optimal portfolio:
\begin{align}\label{Def:pseudoinv-estimator}
\hatbomeastpseudo = \dfrac{\sigma}{\sqrt{\htheta}}\hatbSig\pseudoinv\hatbmu,
\end{align}
where $\hatbSig\pseudoinv$ is the pseudoinverse of $\hatbSig$, and $\htheta$ is a consistent estimator to $\theta$ that is detailed in Proposition \ref{Prop:htheta} and Proposition \ref{Prop:htheta_sfm}. If the eigen-decomposition of the sample covariance matrix $\hatbSig$ is given by $\sum_{i=1}^{N}\hattau_i\hatbupsilon_i\hatbupsilon_i\T$, where $\hatbupsilon_i$ are the eigenvectors and $\hattau
_i$ are the eigenvalues, satisfying $\hattau_1\geq\hattau_2\geq\dots\geq\hattau_N$. Then the pseudoinverse $\hatbSig\pseudoinv=\sum_{i=1}^{K}\hattau_i\inv\hatbupsilon_i\hatbupsilon_i\T$, where $K=min(T-1,N)$.
\end{definition}

Our estimator coincides with the standard plug-in estimator in the low-dimension case, and in the high-dimensional case, it replaces the inverse of the sample covariance matrix with its pseudoinverse. We have three main reasons for choosing this estimator:
\begin{enumerate}[labelindent=0pt,label=\roman*)]
\item  The pseudoinverse estimator is well-defined in the over-parameterization regime, which is important when we evaluate the performance of high-dimensional models.
\item Traditional regularization techniques, such LASSO or ridge 
regularization, explicitly reduce model complexity to prevent overfitting. Since our goal is to study model complexity, we want to use an estimator without explicit regularization. 
This is crucial for studying “benign overfitting” properties, as discussed in works like \cite{hastie2022surprises} and \cite{bartlett2020benign}. 
\item Our estimator aligns with the pseudoinverse estimator in least squares problems. The pseudoinverse estimator is linked to gradient descent methods, which are commonly used to train neural networks. It has been shown that gradient descent on least squares problems with zero initialization converges to the minimum norm solution (\cite{tibshiranioverparametrized}.\footnote{In least squares regression, let \(\by\) be the response vector and \(\bX\) the predictor matrix. The minimum-norm least squares estimator is given by: 
$\widehat{\bm{\beta}}=\left(\bX\T\bX\right)\pseudoinv\bX\T\by.$
Consider running gradient descent on the least squares loss function, initializing ${\bm{\beta}}^{(0)}=0$. The iterative updates are then:
\[
{\bm{\beta}}^{(k)}={\bm{\beta}}^{(k-1)}+\eta\bX\T\left(\by-\bX{\bm{\beta}}^{(k-1)}\right),\quad k=1,2,3,...,
\]
where $\eta$ is the step size. Gradient descent converges to the above minimum-norm solution, $\lim_{k\rightarrow\infty}{\bm{\beta}}^{(k)}=\widehat{\bm{\beta}}$. This convergence occurs because each iterate ${\bm{\beta}}^{(k)}$ lies in the row space of $\bX$, and the gradient updates preserve this property. Therefore, the limit (guaranteed to exist for a sufficiently small $\eta>0$) must also reside in the row space of $\bX$. The minimum-norm least squares solution is unique among all least squares solutions in that it lies entirely within the row space of \(\bX\), making it the solution to which gradient descent converges when initialized at zero.} This connection further relates our estimator to neural network models used in portfolio optimization, for example \cite{du2022mean} and \cite{snow2020machine}.
\end{enumerate}


\subsection{Performance evaluation metrics}

In regression settings, the common metric used to evaluate the performance of a model is the mean squared error (MSE). Portfolio optimization problems are different. We use two different metrics to evaluate portfolio optimization models. The first one is closely related to the MSE. The second one is the portfolio's Sharpe ratio. Both metrics are measured out-of-sample, assuming the out-of-sample data is independent from the training data. 

Our first performance metric is defined as the expected squared distance between the estimated portfolio return and the true optimal portfolio return. This metric is essentially the
MSE of the estimated portfolio on a test sample - a fundamental measure of ``risk” in statistical theory. This metric is also closely related to the tracking error of the estimated portfolio, an important performance metric in finance.

\begin{definition}
For an estimator $\hatbomeast$, which is a function of the observed data $\bR$, the out-of-sample prediction loss is measured by MSE, i.e.
\begin{align}\label{Def:prediction_loss}
L_{\bR}\left[\hatbomeast;\bomeast\right] &= \mathbb{E}\left[\left(\hatbomeast\T\br_0-{\bomeast}\T\br_0\right)^2\Big\vert\bR\right].
\end{align}
where $\bomeast$ is the true optimal portfolio in Equation (\ref{omega_star}) and $\textbf{r}_0$ is asset return independent of the observed data $\textbf{R}$.
\end{definition}

In simpler terms, this loss metric indicates that the closer an estimated portfolio's return is to the true optimal portfolio, the lower the associated risk. Note that while the loss is conditional on the sample realization $\bR$, as indicated by the subscript in $L_{\bR}$, we will show below that its limit is deterministic (non-random). 


Our second performance metric is the Sharpe ratio, a key objective in portfolio selection. The definition is as follows:
\begin{definition}
For an estimator $\hatbomeast$, which is a function of the observed data $\bR$, the out-of-sample Sharpe ratio is 
\begin{equation}
SR\left[\hatbomeast\right]=\dfrac{\hatbomeast\T\bmu}{\sqrt{\hatbomeast\T\bSig\hatbomeast}}.\nonumber   
\end{equation}
\end{definition}

It’s important to note that the loss metric in Equation (\ref{Def:prediction_loss}) and the Sharpe ratio, although are closely related, are distinct measures. Each 
metric captures different aspects of portfolio performance from both statistical and financial perspectives.

\section{Theoretical Results}\label{Sec:Main_Results}

This section presents our primary theoretical findings, focusing on two different scenarios. In the first scenario, we assume that asset returns are uncorrelated, the simplest case possible. In the second scenario, we extend the analysis by introducing a factor structure among the asset returns. Building on the theoretical findings, we present a case designed to mimic real-world conditions in subsection \ref{Subsection:Double_Descent}, providing a theoretical explanation for the emergence of the double ascent (in Sharpe ratio) phenomenon.


Our asymptotic framework deviates from the traditional setting where the number of observations $T\rightarrow\infty$. Instead, we consider a high-dimensional regime where both the number of assets $N$ and observations $T$ approach infinity. The asymptotic results are grounded in key concepts from random matrix theory (RMT), which provides a rigorous framework to establish the asymptotic relationship between the true covariance matrix and its sample estimators. The detailed proofs of our results are provided in the supplementary material available online.

\subsection{Scenario 1: uncorrelated assets} 
We begin by considering the simpler case in which $\bSig=\bI_N$, where the assets are uncorrelated.
\subsubsection{Assumptions}
\begin{assumption}[High Dimensionality]\label{ASSU1:HD}
We consider asymptotic setup where $T, N\rightarrow\infty$, and $\rho_T\coloneq N/T\rightarrow\rho\in(0,1)\cup (1,\infty)$. 
\end{assumption}
The case $\rho=1$ is excluded from theoretical analysis due to technical constraints, as $1-\rho$ appears in the denominator of certain limiting expressions. However, this scenario works effectively in practical applications, as evidenced by simulation and empirical studies.

\begin{assumption}[Factorless Model]\label{ASSU2:DGP}
The excess returns vector at time $t$ is generated as $\br_t=\bmu+\by_t$, where $\by_t$ is a vector of $i.i.d.$ random variables with zero mean, unit variance and bounded $4^{th}$ moment. 
\end{assumption}

This assumption is common in financial and statistical problems and does not impose any strong restrictions. Asset returns are generally robust to deviations from normality. The requirement of the existence of the $4^{th}$ moment is a technical necessity. 

\begin{assumption}[Constant Scale]\label{ASSU3:MSR}
Assume that the square of theoretical Sharpe ratio, ${\theta}$ (equivalently, ${\bmu\T\bSig\inv\bmu}$), converges to ${\ttheta}$ as $N$ increases to infinity, where ${\ttheta}$ is bounded away from 0 and infinity.
\end{assumption}

This is a technical assumption that bridges the gap between finite sample observations and asymptotic limits. Since our asymptotic theory is based on the idea that both $N$ and $T$ go to infinity (while keeping the ratio constant), to ensure the asymptotic limit exists, we require the squared Sharpe ratio $\theta$ converges to a constant in the limit as $N\rightarrow\infty$. The easiest way to imagine this is that suppose we run a series of simulation experiments with ever increasing $N$ and $T$, in each experiment, we keep ${\bmu\T\bSig\inv\bmu}$ fixed at ${\ttheta}$. One should not confuse the $\theta$ in this simulation exercise with the actual clairvoyant Sharpe ratio in the real world, which depends on the number of assets.   


\subsubsection{\texorpdfstring{\textcolor{black}{ Asymptotic results for uncorrelated assets}}%
{Case 1 }}

\begin{theorem}[Sharpe ratio]
\label{Thm:sr}
Under Assumptions \ref{ASSU1:HD}-\ref{ASSU3:MSR}, it holds {in probability} that
\begin{equation}\label{ThmRslt:SR}
SR\left[\hatbomeastpseudo\right] \rightarrow SR\left(\ttheta,\rho\right)=
\left\{
    \begin{aligned}
    &\sqrt{\dfrac{1-\rho}{\ttheta+\rho}}\ttheta, & \text{for } \rho < 1, \\
    &\sqrt{\dfrac{\rho-1}{\ttheta+\rho}}\dfrac{\ttheta}{\rho}, & \text{for } \rho > 1.
    \end{aligned}
\right.
\end{equation}
\end{theorem}

\begin{figure}[H]
    \centering    \includegraphics[trim=0 0 0 0,width=0.7\textwidth]{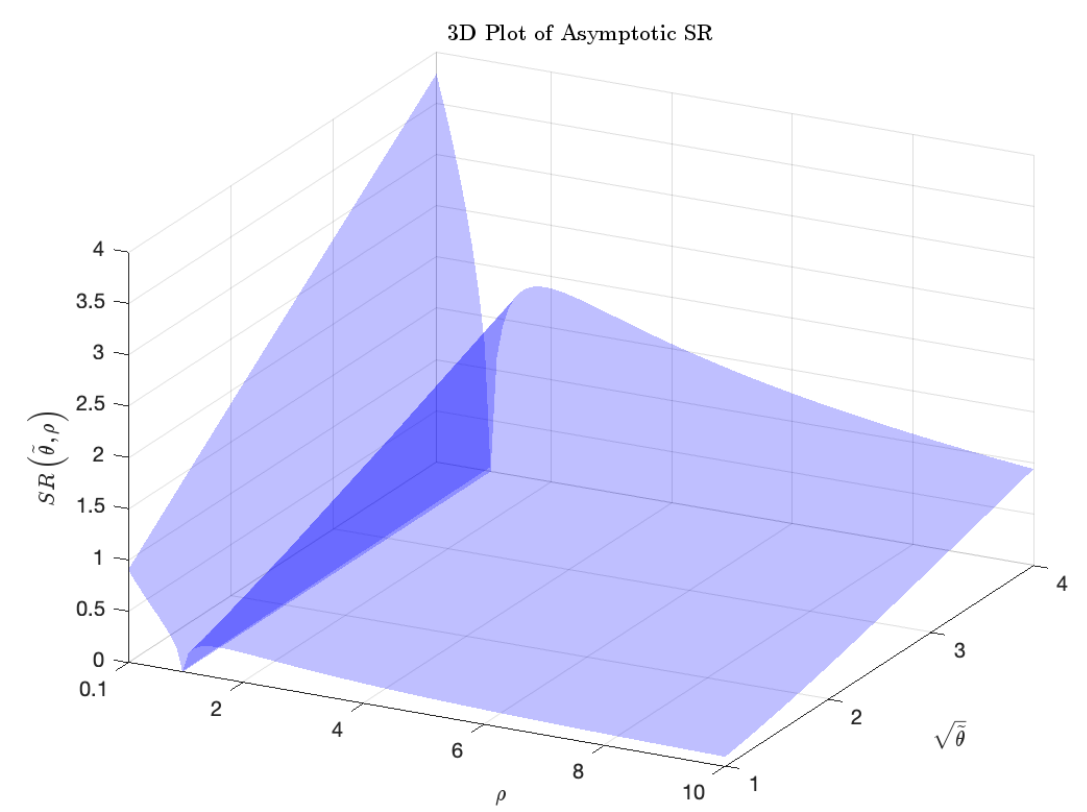}
    \caption{Asymptotic Sharpe ratio 3D heatmap in Theorem \ref{Thm:sr}, with $\sqrt{\ttheta}$ ranging from 1 to 4 and $\rho \in (0,1)\cup(1,10)$.}
    \label{Fig:Asy_SR}
\end{figure}

As shown, the asymptotic Sharpe ratio depends solely on the two limiting values $\rho$ and $\ttheta$. Figure \ref{Fig:Asy_SR} presents a 3D heatmap illustrating the results of Theorem \ref{Thm:sr}, showing how the asymptotic Sharpe ratio varies across different combinations of $\rho$ and $\sqrt{\ttheta}$. The parameter $\sqrt{\ttheta}$ ranges from 1 up to 4. In the region where $\rho \in (0, 1)$, the asymptotic Sharpe ratio decreases toward zero as $\rho$ increases. In contrast, when $\rho \in (1, \infty)$, the asymptotic Sharpe ratio initially increases, reaching approximately half of the theoretical maximum Sharpe ratio $(\sqrt{\ttheta})$, before gradually decreasing again toward zero. This reveals a surprising ascent in the high-dimensional regime. 

\begin{figure}[H]
    \centering    \includegraphics[trim=0 0 0 0,width=0.5\textwidth]{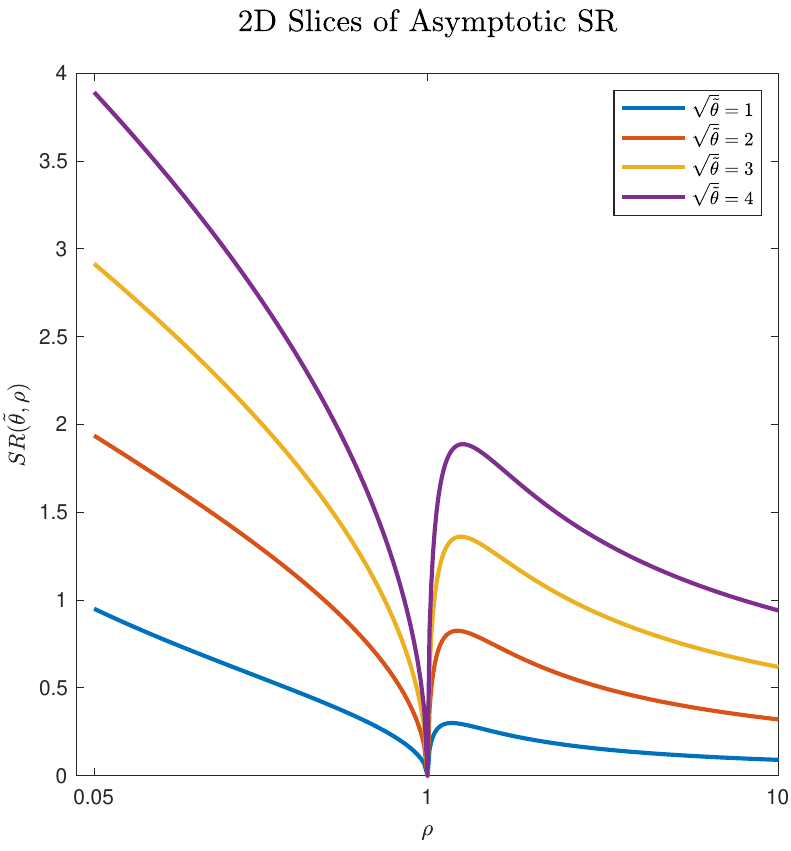}
    \caption{Asymptotic Sharpe ratio curves as a function of $\rho$, for $\sqrt{\ttheta} = 1, 2, 3, 4$.}
    \label{Fig:2D_Asy_SR}
\end{figure}

For better visualization, Figure \ref{Fig:2D_Asy_SR} presents 2D slices from Figure \ref{Fig:Asy_SR}, illustrating how the asymptotic Sharpe ratio varies with $\rho$ for selected values of \(\sqrt{\ttheta}\).

At first glance, Figure \ref{Fig:2D_Asy_SR} suggests that the best outcome is achieved with the simplest model, as the out-of-sample Sharpe ratio is highest when $\rho$ is at its lowest. This occurs because this figure is only about estimation accuracy. When $\rho$ is near zero, the problem approximates a large-sample scenario where estimation error becomes asymptotically negligible. However, as discussed later in Subsection \ref{Subsection:Double_Descent}, we present a case where the highest out-of-sample Sharpe ratio may occur in the $\rho > 1$ regime. The details of this phenomenon are explored there.


What is important in this figure is the ``ascent'' observed in the high-dimensional region, when $\rho>1$. From a finance perspective, the observed ``ascent" suggests that increasing the number of assets beyond the sample size can enhance the out-of-sample risk-adjusted returns of the estimated portfolio, challenging the traditional belief that over-parameterization leads to poorer performance due to estimation errors. Instead, including more assets can improve portfolio performance - even when the number of assets exceeds the sample size. As dimensionality continues to increase, the Sharpe ratio eventually declines, highlighting a threshold beyond which further complexity becomes detrimental due to error accumulation. 

Next, we introduce a consistent estimator for the clairvoyant Sharpe ratio and present the asymptotic results for the prediction loss.

\begin{proposition}
\label{Prop:htheta}
Under Assumptions \ref{ASSU1:HD}-\ref{ASSU3:MSR}, it holds {in probability} that
\begin{equation}
\htheta\coloneq
\left\{
    \begin{aligned}
    &(1-\rho_T)\htheta_s-\rho_T, & \text{for } \rho_T < 1 \\
    &\rho_T\left[(\rho_T-1)\htheta_s-1\right], & \text{for } \rho_T > 1
    \end{aligned}
\right.\rightarrow\ttheta,
\end{equation}
where $\htheta_s=\hatbmu\T\hatbSig\pseudoinv\hatbmu$.
\end{proposition}
The estimator $\htheta$ for the case $\rho_T < 1$ has been discussed in \cite{kan2007optimal}. Here, we extend it to the case $\rho_T > 1$ under our specific setting. And this completes our pseudoinverse estimator (\ref{Def:pseudoinv-estimator}).

\begin{theorem}[Out-of-sample prediction loss]
\label{Thm:pl}
Under Assumptions \ref{ASSU1:HD}-\ref{ASSU3:MSR}, it holds {in probability} that
\begin{equation}
L_{\bR}\left[\hatbomeastpseudo;{\bomeast}\right]\rightarrow L_{\bR}\left(\ttheta,\rho,\sigma^2\right) =
\left\{
    \begin{aligned}
        &\sigma^2\left[\dfrac{\ttheta+\rho}{(1-\rho)^3\ttheta}+\dfrac{\ttheta}{(1-\rho)^2}-\dfrac{2(\ttheta+1)}{1-\rho}+\ttheta+1\right], & \text{for } \rho < 1, \\
        &\sigma^2\left[\dfrac{\ttheta+\rho}{(\rho-1)^3\ttheta}+\dfrac{\ttheta}{\rho^2(\rho-1)^2}-\dfrac{2(\ttheta+1)}{\rho(\rho-1)}+\ttheta+1\right], & \text{for } \rho > 1.
    \end{aligned}
\right.
\end{equation}
\end{theorem}

\begin{figure}[H]
    \centering    \includegraphics[trim=0 0 0 0,width=0.7\textwidth]{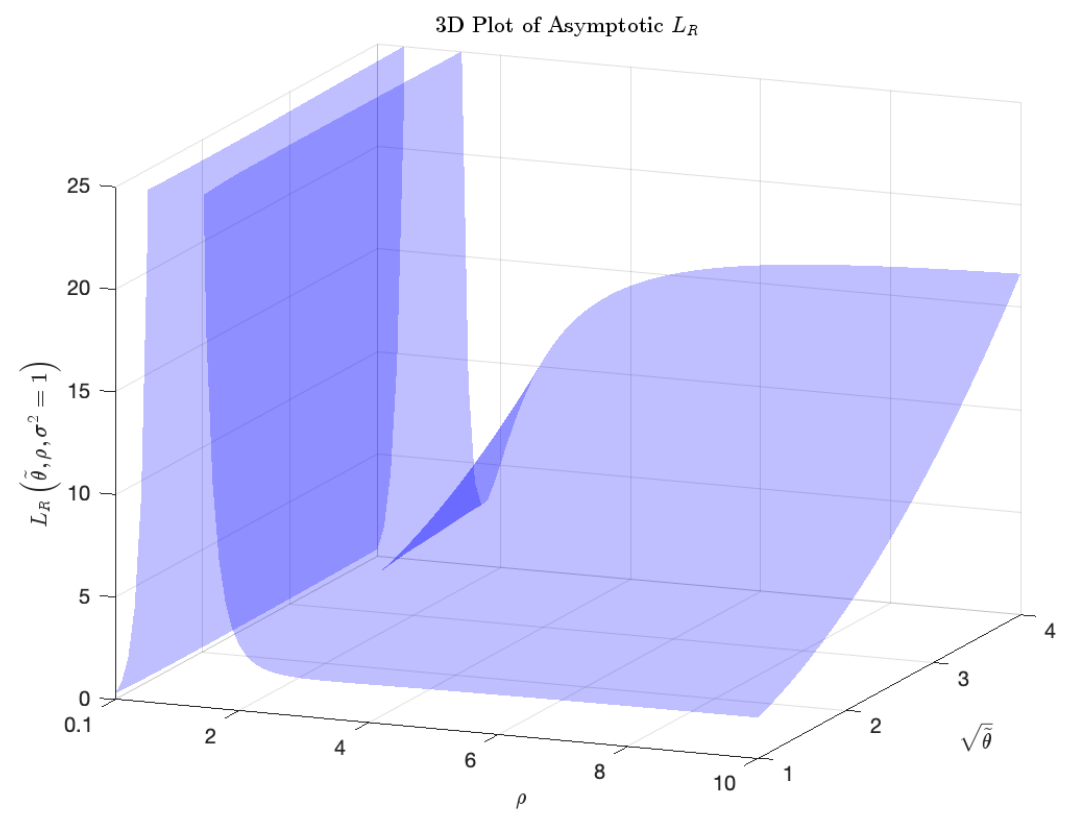}
    \caption{Asymptotic prediction loss 3D heatmap in Theorem \ref{Thm:pl}, with $\sqrt{\ttheta}$ ranging from 1 to 4 and $\rho \in (0,1)\cup(1,10)$.}
    \label{Fig:Asy_PL}
\end{figure}

In our mean-variance portfolio formulation, $\sigma^2$ represents the specified risk constraint. It is straightforward to show that the asymptotic limits of the prediction loss are proportional to $\sigma^2$, given that the optimal portfolio (see Equation (\ref{omega_star})) is proportional to $\sigma$. Similar to Theorem \ref{Thm:sr}, Theorem \ref{Thm:pl} illustrates how the asymptotic limits of the prediction loss is determined by the interaction between $\ttheta$ and $\rho$. These findings are visualized in Figure \ref{Fig:Asy_PL}. Notably, a decrease in prediction loss is observable within the range $\rho \in (1, \infty)$. Furthermore, the prediction loss converges toward $\sigma^2(1 + \ttheta)$.

\begin{figure}[H]
    \centering    \includegraphics[trim=0 0 0 0,width=0.5\textwidth]{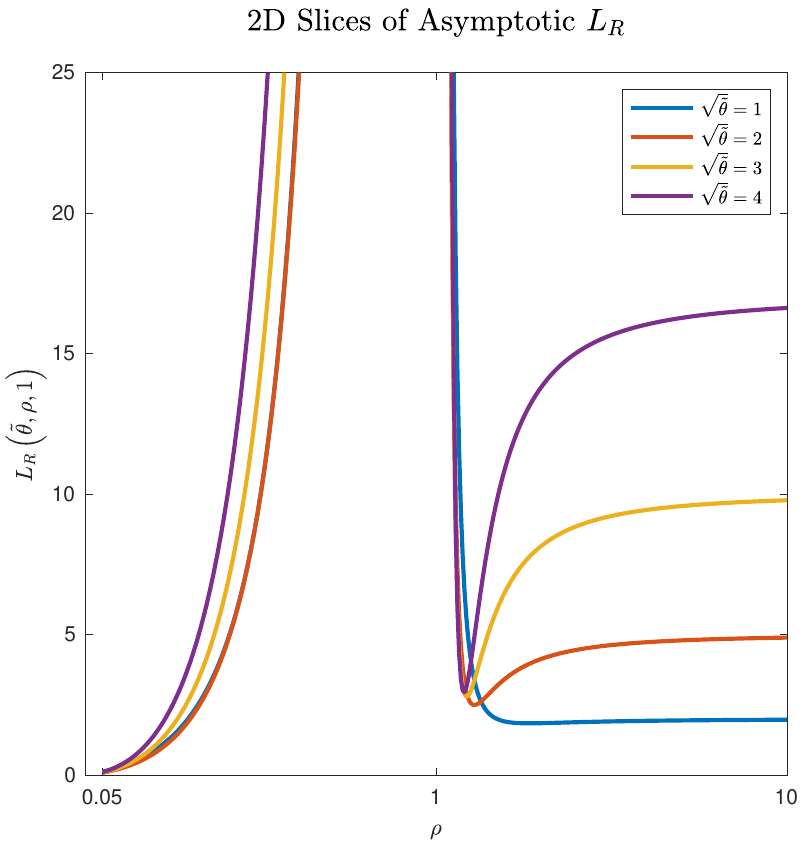}
    \caption{Asymptotic prediction loss curves as a function of $\rho$, for $\sqrt{\ttheta} = 1, 2, 3, 4$.}
    \label{Fig:2D_Asy_PL}
\end{figure}

Figure \ref{Fig:2D_Asy_PL} shows 2D slices from Figure \ref{Fig:Asy_PL}, illustrating how the asymptotic prediction loss varies with $\rho$ for selected values of \(\sqrt{\ttheta}\).

The decrease in prediction loss observed within the range $\rho \in (1, \infty)$ indicates that increasing model complexity by including more assets than observations can enhance the predictive accuracy of portfolio returns - contrary to traditional concerns about overfitting in high-dimensional settings. Statistically, this supports the concept of benign overfitting, where over-parameterized models generalize well despite having more parameters than data points. Financially, improved prediction accuracy means that the estimated portfolio’s returns have lower tracking error with respect to the true optimal portfolio, leading to potentially superior performance.

\subsection{Scenario 2: correlated assets - single factor model}\label{subsec:Scenario2}

In this subsection, we extend our analysis by considering asset returns that exhibit cross-sectional correlations through a factor model. Our primary focus is on understanding how variations in factor strength impact the asymptotic behavior of mean-variance portfolios, especially in the high-dimensional regime. We choose to use a single factor model, as it simplifies the model structure, enhances interpretability and allows for clear visualization, even though the underlying analysis remains technically complex and challenging.

Consider the following model:
\begin{equation}\label{Model:factor_model}
\br_t=\bb f_t+\bepsilon_t,\quad 1\leq t\leq T,
\end{equation}
where $f_t$ is the factor returns at time $t$, $\bb=(b_{1},...,b_N)$ is the $N\times 1$ vector of exposure to the factor risks, and $\bepsilon_t$ is the $N\times 1$ vector of zero-mean idiosyncratic returns. Let $\mu_f$ and $\sigma_f^2$ be the mean and variance of factor returns, respectively, and let $\sigmae^2\bI_N$ represent the covariance structure of idiosyncratic returns. Then the return vector $\br_t$ has the following mean and covariance:
\begin{equation}
\bmu=\bb\mu_f,\quad\bSig=\sigma_f^2\bb\bb\T+\sigmae^2\bI_N.\nonumber
\end{equation}
Under this model, the square of the theoretical Sharpe ratio of the optimal portfolio can be expressed in factor and idiosyncratic components:
\begin{equation}
\theta=\bmu\T\bSig\inv\bmu=\frac{\mu_f^2\left\Vert\bb\right\Vert^2}{\sigmae^2+\sigma_f^2\left\Vert\bb\right\Vert^2},\label{theta_expression_with_factor_components}
\end{equation}
where $\left\Vert\cdot\right\Vert$ is the $\ell_2$ norm.

\subsubsection{Assumptions}
\begin{assumptionB}[High Dimensionality]\label{ASSU_B1:HD}
We consider asymptotic setup where $T, N\rightarrow\infty$, and $\rho_T\coloneq N/T\rightarrow\rho\in(0,1)\cup (1,\infty)$. 
\end{assumptionB}

\begin{assumptionB}[Factor Model]\label{ASSU_B2:DGP}
The excess returns vector at time $t$ can be considered as $\br_t=\bb\mu_f + \bb\sigma_f\cdot z_t + \sigmae\cdot\by_t$, where $z_t$ and $\by_t$ are independent, $\by_t$ is a vector of $i.i.d.$ random variables with zero mean, unit variance and bounded $4^{th}$ moment, $z_t$ is a random variable with zero mean, unit variance and bounded $4^{th}$ moment.
\end{assumptionB}

\begin{assumptionB}[Constant Scale]\label{ASSU_B3:MSR}
Assume that the square of theoretical Sharpe ratio, ${\theta}$ (equivalently, ${\bmu\T\bSig\inv\bmu}$), converges to ${\ttheta}$ as $N$ increases to infinity, where ${\ttheta}$ is bounded away from 0 and infinity.
\end{assumptionB}

Combining Assumption \ref{ASSU_B3:MSR} with Equation (\ref{theta_expression_with_factor_components}) for $\theta$, it is evident that ${\theta} \rightarrow {\ttheta}$ directly implies that the norm $\left\Vert \bb \right\Vert^2$ converges to a constant as $N$ approaches infinity. We denote the limit for $\left\Vert \bb \right\Vert^2$ as $\tilde{b}^2$, i.e., $\left\Vert \bb \right\Vert^2 \rightarrow \tilde{b}^2$, where $\tilde{b}^2$ is bounded away from 0 and infinity. We also define the signal to noise ratio $\phi = \frac{\sigma_f^2\left\Vert \bb \right\Vert^2}{\sigmae^2}$. Therefore, $\phi\rightarrow\tphi = \frac{\sigma_f^2\tilde{b}^2}{\sigmae^2}$.

\subsubsection{Asymptotic behaviours}
\begin{theorem}[Sharpe ratio]
\label{Thm:sr_sfm}
Under Assumptions \ref{ASSU_B1:HD}-\ref{ASSU_B3:MSR}, it holds {in probability} that
\begin{equation}\label{ThmRslt:SR_sfm}
SR\left[\hatbomeastpseudo\right] \rightarrow SR\left(\ttheta,\tphi,\rho\right)=
\left\{
\begin{aligned}
&\sqrt{\dfrac{1-\rho}{\ttheta+\rho}}\ttheta, & \text{for } \rho < 1, \\
&\frac{\ttheta\sqrt{\rho(\rho-1)}}{\sqrt{\left[\frac{\rho+\tphi}{\tphi+1}\right]^4+\ttheta\rho\left[\frac{\rho+\tphi}{\tphi+1}\right]^2+\ttheta\tphi^2\frac{(\rho-1)^2}{(\tphi+1)^2}}}, & \text{for } \rho > 1.
\end{aligned}
\right.
\end{equation}
When $\tphi = 0$, indicating the absence of a common factor, the result simplifies to Theorem \ref{Thm:sr}.
\end{theorem}

\begin{figure}[H]
    \centering    \includegraphics[trim=0 0 0 0,width=0.7\textwidth]{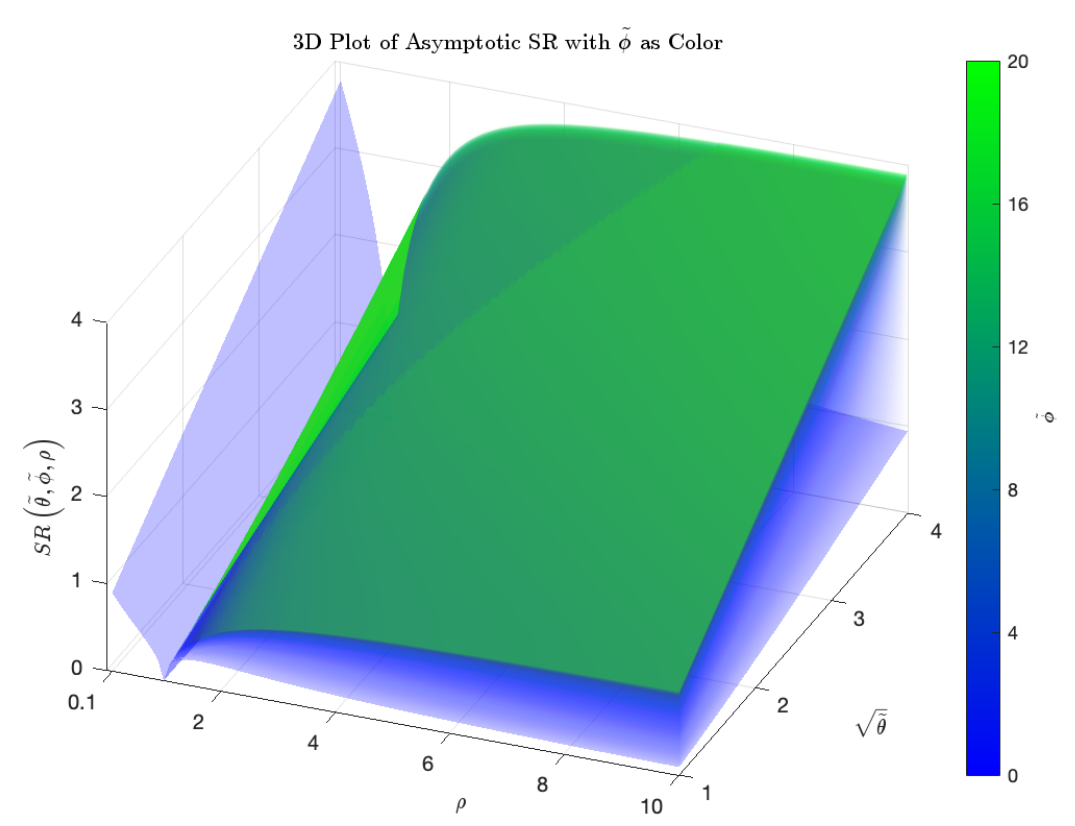}
    \caption{Asymptotic Sharpe ratios 3D heatmap in Theorem \ref{Thm:sr_sfm}, with $\tphi$ ranging from 0 to 20, $\sqrt{\ttheta}$ from 1 to 4 and $\rho \in (0,1)\cup(1,10)$.}
    \label{Fig:Asy_SR_sfm_factor_strength}
\end{figure}

The 3D Figure \ref{Fig:Asy_SR_sfm_factor_strength} reveals a distinct ascent in the $\rho > 1$ regime. Notably, as the factor strength (signal-to-noise ratio) increases, the asymptotic Sharpe ratio rises significantly. When $\tphi$ becomes sufficiently large, the asymptotic Sharpe ratio converges rapidly to the limit of the clairvoyant Sharpe ratio.


\begin{figure}[H]
    \centering    \includegraphics[trim=0 0 0 0,width=0.5\textwidth]{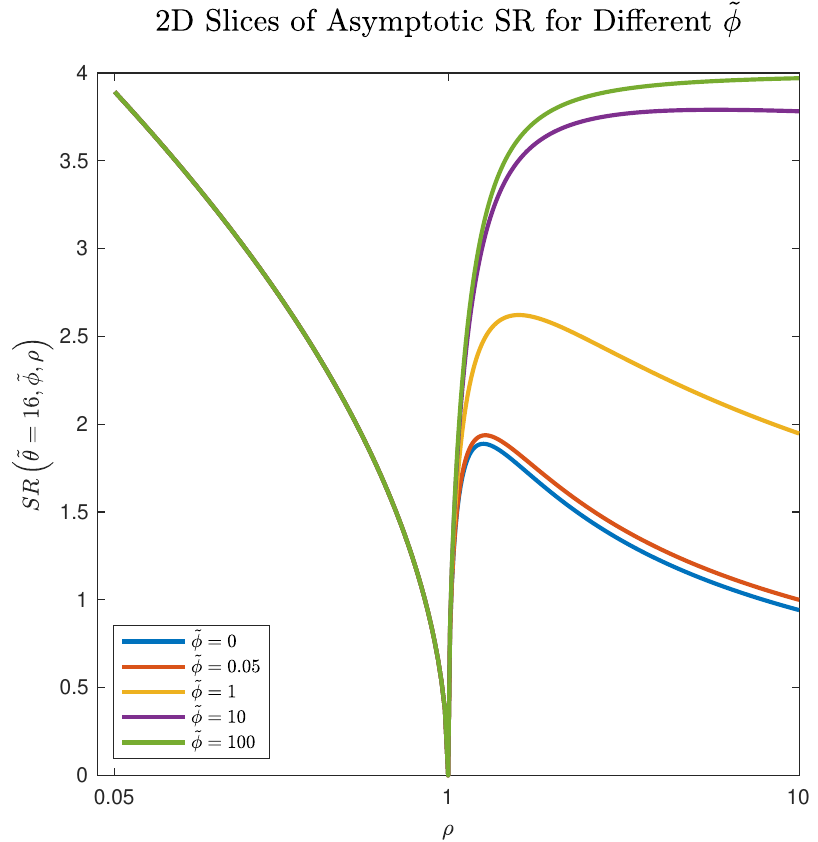}
    \caption{Asymptotic Sharpe ratio curves as a function of $\rho$, for $\tphi=0,0.05,1,10,100$ and $\sqrt{\ttheta}=4$.}
    \label{Fig:2D-Asy_SR_sfm_factor_strength}
\end{figure}

Figure \ref{Fig:2D-Asy_SR_sfm_factor_strength} presents 2D slices of the asymptotic Sharpe ratio for different values of $\tphi$ with a particular value of $\sqrt{\ttheta}$. This figure highlights the variation in the asymptotic Sharpe ratio as the factor strength changes.

From a financial standpoint, the presence and strength of a common factor ($\tphi$) significantly influence a portfolio’s risk-adjusted performance in high-dimensional settings. As $\tphi$ increases, assets become more correlated, allowing the portfolio to better capture systematic risk factors and thereby enhancing the Sharpe ratio. This improvement indicates that incorporating common market influences can lead to superior performance, especially when the number of assets exceeds the sample size.

\begin{remark} \label{Remark1}
    Interestingly, for the case $\rho < 1$, the asymptotic result is identical to that of the no-factor case. From a technical standpoint, this is because $\hatbSig$ is always invertible, so that $\hatbSig\inv=\bSig^{-\frac{1}{2}}\calS\inv\bSig^{-\frac{1}{2}}$ and the terms $\bSig^{-\frac{1}{2}}$ on the left and right will be canceled out by its quadratic terms. For example, a key term involved in the proof, $\hatbsm\T\hatbSig\inv\hatbsm$ (first appears in Appendix~\ref{Appendix:proofs_P1T1T2}), simplifies to $\left(\bSig^{\frac{1}{2}}\bY\T\be/T\right)\T\bSig^{-\frac{1}{2}}\calS\inv\bSig^{-\frac{1}{2}}\left(\bSig^{\frac{1}{2}}\bY\T\be/T\right)=\left(\bY\T\be/T\right)\T\calS\inv\left(\bY\T\be/T\right)$. Thus, regardless of the form of $\bSig$, we obtain the same asymptotic result. For the case $\rho > 1$, Theorem \ref{Thm:sr} is a special case of Theorem \ref{Thm:sr_sfm} when $\tphi = 0$. The same applies to the prediction loss asymptotics in Theorem \ref{Thm:pl_sfm}.
\end{remark}

The following proposition provides a consistent estimator for the clairvoyant Sharpe ratio under the factor model. Subsequently, we present asymptotic results for the prediction loss within the framework of the factor model.
\begin{proposition}
\label{Prop:htheta_sfm}
Under Assumptions \ref{ASSU_B1:HD}-\ref{ASSU_B3:MSR}, it holds {in probability} that
\begin{equation}\label{Eqn:htheta}
\htheta\coloneq
\left\{
    \begin{aligned}
    &(1-\rho_T)\htheta_s-\rho_T, & \text{for } \rho_T < 1 \\
    &\frac{(\phi+\rho_T)^2}{(\phi+1)^2\rho_T}\left[(\rho_T-1)\htheta_s-1\right], & \text{for } \rho_T > 1
    \end{aligned}
\right.\rightarrow\ttheta,
\end{equation}
where 
$
\htheta_s=\hatbmu\T\hatbSig\pseudoinv\hatbmu.
$
\end{proposition}

\begin{theorem}[Out-of-sample prediction loss]
\label{Thm:pl_sfm}
Under Assumptions \ref{ASSU_B1:HD}-\ref{ASSU_B3:MSR}, it holds {in probability} that
\begin{equation}
L_{\bR}\left[\hatbomeastpseudo;{\bomeast}\right] \rightarrow L_{\bR}\left(\ttheta,\tphi,\rho,\sigma^2\right)=
\left\{
\begin{aligned}
&\sigma^2\left[\dfrac{\ttheta+\rho}{(1-\rho)^3\ttheta}+\dfrac{\ttheta}{(1-\rho)^2}-\dfrac{2(\ttheta+1)}{1-\rho}+\ttheta+1\right],\\
& \qquad\qquad\qquad\qquad\qquad\qquad\qquad\qquad\qquad\qquad\text{for } \rho < 1, \\
&\sigma^2\left[\frac{(\tphi+1)^4\rho^2\ttheta}{(\tphi+\rho)^4(\rho-1)^2}+\frac{(\tphi+1)^2\tphi^2\rho}{(\tphi+\rho)^4(\rho-1)}+\frac{(\tphi+1)^2\rho^2}{(\tphi+\rho)^2(\rho-1)^3}\right.\\
& \left.\quad+\frac{\rho}{\ttheta(\rho-1)^3} -\frac{2(\tphi+1)^2\rho(\ttheta+1)}{(\tphi+\rho)^2(\rho-1)}+1+\ttheta\right],\\
& \qquad\qquad\qquad\qquad\qquad\qquad\qquad\qquad\qquad\qquad\text{for } \rho > 1.
\end{aligned}
\right.
\end{equation}
When $\tphi = 0$, indicating the absence of a common factor, the result simplifies to Theorem \ref{Thm:pl}.
\end{theorem}

\begin{figure}[H]
    \centering    \includegraphics[trim=0 0 0 0,width=0.7\textwidth]{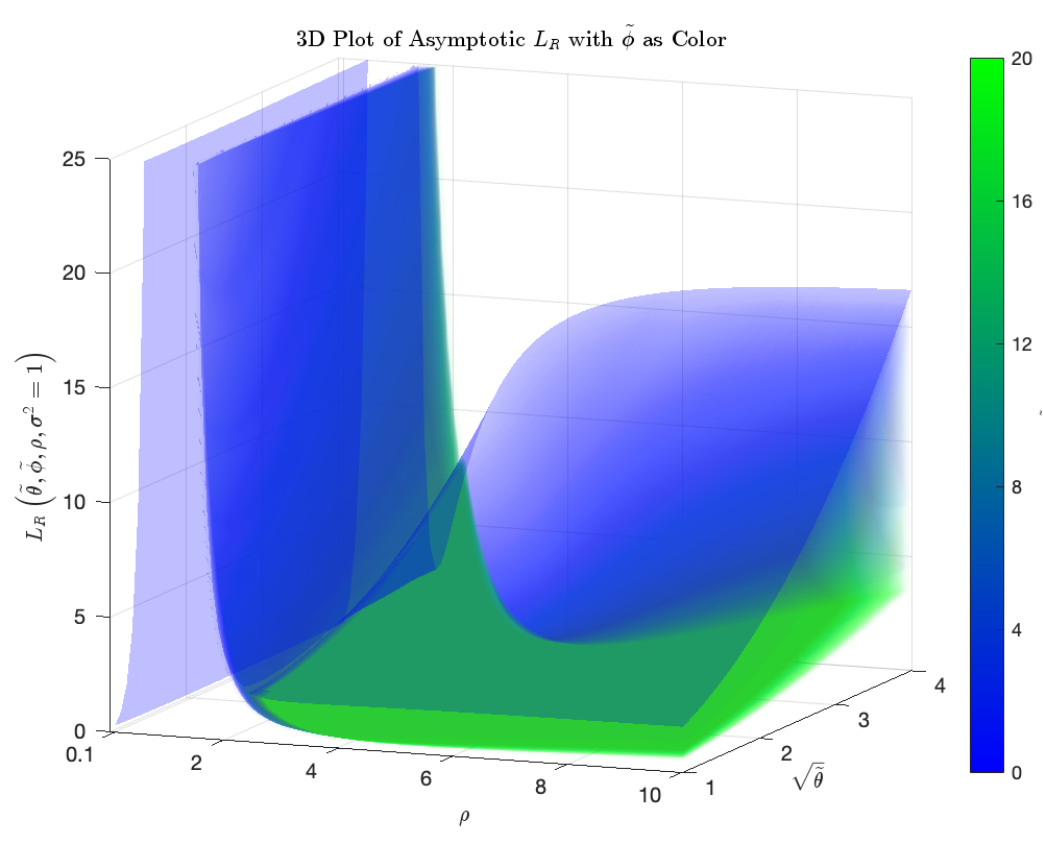}
    \caption{Asymptotic prediction loss 3D heatmap in Theorem \ref{Thm:pl_sfm}, with $\tphi$ ranging from 0 to 20, $\sqrt{\ttheta}$ from 1 to 4 and $\rho \in (0,1)\cup(1,10)$.}
    \label{Fig:Asy_LR_sfm_factor_strength}
\end{figure}

\begin{figure}[H]
    \centering    \includegraphics[trim=0 0 0 0,width=0.5\textwidth]{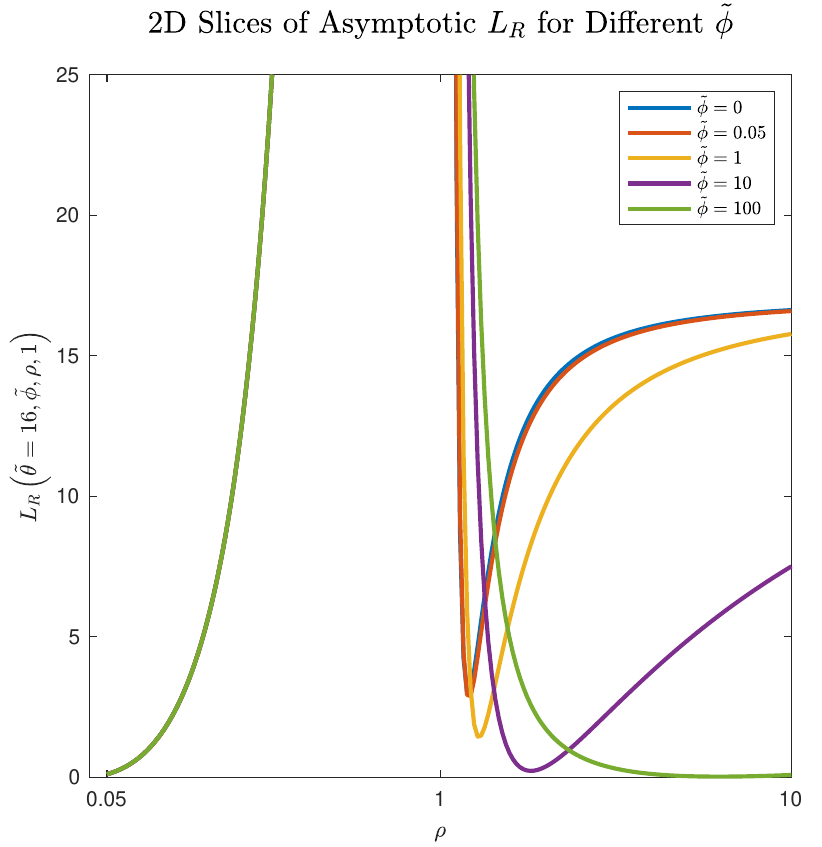}
    \caption{Asymptotic prediction loss curves as a function of $\rho$, for $\tphi=0,0.05,1,10,100$ and $\sqrt{\ttheta}=4$.}
    \label{Fig:2D-Asy_LR_sfm_factor_strength}
\end{figure}

The findings are visualized in Figure \ref{Fig:Asy_LR_sfm_factor_strength} and Figure \ref{Fig:2D-Asy_LR_sfm_factor_strength}. It is obvious that in the large $rho$ regime, stronger factors result in lower prediction loss.


\subsection{\texorpdfstring{Mimicking real-world: the emergence of double ascent}%
{Allowing theta to change with model complexity }}
\label{Subsection:Double_Descent}



The previous sections show how the out-of-sample Sharpe ratio changes with model complexity holding constant the clairvoyant Sharpe ratio. In the real world, the clairvoyant Sharpe ratio also changes with the model complexity, i.e., the nubmer of assets. Here, we present a scenario that mimics real-world cases to illustrate the emergence of the ``double ascent'' phenomenon. In this example, we consider the number of assets increasing from $2$ to $1000$, and assume that the clairvoyant Sharpe ratio, $\theta$, is an increasing function of $N$ that decays exponentially:
\begin{equation}\label{Eqn:simulated_thetaN}
    \theta(N) = 1 + 15\left(1-e^{-0.05N}\right).
\end{equation}

\begin{figure}[H]
    \centering    \includegraphics[trim=0 0 0 0,width=0.5\textwidth]{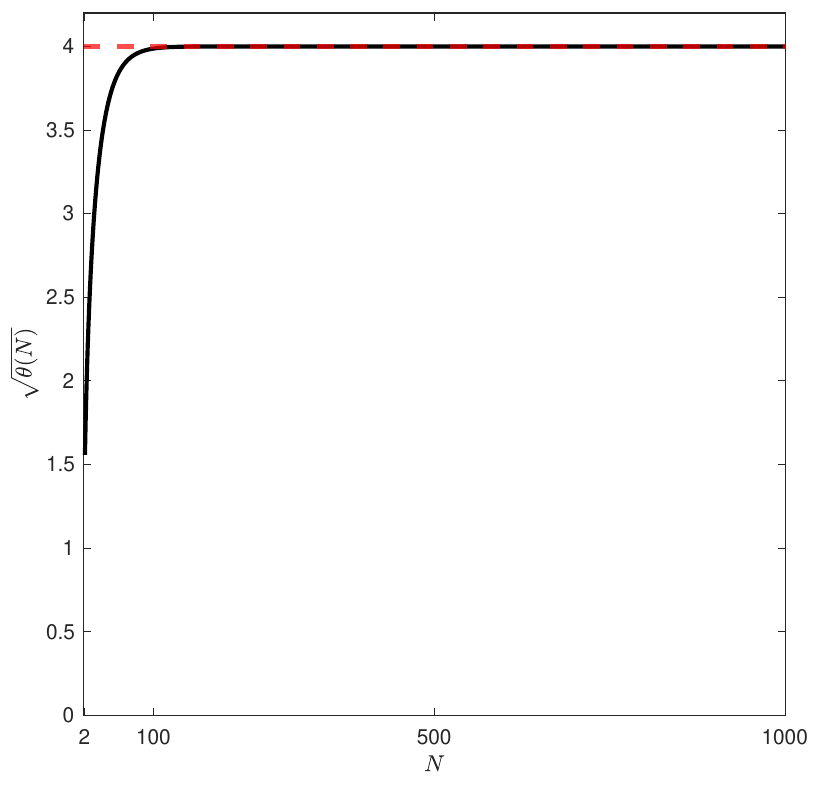}
    \caption{Simulated increase clairvoyant Sharpe ratio to its limit.}
    \label{Fig:theta_p_1}
\end{figure}
This example is only for illustration purpose. Later, we take our model more seriously to estimate the relationship between the clairvoyant Sharpe ratio and the number of assets. 
If we consider a fixed sample size of $T = 100$, the corresponding asymptotic behaviors for $\ttheta$ given by Equation (\ref{Eqn:simulated_thetaN}) and $\rho = 0.05 \text{ to } 10$ are presented below.


\begin{figure}[H]
    \centering    \includegraphics[trim=0 0 0 0,width=1\textwidth]{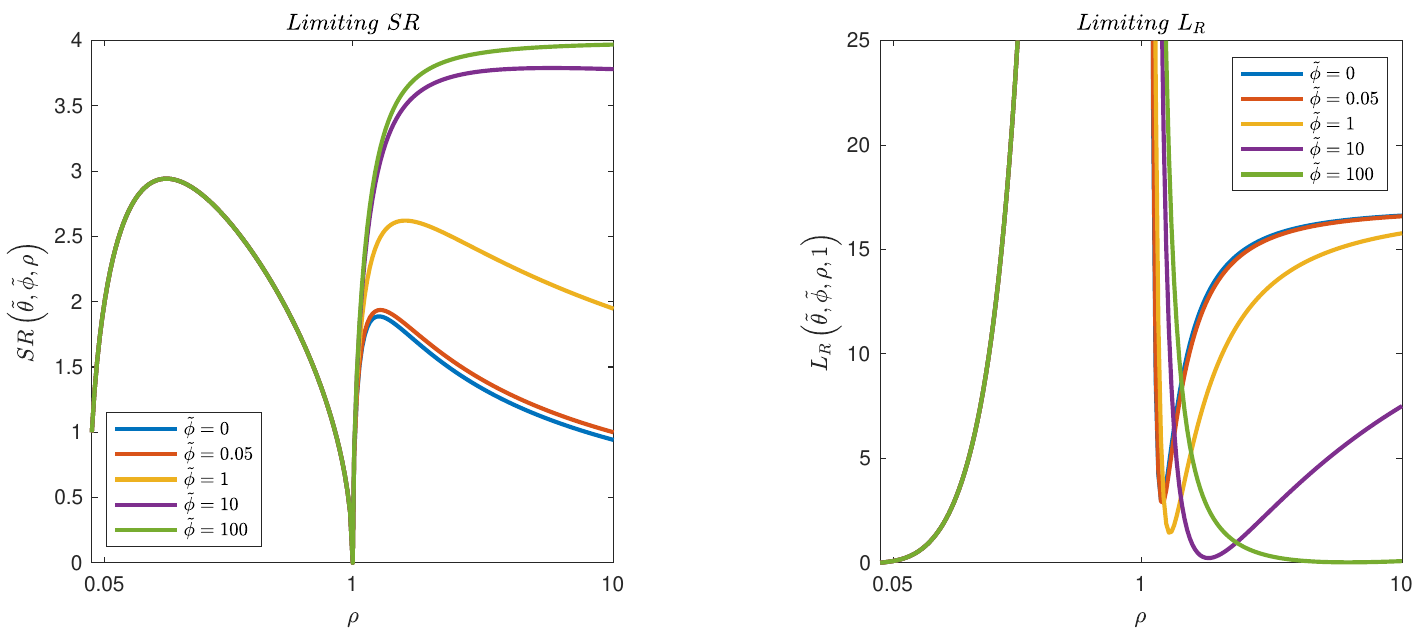}
    \caption{Asymptotic OOS Sharpe ratio and prediction loss with $\sqrt{\ttheta}$ following Figure \ref{Fig:theta_p_1}, when $\tphi=0,0.05,1,10,100$.}
    \label{Fig:theta_p_1_SRLR}
\end{figure}

From Figure \ref{Fig:theta_p_1_SRLR}, we clearly see the double ascent for the limiting Sharpe ratio and a descent in the over-parameterized regime for prediction loss. In terms of the Sharpe ratio, the first ascent is due to the increase in clairvoyant Sharpe ratio. The second ascent is due to the benefit of model complexity in the high-dimensional regime. 

The double ascent  Sharpe ratio curve is analogous to the double descent risk curve in prediction problems. In the low-dimensional regime, there is a ``sweet spot'' that best balances the benefit of having a complex model (i.e., having more assets to invest) and the cost of overfitting the data. The literature has traditionally focused on finding this sweet spot using various techniques. However, there is a second peak in the high-dimensional regime. If the factor structure of asset returns is strong, then there is a wide range of over-parameterization that can lead to even better performance than the left peak. This explains why heavily over-parameterized models work well in the recent machine learning literature.


\subsection{Finite Sample Performance Illustration}
In this subsection, we conduct finite sample simulations based the above designed scenario for showing ``double descent", and validate the consistency of our theoretical results. We generate our data according to the factor model in Equation (\ref{Model:factor_model}). Let \(T=100\), and \(N\) taking values from the set:
\[
N \in \{3, 5, 8,\underbrace{10, 15, 20, \ldots, 95}_{\text{difference = 5}}, 110, 120, \underbrace{150, 200, 250, \ldots, 1000}_{\text{difference = 50}}\}.
\]
Let $\sigmae=1$, $\left\Vert\bb\right\Vert^2=1$, $\sigma_f^2 \in \{0,1/T, 1/\sqrt{T},1,log(T),T\}$, such that we have different signal to noise ratio 
\[
\phi\in\{0,1/T, 1/\sqrt{T},1,log(T),T\}.
\]
Let the clairvoyant Sharpe ratio follow the pattern shown in Figure \ref{Fig:theta_p_1}, where it remains constant as \(N\) increases. To achieve this, we modify \(\mu_f^2\) as follows:
\[
\mu_f^2 = \frac{\theta(\sigmae^2 + \sigma_f^2 \left\Vert \bb \right\Vert^2)}{\left\Vert \bb \right\Vert^2}.
\]
Let $\bb = {\tilde{\bb}}/ {\Vert\tilde{\bb}\Vert}$, where $\tilde{\bb}\sim\NormDis\left(\bm{0},\bI_N\right)$; $f_t\sim\NormDis(\mu_f,\sigma_f^2)$. We exam for each combination of $\rho_T=p/T$ and $\phi$ for 1000 times, and setting $\sigma=1$ for simplicity. For each simulated sample, we then calculate estimations $\hatbmu$ and $\hatbSig\pseudoinv$. Consequently, 
\begin{align}
    \left[\hatbomeastpseudo\right]_i &= \dfrac{1}{\sqrt{\htheta_i}}\hatbSig\pseudoinv_i\hatbmu_i,\nonumber\\ 
    \text{Avg}\left({SR}\left(\hatbomeastpseudo\right)\right)& = \dfrac{1}{1000}\sum_{i=1}^{1000}\dfrac{\hatbmu_i\T\hatbSig_i\pseudoinv\bmu}{\sqrt{\hatbmu_i\T\hatbSig_i\pseudoinv\bSig\hatbSig_i\pseudoinv\hatbmu_i}},\label{bar_SR}\\
    \text{Avg}\left(L_{\bR}\left(\hatbomeastpseudo;{\bomeast}\right)\right)& = \dfrac{1}{1000}\sum_{i=1}^{1000}\left(\left[\hatbomeastpseudo\right]_i-{\bomeast}\right)\T\left(\bSig+\bmu\bmu\T\right)\left(\left[\hatbomeastpseudo\right]_i-{\bomeast}\right),\label{bar_LR}
\end{align}
for $i=1,...,1000$. The calculated values of Equations (\ref{bar_SR}) and (\ref{bar_LR}) are graphically represented in Figures the following Figure \ref{Fig:simulations}.

\begin{figure}[H]
    \centering
    \includegraphics[trim=0 0 0 0,width=1\textwidth]{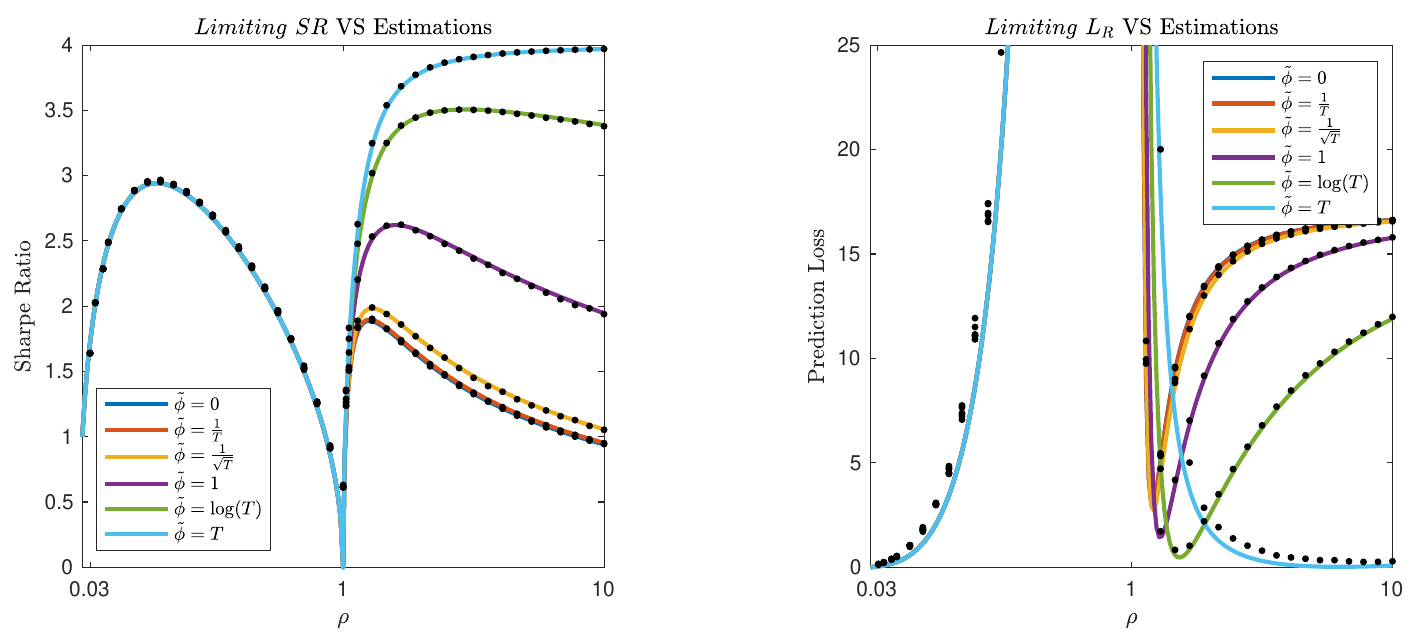}
    \caption{\textbf{Curves}: asymptotic performance for the hypothetical scenario. \textbf{Points}: finite-sample simulations with $T=100$, over various values $\phi$.}
    \label{Fig:simulations}
\end{figure}

\section{Explanation: Eigen-Analysis and Inductive Bias}\label{Section:induction}
This section offers more discussions on the intuition behind the benefit of model complexity. To investigate the underlying reasons behind the surprising phenomena observed, we analyze the estimation accuracy from two viewpoints - eigen-analysis and inductive bias.

\subsection{Eigen Portfolios}

This section shows how the stability of estimated portfolio weights depends on the model complexity. We present the spectral representation of the optimal portfolio weights. This concept has been studied in previous works, such as \cite{lopez2020common}, \cite{chen2016efficient}, and \cite{guo2018eigen}. Let $\tau_i$ and $\bupsilon_i$ denote the eigenvalues and eigenvectors of the covariance matrix $\bSig$, respectively. The optimal Sharpe ratio and portfolio can be expressed as:
\begin{align}
\sqrt{\theta}&=\sqrt{\bmu\T\bSig\inv\bmu}
=\sqrt{\sum_{i=1}^{N}\left(\frac{\bupsilon_i\T\bmu}{\sqrt{\tau_i}}\right)^2}=\sqrt{\sum_{i=1}^{N}\left(\frac{\mathbb{E}[\tilde{\br}_i]}{\sqrt{\text{Var}[\tilde{\br}_i]}}\right)^2}
=\sqrt{\sum_{i=1}^{N}SR_i^2},\nonumber\\
\bomeast\textbf{ }&\propto\textbf{ }\bSig\inv\bmu=\sum_{i=1}^{N}\frac{1}{\tau_i}\bupsilon_i\bupsilon_i\T\bmu=\sum_{i=1}^{N}\frac{\bupsilon_i\T\bmu}{\sqrt{\tau_i}}\frac{1}{\sqrt{\tau_i}}\bupsilon_i=\sum_{i=1}^{N}\frac{\mathbb{E}[\tilde{\br}_i]}{\sqrt{\text{Var}[\tilde{\br}_i]}}\frac{1}{\sqrt{\tau_i}}\bupsilon_i=\sum_{i=1}^{N}\frac{SR_i}{\sqrt{\tau_i}}\bupsilon_i,\nonumber
\end{align}
where $\mathbb{E}[\tilde{\br}_i]=\bupsilon_i^\top \bmu$ represents the expected return of the $i$-th eigenvector as a portfolio, and $\sqrt{\text{Var}[\tilde{\br}_i]}=\sqrt{\tau_i}$ is its risk. Thus, the term $\frac{\bupsilon_i^\top \boldsymbol{\mu}}{\sqrt{\tau_i}}$ is the Sharpe ratio of the $i$-th eigen portfolio. 

\begin{figure}[H]
    \centering    \includegraphics[trim=0 0 0 0,width=1\textwidth]{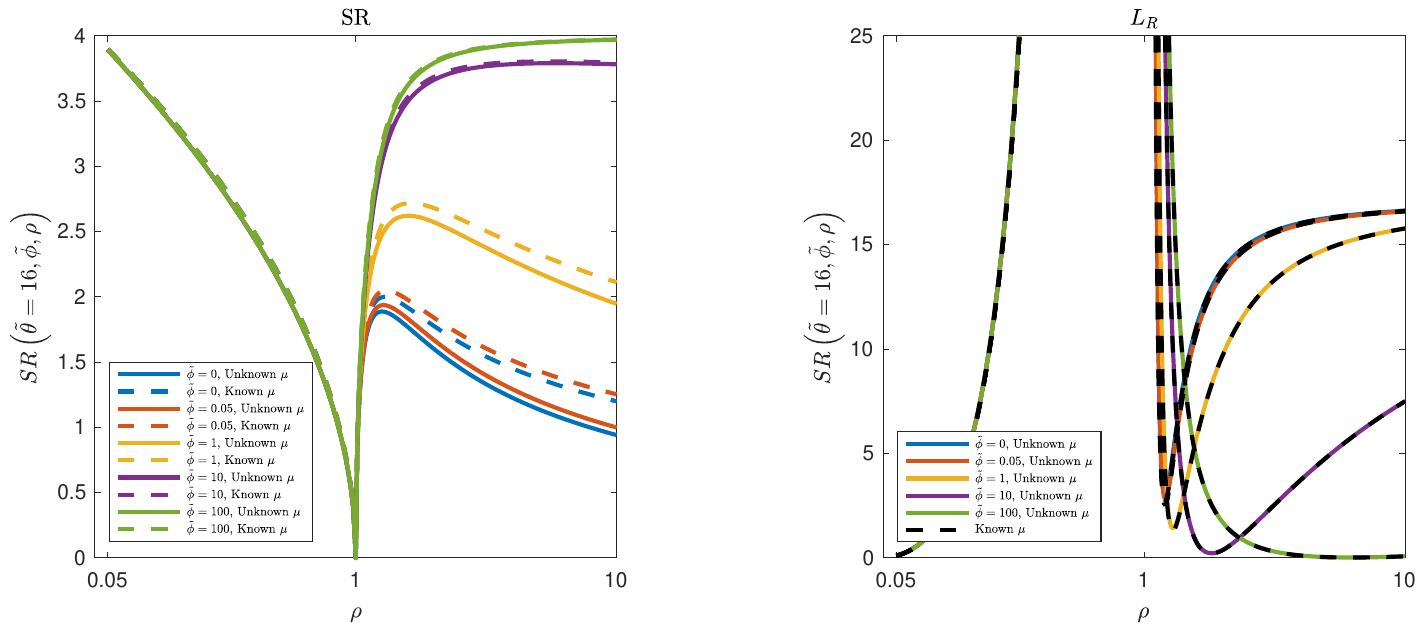}
    \caption{Asymptotic Sharpe ratio and prediction loss with true and estimated $\bmu$, $\sqrt{\ttheta}=4$ and diverse $\tphi$ values.}
    \label{Fig:theta_p_1_SRLR_compareMu}
\end{figure}

This spectral representation reveals two important insights: First, the squared Sharpe ratio of the optimal portfolio is the sum of the squared Sharpe ratios from all principal component directions. Second, the optimal portfolio weights are a linear combination of eigen portfolios, i.e. the eigenvectors $\{\bupsilon_i, i=1, \ldots, N\}$, where each eigenvector is weighted by $\frac{SR_i}{\sqrt{\tau_i}}$. 
Thus, to accurately estimate the optimal portfolio weight, it requires precise estimates of the eigenvalues $\tau_i$, eigenvectors $\bupsilon_i$ and the expected return vector $\bmu$.  
Figure \ref{Fig:theta_p_1_SRLR_compareMu} demonstrates the influences of the estimation accuracy of the mean vector $\bmu$, showing that its effects are relatively limited.

\begin{figure}[H]
    \centering    \includegraphics[trim=0 0 0 0,width=1\textwidth]{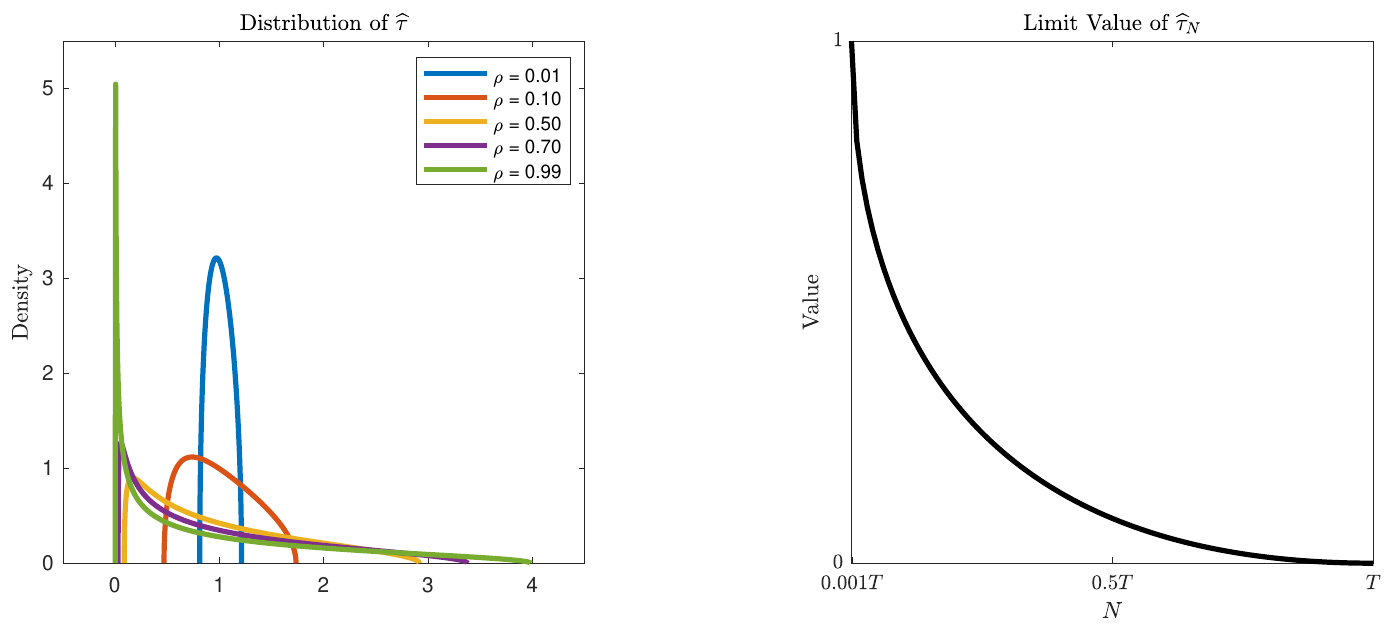}
    \caption{Left: Mar\v enko-Pastur distributions; Right: limits of the smallest sample eigenvalue.}
    \label{Fig:MP_Law}
\end{figure}

Next we analyze the estimation accuracy of eigenvalues based on fundamental results from random matrix theory. 
When $N<T$, the optimal portfolio weight estimator is 
\begin{align}
\hatbomeastpseudo\textbf{ }\propto\textbf{ }\hatbSig\inv\hatbmu=\sum_{i=1}^{N}\frac{\hatbupsilon_i\T\hatbmu}{\sqrt{\hattau_i}}\frac{\hatbupsilon_i}{\sqrt{\hattau_i}}.\nonumber
\end{align}
Recall from Remark \ref{Remark1} that when the ratio  $\rho < 1$, it is sufficient to consider the case where \(\bSig = \bI_N\). In this setting, the empirical spectral distribution of the sample covariance matrix follows the Mar\v enko-Pastur law. As shown in the left panel of Figure \ref{Fig:MP_Law}, as $\rho$ increases, the eigenvalues of the sample covariance matrix deviate further from 1, introducing more estimation errors in \(\frac{1}{\hattau_i}\), which are used to compute the estimated portfolio weights. The smallest eigenvalue, converges to $(1-\sqrt{\rho})^2$ (as shown in the right panel of Figure \ref{Fig:MP_Law}). When $N$ approaches $T$, \(\hattau_N\) approaches 0, making its inverse asymptotically ill-defined and unstable, thereby significantly amplifying the estimation error. This explains why the asymptotic Sharpe ratio and prediction loss performance monotonically decrease in the $\rho < 1$ regime.

When $N>T$, the covariance structure $\bSig$ affects eigen portfolios.  In Scenario 2, we account for a common factor, which introduces a spiked component in covariance matrix. We divide estimated eigen-portfolio into three parts:
\begin{align}
\hatbomeastpseudo\textbf{ }\propto\textbf{ }\hatbSig\pseudoinv\hatbmu=\underbrace{\frac{\hatbupsilon_1\T\hatbmu}{\sqrt{\hattau_1}}\frac{\hatbupsilon_1}{\sqrt{\hattau_1}}}_{\text{accurate due to strong factor}}
+\underbrace{\sum_{i\in\mathcal{G}_2}\frac{\hatbupsilon_i\T\hatbmu}{\sqrt{\hattau_i}}\frac{\hatbupsilon_i}{\sqrt{\hattau_i}}}_{\text{inaccurate due to high-dimensionality}}
+\underbrace{\sum_{i\in\mathcal{G}_3}\frac{\hatbupsilon_i\T\hatbmu}{\sqrt{\hattau_i}}\frac{\hatbupsilon_i}{\sqrt{\hattau_i}}}_{\text{asymptotically ill-defined}},\nonumber
\end{align}
where $1+|\mathcal{G}_2|+|\mathcal{G}_3|=T-1$, and $|\cdot|$ represents the cardinality.

The first eigen-portfolio corresponds to the largest eigenvalue $\tau_1$ (strong factor), where $\hattau_1$ is a relatively accurate estimate. Group $\mathcal{G}_2$ corresponds to the most of the eigen-portfolios that $\tau$ can not be accurately estimated due to the challenges of high-dimensionality. Group $\mathcal{G}_3$ includes $\hattau_j$ for $j\in(T-\delta,T-1)$, where $\delta$ is small. This group has small sample eigenvalues asymptotically approaching zero, so inverting these values amplifies estimation noise.

\begin{figure}[H]
    \centering    \includegraphics[trim=0 0 0 0,width=0.5\textwidth]{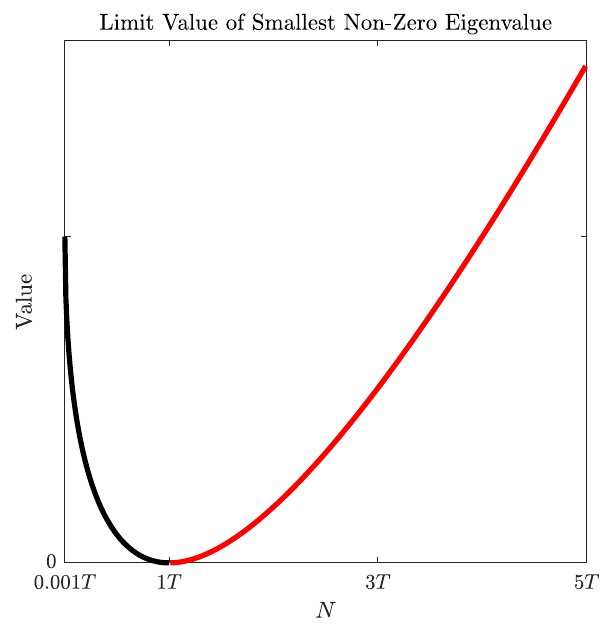}
    \caption{limits of the smallest non-zero empirical eigenvalue ($\bSig=\sigma_f^2\bb\bb\T+\sigmae^2\bI_N$) for wide range of $\rho$.}
    \label{Fig:MP_Law_SmallestEigRho}
\end{figure}

When $\rho$ is slightly greater than 1, group $\mathcal{G}_3$ dominates due to the inverse of near-zero eigenvalues. As $\rho$ increases, small eigenvalues move away from zero, reducing the impact of group $\mathcal{G}_3$. This is illustrated in Figure \ref{Fig:MP_Law_SmallestEigRho}. This shift explains the initial improvement in the asymptotic Sharpe ratio and prediction loss as $\rho$ increases. However, as $\rho$ continues to increase, group $\mathcal{G}_2$ becomes more dominant, resulting in a gradual decline in performance. Finally, for the same $\rho$ but with increasing factor strength, the estimation accuracy for the first eigen-portfolio improves. 

\subsection{Inductive bias and the equal-weighted portfolio}
We have shown thus far that when $\rho>1$, model performance starts to improve with model complexity. What is special about this threshold? This threshold is known as the interpolation threshold in the machine learning literature (\cite{belkin2019reconciling}). Similarly in our setting, when $\rho>1$, there are more assets than the number of observations in the data, which means one can form a portfolio with these assets to generate positive in-sample excess return without any risk. In fact, when $\rho>1$, there are multiple portfolios that perfectly fit the in-sample data (i.e., generating positive return without risk).\footnote{Imagine solving a system of equations with more unknowns than the number of equations. The solutions are not unique. In this case, the unknowns are portfolio weights. An equation corresponds to an observation.} In the high-dimensional regime, data can no longer tell which portfolio is the best, since many deliver the same in-sample result. Therefore, the modeler has some discretion to choose the best portfolio. This discretion is known as inductive bias in machine learning and plays a key role in the success of the model. 

In our pseudoinverse portfolio estimator, defined in Equation \ref{Def:pseudoinv-estimator}, the inductive bias is to choose the portfolio with the smallest $\ell_2$-norm. In fact, when $\rho>1$, we can rewrite our portfolio optimization problem as
\begin{equation}\label{Eqn:min-norm}
    {\hatbomeastpseudo}=\argmin_{\bome}\Big\{\left\Vert\bome\right\Vert_2 \text{subject to }\bR\bome=\br^c\Big\},
\end{equation}
where $\br^c$ is a constant vector with each element equals to $\frac{\sigma}{\sqrt{\htheta}}$ and $\bR$ is the training data (\cite{tibshiranioverparametrized}). As the number of assets increases, there are more possible portfolios that meet the interpolation constraint. How to choose from these feasible portfolios? The machine learning literature has often found that using the $\ell_2$-norm as the inductive bias often produces good out-of-sample predictions. We find that the portfolio that perfectly fits the training data while having the smallest $\ell_2$-norm also produces a good out-of-sample Sharpe ratio. The $\ell_2$-norm of a portfolio is closely related to the variance of its weights. By minimizing the $\ell_2$-norm, we are choosing the portfolio with small variations in weights. The limit of this is the equal-weighted portfolio, which has been shown in the finance literature to perform well out-of-sample (\cite{demiguel2009optimal}). The connection between the $\ell_2$-norm and the equal-weighted portfolio is another reason that our estimator performs well in the high-dimensional regime.

\section{An Empirical Study of U.S. Equities}\label{Sec:empirical_rolling}
The goal of this section is to assess the out-of-sample performance of portfolios constructed using the pseudoinverse estimator, particularly in the context of increasing dimensionality while maintaining a fixed sample size. We find the empirical results are consistent with our theory prediction. Based on the empirical performance of portfolio models with different complexity, we can also estimate the relationship between the clairvoyant Sharpe ratio and the number of assets.

\subsection{Data}\label{sec:data}
Our study utilizes a comprehensive dataset of monthly total individual equity returns sourced from the Center for Research in Security Prices (CRSP). This dataset includes firms listed on the New York Stock Exchange (NYSE), the American Stock Exchange (AMEX), and NASDAQ. The analysis spans from January 1967 to June 2019, covering a total of 630 months. Our sample comprises nearly 30,000 stocks, with an average of over 6,200 stocks per month. In Subsection~\ref{Subsection:1500}, we focus specifically on the largest 1,500 stocks each month, which are the most liquid stocks in the market. Additionally, in the robustness check, we use all stocks and categorize them them into four size groups.

We collected 70 stock-level characteristics as documented in \cite{gu2020empirical}. For each month's dataset of individual stock returns, we commenced by selecting the top 1,500 stocks, prioritized according to their market capitalization. These stocks were then ascendingly reordered based on the values of one of the 70 characteristics. Following this, the 1,500 stocks were divided into 10 groups.\footnote{If the characteristic value for some stocks is unavailable, only the common set is retained, meaning stocks that are both in the top 1,500 and have characteristic values.} The segmentation was sequential: the first group included the top one-fifteenth of ranked stocks, the second group included the next one-fifteenth, and so on, resulting in 10 characteristic-ranked portfolios. The average return of the stocks within each group was computed to represent the return of the new asset formed by that group. With 70 characteristics at hand, this methodology constructed 700 characteristic-ranked portfolios.

The following Figure \ref{Fig:screeplot} shows the largest 20 eigenvalues of the covariance matrix of the dataset. The presence of a single spiked eigenvalue, followed by a rapid decrease to smaller eigenvalues, suggests the existence of one strong factor in the data. Therefore, we assume that the returns of the constructed assets follow a single-factor model, as discussed in Section \ref{subsec:Scenario2}.

\begin{figure}[H]
    \centering    \includegraphics[trim=0 0 0 0,width=0.5\textwidth]{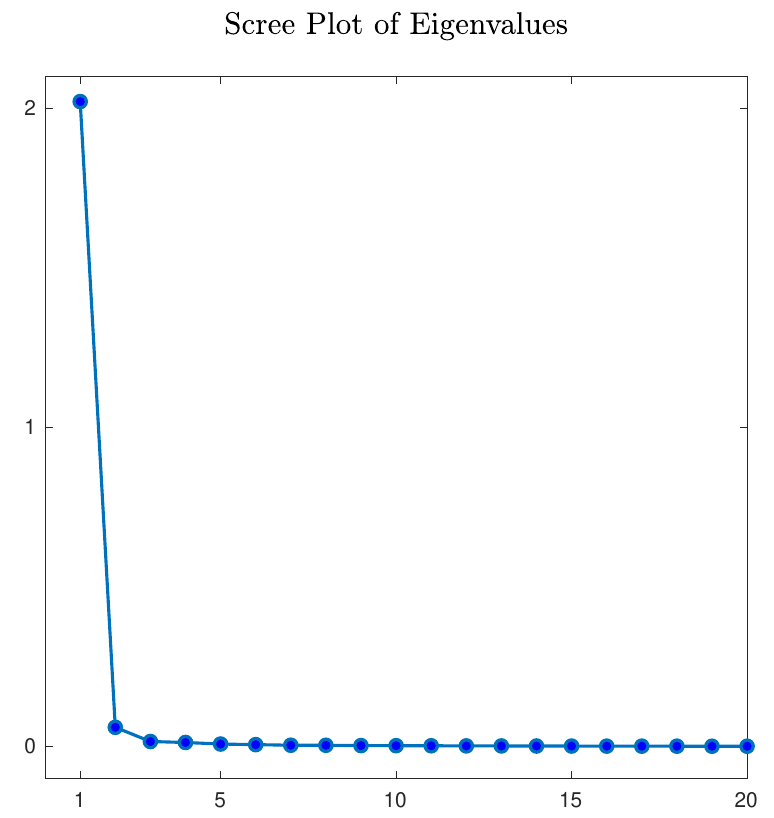}
    \caption{20 largest eigenvalues of the covariance matrix estimated from the $700\times630$ dataset.}
    \label{Fig:screeplot}
\end{figure}


\subsection{Portfolio-formation rules}

All mean variance efficient portfolios of characteristic-ranked portfolios in this study are constructed and updated on a monthly basis, utilizing the most recent \textcolor{black}{120} monthly returns, denoted as \textcolor{black}{$T=120$}.\footnote{A study for $T=60$ is provided in the supplementary material.} Therefore, the out-of-sample period ranges from \textcolor{black}{01/1977} through \textcolor{black}{06/2019}, resulting in a total of \textcolor{black}{510} months. We denote the investment dates by \textcolor{black}{$h=121(T+1),...,630$}. We explore a comprehensive array of portfolio sizes, specifically choosing $N$ values within the set:
\[
N\in\{\underbrace{2, 3, \ldots, 24}_{\text{difference = 1}}, \underbrace{27,30, \ldots,120}_{\text{difference = 3}}, \underbrace{126, 132, \ldots, 696}_{\text{difference = 6}}\},
\]
which encompasses a total of 151 distinct $N$ values.

For each $N$ case, we randomly select $N$ characteristic-ranked portfolios, and use the same selection for all $h=121,...,630$ (we repeat this process 100 times). For each specified combination of $(h,N)$, we use data from $h-T$ to $h-1$ to estimate the mean-variance portfolio weight for time $h$. We define $\br_{t,N}$ as the vector representing the returns of the selected $N$ characteristic-ranked portfolios at the  $t^{th}$ month. Please note that assessing the prediction loss defined in Equation (\ref{Def:prediction_loss}) is infeasible, as it requires information from the population, such as the true portfolio weight $\bomeast$, the population covariance matrix $\bSig$, and the population mean vector $\bmu$. However, the Sharpe ratio can be calculated because it is empirically equivalent to the mean of portfolio returns divided by the standard error. Thus in this section, we only present empirical results for Sharpe ratio. For each investment date $h$, the corresponding portfolio weight is calculated by the formula (In this study, we fix $\sigma=0.03$):
\begin{equation}
\left[\hatbomeastpseudo\right]^{h,N}=\frac{\sigma\hatbSig\pseudoinv_{h,N}\hatbmu_{h,N}}{\sqrt{\htheta_{h,N}}},
\end{equation}
where 
\begin{align}
    \hatbmu_{h,N}&=\dfrac{1}{T}\sum^{h-1}_{t=h-T}\br_{t,N},\nonumber\\
    \hatbSig_{h,N} &= \dfrac{1}{T-1}\sum^{h-1}_{t=h-T}\left(\br_{t,N}-\hatbmu_{h,N}\right)\left(\br_{t,N}-\hatbmu_{h,N}\right)\T,\nonumber\\
    \htheta_{h,N}&=
\left\{
    \begin{aligned}
    &\left(1-\frac{N}{T}\right)\htheta_{s,h,N}-\frac{N}{T}, & \text{when } N < T, \\
    &\frac{\left(\hphi_{h,N}+\frac{N}{T}\right)^2}{\left(\hphi_{h,N}+1\right)^2\frac{N}{T}}\left[\left(\frac{N}{T}-1\right)\htheta_{s,h,N}-1\right], & \text{when } N > T,\nonumber
    \end{aligned}
\right.\\
\htheta_{s,h,N}&=\hatbmu_{h,N}\T\hatbSig\pseudoinv_{h,N}\hatbmu_{h,N},\nonumber\\
\hphi_{h,N}&=\frac{\hattau_1\left(\hatbSig_{h,N}\right)}{\widehat{\sigmae}^2}.\nonumber
\end{align}

From Figure \ref{Fig:screeplot}, we observe the presence of a single strong factor, leading us to assume that the data follows the model specified in Equation (\ref{Model:factor_model}). Also, when a strong factor is present, we can accurately estimate the signal-to-noise ratio $\phi$, where $\hattau_1(\bA)$ is the largest eigenvalue of $\bA$, and $\widehat{\sigmae}$ is the variance of the asset returns after removing factor components\footnote{When strong factors are present in the data, we can separate the factor components from the error components \cite{stock2002forecasting,bai2003inferential,fan2013large}. The factor $f_t$ can be estimated by the eigenvector associated with the largest eigenvalue; the variance of the factor, $\sigma_f^2$, can be estimated by the largest eigenvalue itself; and the factor loadings $\bb$ can be estimated using the scores corresponding to the first eigenvector. Additionally, we assume that the errors are independent and identically distributed. Therefore, we flatten the residual matrix - obtained by removing the factor components from the return data - into a vector to calculate for $\sigmae^2$.}.

\subsection{Out-of-sample performances}\label{Subsection:1500}
Let us denote the out-of-sample portfolio return for a combination $(h,N)$ as
\[
\left[R\right]_{h,N} = \br_{h,N}\T\left[\hatbomeastpseudo\right]^{h,N},\quad\text{for } h = 121,...,630.
\]
Subsequently, we calculate the average and standard deviation of out-of-sample portfolio return as
\[
\text{Avg}\left({\left[R\right]_N}\right) = \dfrac{1}{\textcolor{black}{510}}\sum^{630}_{h=\textcolor{black}{121}}\left[R\right]_{h,N},\quad\text{and}\quad \text{Std}\left(\left[R\right]_N\right) = \sqrt{\dfrac{1}{\textcolor{black}{510}}\sum^{630}_{h=121}\left(\left[R\right]_{h,N}-\text{Avg}\left({\left[R\right]_N}\right)\right)^2}.
\] 
Hence, the estimated Sharpe Ratio for each dimensionality $N$ is given by
\[
\left[SR\right]_N = \dfrac{\text{Avg}\left({\left[R\right]_N}\right)}{\text{Std}\left({\left[R\right]_N}\right)}.
\]

\begin{figure}[H]
\centering
\includegraphics[trim=0 0 0 0,width=0.5\textwidth]{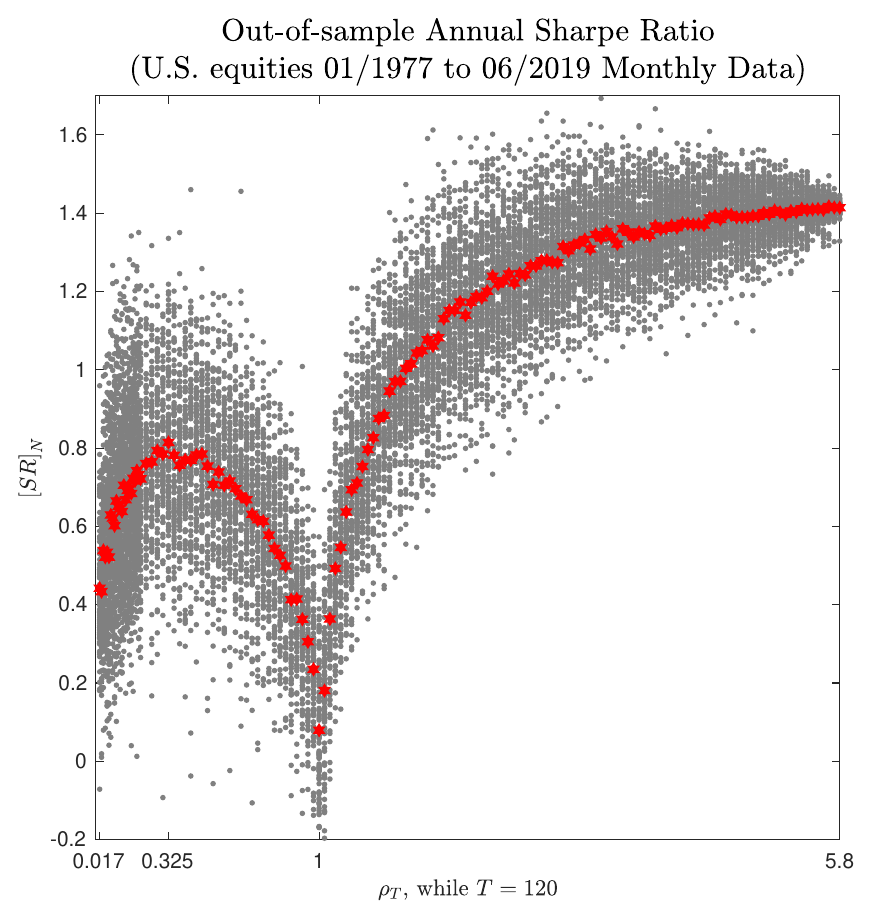}
\caption{Out-of-sample annualized Sharpe ratio based on \textcolor{black}{510} monthly returns, with the most recent \textcolor{black}{$T=120$} for training. The grey dots are 100 replications for each $N$, while the red asterisks indicate the average.}
\label{Fig:SR_emp}
\end{figure}

We repeat the process 100 times, and for each replication of the combination $(h,N)$, the same assets are selected across all 510 testing procedures. Figure \ref{Fig:SR_emp} presents the variations in the out-of-sample annual Sharpe ratio. Each grey point represents one value of $\left[SR\right]_N$. For each $N$, there are 100 grey points corresponding to the 100 replications, with the red point indicating the average of these 100 values. Notably, the Sharpe ratio exhibits a minimum near $\rho_T=1$. Within the domain of $\rho_T\in (0,1)$, a peak of average emerges around \textcolor{black}{$\rho_T=0.3250$ $(N=39)$}, highlighting an optimal equilibrium between the number of stocks and the size of the training sample. As $\rho_T$ extends into the $(1,\infty)$ range, The Sharpe ratio shows a significant upward trend, eventually stabilizing around a value of \textcolor{black}{1.4}. This is a high Sharpe ratio, especially when compared to the best neural network performance reported in \cite{gu2020empirical}, which achieved a Sharpe ratio of 1.35. This second ascent, markedly higher than that observed within the $(0,1)$ range, highlights enhanced portfolio performance with increasing dimensionality. The figure is actually very similar to the case with a strong factor that we see in Figure \ref{Fig:theta_p_1_SRLR}, where we assume the theoretical Sharpe ratio increases first and quickly reaches its maximum as $N$ increases. We emphasize that this does not imply our collected data follows the distribution in theorem model exactly. However, when the data exhibits properties similar to those in our theorem model - such as a strong factor structure and a theoretical Sharpe ratio that increases and then plateaus - the asymptotic behavior will show similar patterns.

In Table \ref{Table:1500_equal}, we present the mean of 100 replications for various portfolio performance metrics. Each row shows the results with fixed $N$ across all replications. For columns, $\bm{CER}$ represents the certainty equivalent return, computed as:
\[
CER = \text{Avg}\left({\left[R\right]_N}\right)-\frac{\gamma}{2}\text{Var}\left({\left[R\right]_N}\right),
\]
$\bm{CAPM}$ $\bm{\alpha}$ is the alpha value regressed on excess return on the market.
The last five columns represent Sharpe ratio and $\bm{CER}$ for the minimum variance portfolio, equal-weighted portfolio, and in-sample Sharpe ratio.\footnote{We use the in-sample mean and covariance vector of all sample periods to measure the in-sample Sharpe ratio.}

From the table, across the dimension $N$ from 2 to 696, we observe a double ascent not only in the Sharpe ratio, as shown in Figure \ref{Fig:SR_emp}, but also in $\bm{CER}$. In the $N < T$ regime, values for $\bm{Avg}$, $\bm{Std}$, and $\bm{CAPM}$ $\bm{\alpha}$ increase, while in the $N > T$ regime, they decrease\footnote{Pattern for $\bm{Avg}$, $\bm{Std}$ can be confirmed from our theoretical results presented in Supplementary Subsection \ref{SubSection:Mean_SD}.}. For the minimum variance portfolio, we also observe a double ascent in both $\bm{SR}$ and $\bm{CER}$. For the equal-weighted portfolio, we observe a flat pattern. The Sharpe ratio and CER of the equal-weighted strategy is almost constant across different number of assets.\footnote{There is a slight improvement when $N$ is small.} For the in-sample Sharpe ratio, we see it always increases with the number of assets. However, in-sample Sharpe ratio explodes in the high-dimensional regime, e.g. when $N=600$, the monthly in-sample Sharpe ratio is 5.89 (and 20.4, if annualized). This shows that the in-sample Sharpe ratio is not a reliable estimate of the out-of-sample Sharpe ratio nor the clairvoyant Sharpe ratio.

\bigskip
\begin{table}[H]
\centering
\scriptsize
\caption{Monthly portfolio performance}
\label{Table:1500_equal}
\scalebox{0.8}{
\begin{tabular}{l|l|lllll|lllll}
    \Xhline{5\arrayrulewidth}
    & $\bm{N}$ & $\bm{SR}$ & $\bm{CER}$ & $\bm{Avg}[R]_N$ & $\bm{Std}[R]_N$ & $\bm{CAPM}$ $\bm{\alpha}$ & $\bm{Minvar}$ $\bm{SR}$ & $\bm{Minvar}$ $\bm{CER}$& $\bm{1/N}$ $\bm{SR}$ & $\bm{1/N}$ $\bm{CER}$ & \textbf{In-sample SR} \\
    \hline\hline
    \multirow{9}{*}{\textbf{\makecell[l]{Low \\ Dimensional \\ Regime}}} 
    & 2 & 0.13 & 0.25 & 0.39 & 3.03 & 0.09 & 0.17 & 0.44 & 0.15 & 0.38 & 0.18 \\
& 5 & 0.15 & 0.33 & 0.48 & 3.21 & 0.28 & 0.18 & 0.48 & 0.15 & 0.38 & 0.23 \\
& 10 & 0.17 & 0.42 & 0.60 & 3.46 & 0.45 & 0.19 & 0.51 & 0.15 & 0.37 & 0.29 \\
& 15 & 0.20 & 0.55 & 0.75 & 3.68 & 0.64 & 0.19 & 0.52 & 0.15 & 0.38 & 0.35 \\
& 20 & 0.21 & 0.59 & 0.83 & 3.96 & 0.73 & 0.18 & 0.47 & 0.15 & 0.39 & 0.39 \\
& 30 & 0.22 & 0.69 & 1.00 & 4.52 & 0.91 & 0.17 & 0.43 & 0.15 & 0.39 & 0.46 \\
& \textbf{39} & \textbf{0.24} & \textbf{0.81} & \textbf{1.19} & \textbf{5.05} & \textbf{1.12} & \textbf{0.16} & \textbf{0.40} & \textbf{0.15} & \textbf{0.39} & \textbf{0.52} \\
& 60 & 0.22 & 0.62 & 1.62 & 7.68 & 1.54 & 0.14 & 0.27 & 0.15 & 0.38 & 0.62 \\
& 90 & 0.18 & -4.23 & 2.73 & 16.52 & 2.65 & 0.09 & -0.15 & 0.15 & 0.38 & 0.75 \\
& 108 & 0.12 & -27.91 & 4.50 & 39.51 & 4.30 & 0.06 & -0.73 & 0.15 & 0.39 & 0.82 \\
    \hline
    \multirow{8}{*}{\textbf{\makecell[l]{High \\ Dimensional \\ Regime}}} 
    & 132 & 0.10 & -66.42 & 6.55 & 65.98 & 6.45 & 0.04 & -1.17 & 0.15 & 0.39 & 0.92 \\
& 150 & 0.18 & -2.59 & 3.72 & 20.42 & 3.64 & 0.08 & -0.03 & 0.15 & 0.39 & 0.99 \\
& 180 & 0.24 & 0.76 & 2.70 & 11.36 & 2.64 & 0.12 & 0.22 & 0.15 & 0.39 & 1.09 \\
& 210 & 0.28 & 1.29 & 2.39 & 8.54 & 2.33 & 0.14 & 0.30 & 0.15 & 0.39 & 1.21 \\
& 240 & 0.31 & 1.45 & 2.22 & 7.16 & 2.17 & 0.14 & 0.31 & 0.15 & 0.38 & 1.33 \\
& 480 & 0.39 & 1.43 & 1.73 & 4.44 & 1.67 & 0.18 & 0.39 & 0.15 & 0.39 & 4.30 \\
& 600 & 0.40 & 1.41 & 1.67 & 4.14 & 1.61 & 0.19 & 0.41 & 0.15 & 0.39 & 5.89 \\
& \textbf{684} & \textbf{0.41} & \textbf{1.40} & \textbf{1.64} & \textbf{4.01} & \textbf{1.58} & \textbf{0.19} & \textbf{0.42} & \textbf{0.15} & \textbf{0.39} & \textbf{4.05} \\
    \Xhline{5\arrayrulewidth}
\end{tabular}}
\captionsetup{justification=raggedright, singlelinecheck=false,font=scriptsize} 
\caption*{$\bm{SR}$: monthly Sharpe ratio; $\bm{CER}$: certainty equivalent return with $\gamma=3$; \textbf{Low Dimensional Regime}: Scenarios where $N<T$; \textbf{High Dimensional Regime}: Scenarios where $N>T$; The bolded rows correspond to $N$ where the Sharpe ratios reach optimal values on either the left or right.}
\end{table}

\subsection{Sharpe ratio upper bound and asset pricing theory}
Our theory is also useful in informing us the clairvoyant Sharpe ratio of the underlying assets (i.e., the true Sharpe ratio one would achieve if the true distribution of these assets is known). We estimate the clairvoyant Sharpe ratio as a function of the number of assets $N$ based on the empirical performance in Figure \ref{Fig:SR_emp}. 

To proceed, we first make an assumption about the functional form of $\theta$. We assume that the clairvoyant Sharpe ratio of $N$ assets is
\begin{equation}
    \sqrt{\theta(N)}=\sqrt{\theta_1}e^{-\lambda(N-1)} + \sqrt{\bar{\theta}}(1-e^{-\lambda(N-1)})\label{Eqn:model_theta(N)}
\end{equation}
where $\sqrt{\theta_1}$ is the Sharpe ratio of a single asset, $\sqrt{\bar{\theta}}$ is the maximum clairvoyant Sharpe ratio in the economy, and $\lambda$ is the speed parameter. We estimate these three parameters by fitting the empirical performance curve in Figure \ref{Fig:SR_emp} based on Equation (\ref{ThmRslt:SR_sfm}). To accomplish this task, we also need to estimate the factor strength of our test assets, which is accomplished in Section \ref{sec:data}.

\begin{figure}[H]
    \centering    \includegraphics[trim=0 0 0 0,width=0.6\textwidth]{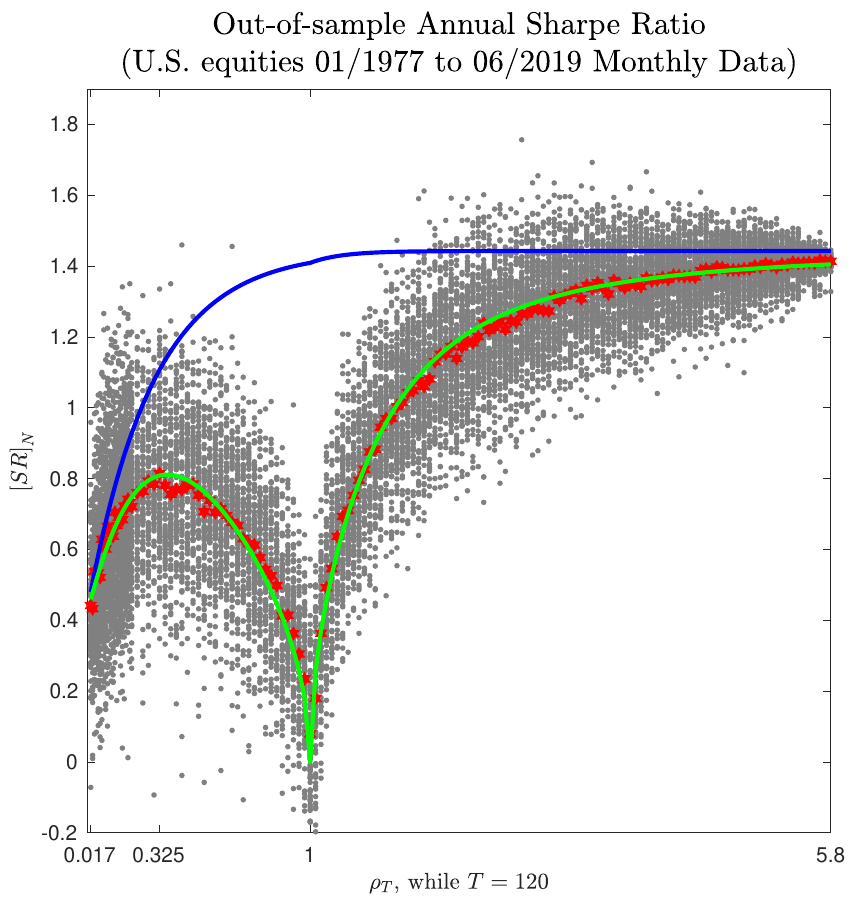}
    \caption{Blue: clairvoyant Sharpe ratio $\sqrt{\theta(N)}=0.4526e^{-0.0285(N-1)} + 1.4431(1-e^{-0.0285(N-1)})$. Green: Asymptotic limits of out-of-sample Sharpe ratio. Red dots: out-of-sample Sharpe ratio in the data.}
    \label{Fig:backedout_theta(N)}
\end{figure}

Figure \ref{Fig:backedout_theta(N)} illustrates the calibrated clairvoyant Sharpe ratio (the blue line), the implied model performance from Theorem \ref{Thm:sr_sfm} (the green line), and the actual model performance. As we can see, it completely meets our expectation that it first quickly increases with $N$ and then holds constant. The three parameters values are: $\sqrt{\theta_1}=0.4526$, $\sqrt{\bar{\theta}}=1.4431$, and $\lambda=0.0285$. 


This calibration exercise shows that the upper limit of the clairvoyant Sharpe ratio among our test assets is 1.44, which is informative about the asset pricing model needed to explain these test assets. An important question in the asset pricing literature is how many factors should be in an asset pricing mode. The CAPM has a single factor. The Fama-French three and five factor models have 3 and 5 factors, respectively. Given the Sharpe ratio upper bound of 1.44, we can back out the number of independent factors needed to generate this high Sharpe ratio. Specifically, the market factor has a Sharpe ratio of 0.42 during our sample period. The Sharpe ratio upper bound is 3.5 times that of the market factors. If we assume the asset pricing factors have the same Sharpe ratio as the market, then we need 12 independent asset pricing factors to reach the theoretical Sharpe ratio upper bound. Interestingly, our estimate, based on an entirely different approach, is similar to some other estimates in the asset pricing literature. For example, \cite{freyberger2020dissecting} find 13 firm characteristics have incremental explanatory power of the cross-section of expected returns. \cite{demiguel2020transaction} find that in the presence of transaction costs, 15 firm characteristics are jointly significant.

The speed parameter $\lambda$ informs us how fast the clairvoyant Sharpe ratio increases with the number of assets. To reach half of the Sharp ratio upper bound, the portfolio needs to have around 13 assets. At the optimal low-dimensional point, when $N=39$, the clairvoyant Sharpe ratio is 1.11, which is about three quarters of the upper limit. However, Figure \ref{Fig:backedout_theta(N)} shows that we cannot achieve such a theoretical Sharpe ratio empirically due to estimation errors in the low-dimensional regime. When $N=121$, entering the high-dimensional regime, the clairvoyant Sharpe ratio is 1.42, reaching 98\% of its upper limit. This indicates that the improvement in the empirical out-of-sample Sharpe ratio in the high-dimensional regime is largely due to the reduction in estimation error instead of the increase in the clairvoyant Sharpe ratio.

\section{Conclusion and Future Work}\label{Section:Conclusion}
In this study, we explored several factors affecting the out-of-sample performance of mean-variance portfolios. By quantifying these key parameters, we especially unveiled the pivotal role of clairvoyant Sharpe ratio and factor strength in portfolio optimization. Our findings highlight that stronger factors can lead to better performance in an overparameterization regime compared to an underparameterization regime. This counterintuitive result suggests that embracing higher complexity, under certain conditions, can be beneficial. Our work provides a framework for practitioners to carefully consider the trade-off between the amount of available data and the flexibility of incorporating information in portfolio construction. Looking ahead, there are several promising avenues for further research. One direction is to extend our quantitative analysis to other types of portfolios, such as the minimum variance portfolio. Investigating whether similar patterns of overparameterization and factor strength influence their performance could yield valuable insights. Another exciting prospect is to delve into the ``double descent" in the context of applying artificial intelligence to large-scale portfolio theory. Understanding how AI-driven models can harness complexity without succumbing to overfitting could revolutionize portfolio management practices. Expanding our study in these directions may contribute significantly to both theoretical advancements and practical applications in financial portfolio optimization.

\newpage
\bibliographystyle{apalike} 
\bibliography{references.bib}

\begin{thebibliography}{}

\bibitem[Anderson and Cheng, 2022]{anderson2022portfolio}
Anderson, E. and Cheng, A.-r. (2022).
\newblock Portfolio choices with many big models.
\newblock {\em Management Science}, 68(1):690--715.

\bibitem[Ao et~al., 2019]{ao2019approaching}
Ao, M., Yingying, L., and Zheng, X. (2019).
\newblock Approaching mean-variance efficiency for large portfolios.
\newblock {\em The Review of Financial Studies}, 32(7):2890--2919.

\bibitem[Avramov et~al., 2023]{avramov2023machine}
Avramov, D., Cheng, S., and Metzker, L. (2023).
\newblock Machine learning vs. economic restrictions: Evidence from stock return predictability.
\newblock {\em Management Science}, 69(5):2587--2619.

\bibitem[Bai, 2003]{bai2003inferential}
Bai, J. (2003).
\newblock Inferential theory for factor models of large dimensions.
\newblock {\em Econometrica}, 71(1):135--171.

\bibitem[Bai and Silverstein, 2010]{bai2010spectral}
Bai, Z. and Silverstein, J.~W. (2010).
\newblock {\em Spectral analysis of large dimensional random matrices}, volume~20.
\newblock Springer.

\bibitem[Bai et~al., 2007]{10.1214/009117906000001079}
Bai, Z.~D., Miao, B.~Q., and Pan, G.~M. (2007).
\newblock {On asymptotics of eigenvectors of large sample covariance matrix}.
\newblock {\em The Annals of Probability}, 35(4):1532 -- 1572.

\bibitem[Barroso and Saxena, 2022]{barroso2022lest}
Barroso, P. and Saxena, K. (2022).
\newblock Lest we forget: Learn from out-of-sample forecast errors when optimizing portfolios.
\newblock {\em The Review of Financial Studies}, 35(3):1222--1278.

\bibitem[Bartlett et~al., 2020]{bartlett2020benign}
Bartlett, P.~L., Long, P.~M., Lugosi, G., and Tsigler, A. (2020).
\newblock Benign overfitting in linear regression.
\newblock {\em Proceedings of the National Academy of Sciences}, 117(48):30063--30070.

\bibitem[Bartlett et~al., 2021]{bartlett2021deep}
Bartlett, P.~L., Montanari, A., and Rakhlin, A. (2021).
\newblock Deep learning: a statistical viewpoint.
\newblock {\em Acta numerica}, 30:87--201.

\bibitem[Belkin et~al., 2019]{belkin2019reconciling}
Belkin, M., Hsu, D., Ma, S., and Mandal, S. (2019).
\newblock Reconciling modern machine-learning practice and the classical bias--variance trade-off.
\newblock {\em Proceedings of the National Academy of Sciences}, 116(32):15849--15854.

\bibitem[Belkin et~al., 2020]{belkin2020two}
Belkin, M., Hsu, D., and Xu, J. (2020).
\newblock Two models of double descent for weak features.
\newblock {\em SIAM Journal on Mathematics of Data Science}, 2(4):1167--1180.

\bibitem[Bogle et~al., 2010]{bogle2010keynes}
Bogle, P., Garlappi, L., Uppal, R., and Wang, T. (2010).
\newblock Keynes meets markowitz: The trade of between familiarity and diversification.
\newblock In {\em AFA Atlanta Meetings Paper}, volume~41.

\bibitem[Brandt et~al., 2009]{brandt2009parametric}
Brandt, M.~W., Santa-Clara, P., and Valkanov, R. (2009).
\newblock Parametric portfolio policies: Exploiting characteristics in the cross-section of equity returns.
\newblock {\em The Review of Financial Studies}, 22(9):3411--3447.

\bibitem[Bryzgalova et~al., 2023]{bryzgalova2019forest}
Bryzgalova, S., Pelger, M., and Zhu, J. (2023).
\newblock Forest through the trees: Building cross-sections of stock returns.
\newblock {\em Available at SSRN 3493458}.

\bibitem[Carlson et~al., 2004]{carlson2004corporate}
Carlson, M., Fisher, A., and Giammarino, R. (2004).
\newblock Corporate investment and asset price dynamics: Implications for the cross-section of returns.
\newblock {\em The Journal of Finance}, 59(6):2577--2603.

\bibitem[Chen and Yuan, 2016]{chen2016efficient}
Chen, J. and Yuan, M. (2016).
\newblock Efficient portfolio selection in a large market.
\newblock {\em Journal of Financial Econometrics}, 14(3):496--524.

\bibitem[Chen et~al., 2024]{chen2024deep}
Chen, L., Pelger, M., and Zhu, J. (2024).
\newblock Deep learning in asset pricing.
\newblock {\em Management Science}, 70(2):714--750.

\bibitem[Chen et~al., 2011]{chen2011regularized}
Chen, L.~S., Paul, D., Prentice, R.~L., and Wang, P. (2011).
\newblock A regularized hotelling’s t 2 test for pathway analysis in proteomic studies.
\newblock {\em Journal of the American Statistical Association}, 106(496):1345--1360.

\bibitem[Chopra and Ziemba, 1993]{chopra1993aathe}
Chopra, V. and Ziemba, W. (1993).
\newblock The effect of errors in means, variances, and covariances on optimal portfolio choice.
\newblock {\em The Journal of Portfolio Management}, 19:6--11.

\bibitem[Cong et~al., 2023]{cong2023asset}
Cong, L.~W., Feng, G., He, J., and He, X. (2023).
\newblock Asset pricing with panel tree under global split criteria.
\newblock Technical report, National Bureau of Economic Research.

\bibitem[Cong et~al., 2022]{cong2021alphaportfolio}
Cong, L.~W., Tang, K., Wang, J., and Zhang, Y. (2022).
\newblock Alphaportfolio: Direct construction through deep reinforcement learning and interpretable ai.
\newblock {\em Available at SSRN 3554486}.

\bibitem[DeMiguel et~al., 2009a]{demiguel2009generalized}
DeMiguel, V., Garlappi, L., Nogales, F.~J., and Uppal, R. (2009a).
\newblock A generalized approach to portfolio optimization: Improving performance by constraining portfolio norms.
\newblock {\em Management Science}, 55(5):798--812.

\bibitem[DeMiguel et~al., 2009b]{demiguel2009optimal}
DeMiguel, V., Garlappi, L., and Uppal, R. (2009b).
\newblock Optimal versus naive diversification: How inefficient is the 1/{N} portfolio strategy?
\newblock {\em The Review of Financial studies}, 22(5):1915--1953.

\bibitem[DeMiguel et~al., 2020]{demiguel2020transaction}
DeMiguel, V., Martin-Utrera, A., Nogales, F.~J., and Uppal, R. (2020).
\newblock A transaction-cost perspective on the multitude of firm characteristics.
\newblock {\em The Review of Financial Studies}, 33(5):2180--2222.

\bibitem[Didisheim et~al., 2024]{didisheim2024apt}
Didisheim, A., Ke, S.~B., Kelly, B.~T., and Malamud, S. (2024).
\newblock {APT} or “{AIPT}”? the surprising dominance of large factor models.
\newblock Technical report, National Bureau of Economic Research.

\bibitem[Dong et~al., 2022]{dong2022anomalies}
Dong, X., Li, Y., Rapach, D.~E., and Zhou, G. (2022).
\newblock Anomalies and the expected market return.
\newblock {\em The Journal of Finance}, 77(1):639--681.

\bibitem[Du, 2022]{du2022mean}
Du, J. (2022).
\newblock Mean--variance portfolio optimization with deep learning based-forecasts for cointegrated stocks.
\newblock {\em Expert Systems with Applications}, 201:117005.

\bibitem[Fan et~al., 2013]{fan2013large}
Fan, J., Liao, Y., and Mincheva, M. (2013).
\newblock Large covariance estimation by thresholding principal orthogonal complements.
\newblock {\em Journal of the Royal Statistical Society Series B: Statistical Methodology}, 75(4):603--680.

\bibitem[Freyberger et~al., 2020]{freyberger2020dissecting}
Freyberger, J., Neuhierl, A., and Weber, M. (2020).
\newblock Dissecting characteristics nonparametrically.
\newblock {\em The Review of Financial Studies}, 33(5):2326--2377.

\bibitem[Garlappi et~al., 2007]{garlappi2007portfolio}
Garlappi, L., Uppal, R., and Wang, T. (2007).
\newblock Portfolio selection with parameter and model uncertainty: A multi-prior approach.
\newblock {\em The Review of Financial Studies}, 20(1):41--81.

\bibitem[Giglio et~al., 2022]{giglio2022factor}
Giglio, S., Kelly, B., and Xiu, D. (2022).
\newblock Factor models, machine learning, and asset pricing.
\newblock {\em Annual Review of Financial Economics}, 14(1):337--368.

\bibitem[Gu et~al., 2020]{gu2020empirical}
Gu, S., Kelly, B., and Xiu, D. (2020).
\newblock Empirical asset pricing via machine learning.
\newblock {\em The Review of Financial Studies}, 33(5):2223--2273.

\bibitem[Gunjan and Bhattacharyya, 2023]{gunjan2023brief}
Gunjan, A. and Bhattacharyya, S. (2023).
\newblock A brief review of portfolio optimization techniques.
\newblock {\em Artificial Intelligence Review}, 56(5):3847--3886.

\bibitem[Guo et~al., 2018]{guo2018eigen}
Guo, D., Boyle, P.~P., Weng, C., and Wirjanto, T.~S. (2018).
\newblock Eigen portfolio selection: A robust approach to sharpe ratio maximization.
\newblock {\em Available at SSRN 3070416}.

\bibitem[Hastie et~al., 2022]{hastie2022surprises}
Hastie, T., Montanari, A., Rosset, S., and Tibshirani, R.~J. (2022).
\newblock Surprises in high-dimensional ridgeless least squares interpolation.
\newblock {\em Annals of Statistics}, 50(2):949.

\bibitem[Jagannathan and Ma, 2003]{jagannathan2003risk}
Jagannathan, R. and Ma, T. (2003).
\newblock Risk reduction in large portfolios: Why imposing the wrong constraints helps.
\newblock {\em The Journal of Finance}, 58(4):1651--1683.

\bibitem[Jensen et~al., 2024]{jensen2024machine}
Jensen, T.~I., Kelly, B.~T., Malamud, S., and Pedersen, L.~H. (2024).
\newblock Machine learning and the implementable efficient frontier.
\newblock {\em Review of Financial Studies}.
\newblock Forthcoming.

\bibitem[Kacperczyk, 2004]{kacperczyk2004asset}
Kacperczyk, M. (2004).
\newblock Asset allocation under distribution uncertainty.
\newblock {\em Ann Arbor}, 1001:48109.

\bibitem[Kan and Wang, 2024]{kan2024optimal}
Kan, R. and Wang, X. (2024).
\newblock Optimal portfolio choice with unknown benchmark efficiency.
\newblock {\em Management Science}, 70(9):6117--6138.

\bibitem[Kan et~al., 2022]{kan2022optimal}
Kan, R., Wang, X., and Zhou, G. (2022).
\newblock Optimal portfolio choice with estimation risk: No risk-free asset case.
\newblock {\em Management Science}, 68(3):2047--2068.

\bibitem[Kan and Zhou, 2007]{kan2007optimal}
Kan, R. and Zhou, G. (2007).
\newblock Optimal portfolio choice with parameter uncertainty.
\newblock {\em Journal of Financial and Quantitative Analysis}, 42(3):621--656.

\bibitem[Kelly et~al., 2024a]{kelly2024virtue}
Kelly, B., Malamud, S., and Zhou, K. (2024a).
\newblock The virtue of complexity in return prediction.
\newblock {\em The Journal of Finance}, 79(1):459--503.

\bibitem[Kelly et~al., 2023]{kelly2023financial}
Kelly, B., Xiu, D., et~al. (2023).
\newblock Financial machine learning.
\newblock {\em Foundations and Trends{\textregistered} in Finance}, 13(3-4):205--363.

\bibitem[Kelly et~al., 2024b]{kelly2024universal}
Kelly, B.~T., Malamud, S., Pourmohammadi, M., and Trojani, F. (2024b).
\newblock Universal portfolio shrinkage.
\newblock Technical report, National Bureau of Economic Research.

\bibitem[Knowles and Yin, 2017]{knowles2017anisotropic}
Knowles, A. and Yin, J. (2017).
\newblock Anisotropic local laws for random matrices.
\newblock {\em Probability Theory and Related Fields}, 169:257--352.

\bibitem[Ledoit and P{\'e}ch{\'e}, 2011]{ledoit2011eigenvectors}
Ledoit, O. and P{\'e}ch{\'e}, S. (2011).
\newblock Eigenvectors of some large sample covariance matrix ensembles.
\newblock {\em Probability Theory and Related Fields}, 151(1-2):233--264.

\bibitem[Ledoit and Wolf, 2017]{ledoit2017nonlinear}
Ledoit, O. and Wolf, M. (2017).
\newblock Nonlinear shrinkage of the covariance matrix for portfolio selection: Markowitz meets goldilocks.
\newblock {\em The Review of Financial Studies}, 30(12):4349--4388.

\bibitem[Li et~al., 2023]{li2023benign}
Li, Z., Su, W.~J., and Sejdinovic, D. (2023).
\newblock Benign overfitting and noisy features.
\newblock {\em Journal of the American Statistical Association}, 118(544):2876--2888.

\bibitem[Liang and Rakhlin, 2020]{liang2020just}
Liang, T. and Rakhlin, A. (2020).
\newblock Just interpolate: Kernel “ridgeless” regression can generalize.
\newblock {\em The Annals of Statistics}, 48(3):1329--1347.

\bibitem[Liao et~al., 2023]{liao2023economic}
Liao, Y., Ma, X., Neuhierl, A., and Shi, Z. (2023).
\newblock Economic forecasts using many noises.
\newblock {\em arXiv preprint arXiv:2312.05593}.

\bibitem[Lopez-Lira and Roussanov, 2020]{lopez2020common}
Lopez-Lira, A. and Roussanov, N.~L. (2020).
\newblock Do common factors really explain the cross-section of stock returns?
\newblock {\em Jacobs Levy Equity Management Center for Quantitative Financial Research Paper}.

\bibitem[Marchenko and Pastur, 1967]{marchenko1967distribution}
Marchenko, V.~A. and Pastur, L.~A. (1967).
\newblock Distribution of eigenvalues for some sets of random matrices.
\newblock {\em Matematicheskii Sbornik}, 114(4):507--536.

\bibitem[Markowitz, 1952]{Markowitz1952}
Markowitz, H. (1952).
\newblock {Portfolio Selection}.
\newblock {\em Journal of Finance}, 7(1):77--91.

\bibitem[Markowitz, 2010]{markowitz2010portfolio}
Markowitz, H.~M. (2010).
\newblock Portfolio theory: as i still see it.
\newblock {\em Annu. Rev. Financ. Econ.}, 2(1):1--23.

\bibitem[Mei and Montanari, 2022]{mei2022generalization}
Mei, S. and Montanari, A. (2022).
\newblock The generalization error of random features regression: Precise asymptotics and the double descent curve.
\newblock {\em Communications on Pure and Applied Mathematics}, 75(4):667--766.

\bibitem[Meyer, 1973]{meyer1973generalized}
Meyer, Jr, C.~D. (1973).
\newblock Generalized inversion of modified matrices.
\newblock {\em Siam Journal on Applied Mathematics}, 24(3):315--323.

\bibitem[Mitra, 2019]{mitra2019understanding}
Mitra, P.~P. (2019).
\newblock Understanding overfitting peaks in generalization error: Analytical risk curves for $ l\_2 $ and $ l\_1 $ penalized interpolation.
\newblock {\em arXiv preprint arXiv:1906.03667}.

\bibitem[Muthukumar et~al., 2020]{muthukumar2020harmless}
Muthukumar, V., Vodrahalli, K., Subramanian, V., and Sahai, A. (2020).
\newblock Harmless interpolation of noisy data in regression.
\newblock {\em IEEE Journal on Selected Areas in Information Theory}, 1(1):67--83.

\bibitem[Nakkiran et~al., 2021]{nakkiran2021deep}
Nakkiran, P., Kaplun, G., Bansal, Y., Yang, T., Barak, B., and Sutskever, I. (2021).
\newblock Deep double descent: Where bigger models and more data hurt.
\newblock {\em Journal of Statistical Mechanics: Theory and Experiment}, 2021(12):124003.

\bibitem[Pan and Zhou, 2011]{pan2011central}
Pan, G. and Zhou, W. (2011).
\newblock Central limit theorem for hotelling's t 2 statistic under large dimension.
\newblock {\em The Annals of Applied Probability}, pages 1860--1910.

\bibitem[Shi et~al., 2022]{shi2022}
Shi, Z., Su, L., and Xie, T. (2022).
\newblock l2-relaxation: With applications to forecast combination and portfolio analysis.
\newblock {\em Review of Economics and Statistics}, pages 1--44.

\bibitem[Silverstein, 1995]{silverstein1995strong}
Silverstein, J.~W. (1995).
\newblock Strong convergence of the empirical distribution of eigenvalues of large dimensional random matrices.
\newblock {\em Journal of Multivariate Analysis}, 55(2):331--339.

\bibitem[Snow, 2020]{snow2020machine}
Snow, D. (2020).
\newblock Machine learning in asset management—part 2: Portfolio construction—weight optimization.
\newblock {\em The Journal of Financial Data Science}, 2(2):17--24.

\bibitem[Stock and Watson, 2002]{stock2002forecasting}
Stock, J.~H. and Watson, M.~W. (2002).
\newblock Forecasting using principal components from a large number of predictors.
\newblock {\em Journal of the American statistical association}, 97(460):1167--1179.

\bibitem[Tibshirani, 2023]{tibshiranioverparametrized}
Tibshirani, R. (2023).
\newblock Overparametrized regression: Ridgeless interpolation.
\newblock {\em Lecture Notes from Course of Advanced Topics in Statistical Learning, Spring 2023}.

\bibitem[Tu and Zhou, 2011]{tu2011markowitz}
Tu, J. and Zhou, G. (2011).
\newblock Markowitz meets talmud: A combination of sophisticated and naive diversification strategies.
\newblock {\em Journal of Financial Economics}, 99(1):204--215.

\bibitem[Ullah and Welsh, 2024]{ullah2024effect}
Ullah, I. and Welsh, A. (2024).
\newblock On the effect of noise on fitting linear regression models.
\newblock {\em arXiv preprint arXiv:2408.07914}.

\bibitem[Xidonas et~al., 2020]{xidonas2020robust}
Xidonas, P., Steuer, R., and Hassapis, C. (2020).
\newblock Robust portfolio optimization: a categorized bibliographic review.
\newblock {\em Annals of Operations Research}, 292(1):533--552.

\bibitem[Zhang, 2005]{zhang2005value}
Zhang, L. (2005).
\newblock The value premium.
\newblock {\em The Journal of Finance}, 60(1):67--103.

\end{thebibliography}


\newpage
\appendix
\setcounter{equation}{0}
\setcounter{page}{1}
\renewcommand{\theequation}{\thesection.\arabic{equation}}

\begin{center}
\section*{Supplementary Material to ````Double Descent" in Portfolio Optimization: Dance between Theoretical Sharpe Ratio and Estimation Accuracy"}
\author{{Yonghe Lu}, {Yanrong Yang}, {Terry Zhang}
\\  {\small College of Business and Economics at 
The Australian National University}}
\\ \date{\today}
\end{center}
\addcontentsline{toc}{section}{Appendix} 
\vspace{-4em}
\part{} 
This document provides supplementary materials, including additional explanations, additional empirical results, proofs, theoretical findings for the ridge-regularized estimator, and references.
\parttoc 
\counterwithin{table}{section}
\counterwithin{figure}{section}

\section{Closer Look at Mean and SD of OOS Portfolio Return}\label{SubSection:Mean_SD}

We delve into the theoretical results to explore the underlying reasons behind the surprising ascent in the Sharpe ratio. We separately present the asymptotic results for the mean and standard deviation of out-of-sample return.

\begin{proposition}\label{Prop:LimitingMeanStd}
Under Assumptions \ref{ASSU_B1:HD}-\ref{ASSU_B3:MSR}, it holds {in probability} that
\begin{equation}
\mathbb{E}\left[\hatbomeastpseudo\T\br_0\Big\vert\bR\right] \rightarrow \mu\left(\ttheta,\tphi,\rho,\sigma^2\right)=
\left\{
    \begin{aligned}
    &\frac{\sigma}{\sqrt{\ttheta}}\frac{\ttheta}{1-\rho}, & \text{for } \rho < 1, \\
    &\frac{\sigma}{\sqrt{\ttheta}}\frac{\ttheta\rho(\tphi+1)^2}{\left(\rho-1\right)(\tphi+\rho)^2}, & \text{for } \rho > 1;
    \end{aligned}
\right.
\end{equation}
\begin{equation}
\mathbb{SD}\left[\hatbomeastpseudo\T\br_0\Big\vert\bR\right] \rightarrow \sigma\left(\ttheta,\tphi,\rho,\sigma^2\right)=
\left\{
    \begin{aligned}
    &\frac{\sigma}{\sqrt{\ttheta}}\sqrt{\frac{\ttheta+\rho}{\left(1-\rho\right)^3}}, \\
    & \qquad\qquad\qquad\qquad\qquad\qquad\qquad\qquad\qquad\qquad\text{for } \rho < 1, \\
    &\frac{\sigma}{\sqrt{\ttheta}}\sqrt{\frac{\rho\ttheta\tphi^2(\tphi+1)^2}{(\rho-1)(\tphi+\rho)^4} + \frac{\rho^2\ttheta(\tphi+1)^2}{(\rho-1)^3(\tphi+\rho)^2} + \frac{\rho}{(\rho-1)^3}}, \\
& \qquad\qquad\qquad\qquad\qquad\qquad\qquad\qquad\qquad\qquad\text{for } \rho > 1. \\
    \end{aligned}
\right.
\end{equation}
\end{proposition}

The following figure shows how the mean and standard deviation change with different values of  $\rho$  when \( \ttheta \) is 16 for various factor strengths. For  $\rho < 1$ , both the mean and standard deviation increase with  $\rho$  and are consistently overestimated (the true values of the mean and standard deviation are \( \sigma\sqrt{\ttheta} \) and  $\sigma$, respectively). When  $\rho > 1$ , both the mean and standard deviation decrease. Initially, they are overestimated but then become underestimated as  $\rho$  increases. Overall, the Sharpe ratio rises in the  $\rho > 1$  region because the mean decreases more slowly than the standard deviation. However, as the mean decreases more significantly, the Sharpe ratio eventually falls again after its initial rise.


\begin{figure}[H]
    \centering
    \includegraphics[trim=0 0 0 0,width=1\textwidth]{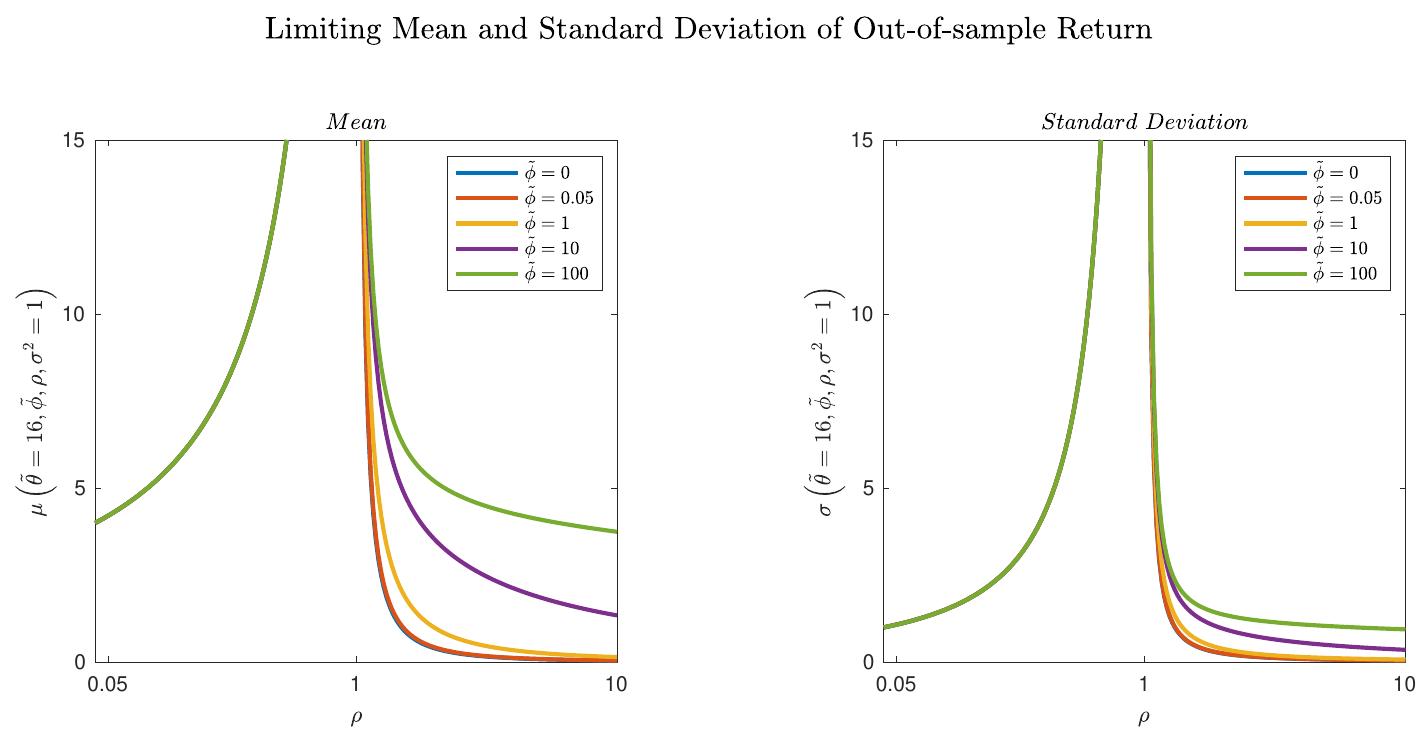}
    \caption{Asymptotic mean and standard deviation of OOS return in Proposition \ref{Prop:LimitingMeanStd}.}
    \label{Fig:Limiting_mean_std}
\end{figure}



\section{Bias-Variance Decomposition}\label{Appendix:decomp}
For an estimator $\hatbomeast$, which is a function of the observed data $\bR$, the out-of-sample prediction loss can be decomposed as:
\begin{align}
L_{\bm{R}}\left(\hatbomeast;\bomeast\right) &= \mathbb{E}\left[\left(\hatbomeast\T\br_0-{\bomeast}\T\br_0\right)^2\Big\vert\bR\right]
\nonumber\\
&=\mathbb{E}\left[\left(\left(\hatbomeast-\bomeast\right)\T\br_0\right)^2\Bigg\vert\bR\right]
\nonumber\\
&=\mathbb{E}\left[\left(\hatbomeast-{\bomeast}\right)\T\br_0\br_0\T\left(\hatbomeast-{\bomeast}\right)\Bigg\vert\bR\right]
\nonumber\\
&=\left(\mathbb{E}\left[\hatbomeast\big\vert \bR\right]-{\bomeast}\right)\T\left(\bSig+\bmu\bmu\T\right)\left(\mathbb{E}\left[\hatbomeast\big\vert \bR\right]-{\bomeast}\right)
\nonumber\\
&\quad \text{     }+\mathbb{E}\left[\left(\hatbomeast-\mathbb{E}\left[\hatbomeast\big\vert \bR\right]\right)\T\left(\bSig+\bmu\bmu\T\right)\left(\hatbomeast-\mathbb{E}\left[\hatbomeast\big\vert \bR\right]\right)\Big\vert\bR\right]
\nonumber\\
&=\underbrace{\left(\mathbb{E}\left[\hatbomeast\big\vert \bR\right]-{\bomeast}\right)\T\left(\bSig+\bmu\bmu\T\right)\left(\mathbb{E}\left[\hatbomeast\big\vert \bR\right]-{\bomeast}\right)}_{B_{\bR}\left[\hatbomeast;{\bomeast}\right]} +\underbrace{ \mathrm{Tr}\left[\mathrm{Cov}\left[\widehat{\bm{\omega^{\ast}}}\big\vert\bR\right]\left(\bSig+\bmu\bmu\T\right)\right]}_{V_{\bR}\left[\hatbomeast;{\bomeast}\right]} 
\nonumber
\end{align}

It is evident in the second-to-last line that  $\mathbb{E}\left[\hatbomeast\big\vert \bR\right]=\hatbomeast$, as $\hatbomeast$ is entirely determined by $\bR$. Consequently, given $\bR$, the expectation of $\hatbomeast$ is the estimator itself. This leads to $\hatbomeast-\mathbb{E}\left[\hatbomeast\big\vert \bR\right]=0$, thereby confirming that ${V_{\bR}\left[\hatbomeast;{\bomeast}\right]}=0$. To sum up, the out-of-sample prediction loss for estimator $\hatbomeast$ can also be expressed as:
\begin{align}
L_{\bR}\left[\hatbomeast;{\bomeast}\right]={B_{\bR}\left[\hatbomeast;{\bomeast}\right]}=\left(\hatbomeast-{\bomeast}\right)\T\left(\bSig+\bmu\bmu\T\right)\left(\hatbomeast-{\bomeast}\right).\nonumber
\end{align}

\section{Proofs}\label{Proofs}

\subsection{Lemma 1}
\begin{lemma}[Quadratic-form-close-to-the-trace, trace lemma, \cite{bai2010spectral} Lemma B.26]\label{lemma:Trace_Lemma}
Let $\bx$ $\in$ $\mathbb{R}^p$ have independent entries $x_i$ of zero mean, unit variance and $\mathbb{E}\left[|x_i|^K\right]\leq c_K$ for some $K\geq 1$. Then for $\bA \in \mathbb{R}^{p\times p}$ and $k\geq 1$,
\begin{equation}
    \mathbb{E}\left[\left\vert \bx\T\bA\bx-\Tr(\bA) \right\vert^k\right]\leq C_k\left[\left(c_4\Tr(\bA\bA\T)\right)^{k/2}+c_{2k}\Tr(\bA\bA\T)^{k/2}\right],\nonumber
\end{equation}
for some constant $C_k>0$ independent of $p$. For generic $\bA$, when the entries of $\bx$ have finite fourth moment, $\Var[\bx\T\bA\bx/p]=O(p^{-1})$.
Consequently,
\begin{equation}
    \dfrac{1}{p}\bx\T\bA\bx-\dfrac{1}{p}\Tr(\bA)=O_p\left(p^{-\frac{1}{2}}\right).
\end{equation}
\end{lemma} 
\setcounter{remark}{0}
\begin{remark}[For Lemma \ref{lemma:Trace_Lemma}]\label{remark_for_lemma1}
Assume we have another vector $\by$ that is independent of $\bx$, but has the same properties with $\bx$, then, $\dfrac{1}{p}\bx\T\bA\by \convergeInP 0 $.
\end{remark}

\bigskip

\subsection{Proofs for Proposition \ref{Prop:htheta}, Theorem \ref{Thm:sr}, Theorem \ref{Thm:pl}} \label{Appendix:proofs_P1T1T2}
\begin{proof}[\textbf{Proof:}]
Let $\hatbsm = \hatbmu-\bmu$ represent the centered sample mean vector. In accordance with the Data Generating Process (DGP) Assumption \ref{ASSU2:DGP}, we know that $\sqrt{T}\hatbsm$ has a zero mean and $\bSig$ (here $\bI_N$) as its covariance matrix. 

To prove for Proposition \ref{Prop:htheta}, we can work out the asymptotic limit of $\htheta_s$ first, then find a consistent estimator for $\ttheta$ reversely. We know that $\htheta_s=\hatbmu\T\hatbSig\pseudoinv\hatbmu$ can be decomposed into:
\begin{align}
\htheta_s &= \left(\bmu+\hatbsm\right)\T\hatbSig\pseudoinv\left(\bmu+\hatbsm\right)\nonumber\\
&=\bmu\T\hatbSig\pseudoinv\bmu+\hatbsm\T\hatbSig\pseudoinv\hatbsm+2\bmu\T\hatbSig\pseudoinv\hatbsm.\nonumber
\end{align}
For Theorem \ref{Thm:pl}, recall that the pseudoinverse estimator is defined by
\begin{align}
\hatbomeastpseudo = \dfrac{\sigma}{\sqrt{\htheta}}\hatbSig\pseudoinv\hatbmu,\nonumber
\end{align} 
then the prediction loss of the pseudoinverse estimator can be expanded using key intermediate terms 
\begin{align}
L_{\bR}\left[\hatbomeastpseudo;{\bomeast}\right]&={B_{\bR}\left(\hatbomeastpseudo;{\bomeast}\right)}
\nonumber\\
&=\left(\hatbomeastpseudo-{\bomeast}\right)\T\left(\bSig+\bmu\bmu\T\right)\left(\hatbomeastpseudo-{\bomeast}\right)
\nonumber\\
&=\sigma^2\left[\dfrac{\hatbmu\T\hatbSig\pseudoinv\bI_N\hatbSig\pseudoinv\hatbmu}{\htheta} - 2\dfrac{\hatbmu\T\hatbSig\pseudoinv{\bmu}}{\sqrt{\htheta}\sqrt{\theta}}+1\right.
\nonumber\\
&\quad \text{     }+\left.\dfrac{\left(\hatbmu\T\hatbSig\pseudoinv\bmu\right)^2}{\htheta} - 2\dfrac{\hatbmu\T\hatbSig\pseudoinv\bmu\sqrt{\theta}}{\sqrt{\htheta}}+{\theta}\right].\nonumber
\end{align}
For Theorem \ref{Thm:sr}, the Sharpe ratio of the pseudoinverse estimator can be expanded as 
\begin{align}
SR\left[\hatbomeastpseudo\right]=\dfrac{\hatbmu\T\hatbSig\pseudoinv\bmu}{\sqrt{\hatbmu\T\hatbSig\pseudoinv\bI_N\hatbSig\pseudoinv\hatbmu}}.\nonumber
\end{align}

Determining the asymptotic limit of the above three statistics simplifies to the task of determining the limits of six specific terms, which we denote as $D1$ to $D6$: $D1=\bmu\T\hatbSig\pseudoinv\bmu$, $D2=\hatbsm\T\hatbSig\pseudoinv\hatbsm$, $D3=\bmu\T\hatbSig\pseudoinv\hatbsm$, $D4=\bmu\T\hatbSig\pseudoinv\bI_N\hatbSig\pseudoinv\bmu$, $D5=\hatbsm\T\hatbSig\pseudoinv\bI_N\hatbSig\pseudoinv\hatbsm$,
$D6=\bmu\T\hatbSig\pseudoinv\bI_N\hatbSig\pseudoinv\hatbsm$. In the following analysis, we will derive the limits of these six terms respectively.

The proof becomes more straightforward when formulated in matrix terms. Consider the $T\times N$ data matrix $\bR$, with its $t^{\text{th}}$ row represents by $\br_t$. Similarly, $\bY$ denotes the $T\times N$ data matrix, where the $t^{\text{th}}$ row is $\by_t$. Then, $\bR=\be\bmu\T+\bY\bI_N$, where $\be$ is a vector of ones, with each entry equal to 1. Additionally, we define $\bH=\bI-\frac{\be\be\T}{T}$, then
\begin{align}
    \hatbSig &= \dfrac{1}{T-1}\left(\bR-\bar{\bR}\right)\T\left(\bR-\bar{\bR}\right)
    \nonumber\\ 
    &=\dfrac{1}{T-1}\left(\bH\bR\right)\T\left(\bH\bR\right)
    \nonumber\\ 
    &=\dfrac{1}{T-1}\left(\bH\bY\bI_N\right)\T\left(\bH\bY\bI_N\right)
    \nonumber\\ 
    &=\dfrac{1}{T-1}\bY\T\bH\bY.
    \nonumber
\end{align}
Note that the matrix $\bH$ is formulated as the identity matrix minus a rank one matrix. When considering the limit as $T\rightarrow\infty$, the difference between $1/T$ and $1/(T-1)$ becomes inconsequential. Hence we define $\calS=T\inv\bY\T\bH\bY$ and consider $\hatbSig=\calS$. Consequently, the pseudoinverse of $\hatbSig$ is given by $\hatbSig\pseudoinv=\calS\pseudoinv$.

\bigskip

\noindent\textbf{\underline{Limit of $D1$}}: Under Assumptions \ref{ASSU1:HD}-\ref{ASSU2:DGP}, we consider the ratio of $D1$ to $\bmu\T\bSig\inv\bmu$. We have
\[
\frac{\bmu\T\hatbSig\pseudoinv \bmu}{\bmu\T\bmu}=\bnu\T
\left(\frac{1}{T}\bY\T\bH\bY\right)\pseudoinv\bnu=\bnu\T
\calS\pseudoinv\bnu,
\]
where $\bnu=\bmu/\left\Vert\bmu\right\Vert_2$ is a vector of $\ell_2$ norm 1.

If $\bO$ is an orthogonal matrix, then $\bO\T\bY\T\bH\bY\bO \equalInLaw \bY\T\bH\bY$. Since $\bY$ is full of $i.i.d.$ random variables with $\mathbb{E}
[y_{11}]=0$, $\mathbb{E}[y_{11}^2]=1$ and $\mathbb{E}[y_{11}^4]<\infty$, it is invariant (in law) by left and right rotation. Also, the eigenvalues and eigenvectors of $\bY\T\bH\bY$ are
independent and its matrix of eigenvectors (i.e., eigenmatrix) is uniformly (i.e., Haar) distributed on the orthogonal group. To see this, we can refer to \cite{10.1214/009117906000001079}, where the authors proved that the eigenmatrix of large sample covariance matrix ($N$ increases with sample size proportionally) is nearly Haar distributed when the population covariance matrix is a multiple of the identity matrix. Let us write a spectral decomposition of $\bY\T\bH\bY$
\[
{\calS}=\frac{1}{T}\bY\T\bH\bY=\sum_{i=1}^N s_i \bu_i
\bu_i\T.
\]
We know that $s_i\neq0$ for all $i$ when $\rho<1$, so
\[
\bnu\T{\calS}\pseudoinv \bnu = \bnu\T{\calS}^{-1} \bnu=\sum_{i=1}^N \frac{1}{s_i} \left(\bnu
\T\bu_i\right)^2 .
\]
Based on Lemma \ref{lemma:Trace_Lemma}, we claim that
\[
\dfrac{1}{N}\left({\sqrt{N}}\bnu\right)\T{\calS}^{-1} \left({\sqrt{N}}\bnu\right)- \frac{1}{N}\sum_{i=1}^N \frac{1}{s_i}\rightarrow 0,\quad\text{in probability}.
\]
To see this, note that $\mathbb{E}((\bnu\T\bu_i)^2)=\| \bnu\|
_2^2/N=1/N$ because $\bu_i$ is uniformly distributed on the unit
sphere when $\bm{\Upsilon}$ (the matrix containing the $\bu_i$) is Haar distributed on the orthogonal group. Also, when $\rho<1$,
\[
\frac{1}{N}\sum_{i=1}^N \frac{1}{s_i} = \int\frac{1}{s}dF_{\calS}(s),
\]
where $F_{\calS}$ is the spectral measure of $T\inv\bY\T\bH\bY$. Recall the Mar\v{c}enko-Pastur equation, the Stieltjes transform of the the spectral distribution of $T\inv\bY\T\bH\bY$ tend to $m_{F_{MP,\rho}}(z)$ in probability and $m_{F_{MP,\rho}}(z)$ satisfies
\[
m_{F_{MP,\rho}}(z) = \dfrac{1}{1-\rho-\rho z m_{F_{MP,\rho}}(z)-z}.
\]
Thus, when $\rho<1$
\[
\frac{\bmu\T\hatbSig\pseudoinv\bmu}{\bmu\T\bI_N\inv\bmu}
\rightarrow m_{F_{MP,\rho}}(0)= \dfrac{1}{1-\rho}.
\]

When $\rho>1$, $s_i\neq0$ for all $i\leq T$ and we now claim that
\[
\dfrac{1}{N}\left({\sqrt{N}}\bnu\right)\T{\calS}\pseudoinv \left({\sqrt{N}}\bnu\right)- \frac{1}{N}\sum_{i=1}^T \frac{1}{s_i}\rightarrow 0,\quad\text{in probability}.
\]
Let $t_i$, $i=1,...,T$ denote the eigenvalues of $\bH\bY\bY\T\bH/N$. Then we may write $s_i=(N/T)t_i$, $i=1,...T$, and 
\[
\frac{1}{N}\sum_{i=1}^T \frac{1}{s_i} =\frac{1}{N}\frac{T}{N}\sum_{i=1}^T \frac{1}{t_i}= \left(\frac{T}{N}\right)^2\int\frac{1}{t}dF_{\bH\bY\bY\T\bH/N}(t),
\]
where $F_{\bH\bY\bY\T\bH/N}$ is the spectral measure of $\bH\bY\bY\T\bH/N$. Now as $T/N\rightarrow 1/\rho<1$, by the same arguments as above, we may conclude that in probability
\[
\frac{\bmu\T\hatbSig\pseudoinv\bmu}{\bmu\T\bI_N\inv\bmu}
\rightarrow \left(\frac{1}{\rho}\right)^2 \dfrac{1}{1-1/\rho}=\dfrac{1}{\rho(\rho-1)}.
\]
To sum up,
\begin{equation}
\frac{\bmu\T\hatbSig\pseudoinv\bmu}{\bmu\T\bI_N\inv\bmu} \convergeInP
\left\{
    \begin{aligned}
    &\dfrac{1}{1-\rho}, & \text{for } \rho < 1, \\
    &\dfrac{1}{\rho(\rho-1)}, & \text{for } \rho > 1.
    \end{aligned}
\right.\nonumber
\end{equation}
Under Assumption \ref{ASSU3:MSR}, where $\bmu\T\bI_N\inv\bmu=\theta\rightarrow\ttheta$, we get
\begin{equation}
\bmu\T\hatbSig\pseudoinv\bmu \convergeInP
\left\{
    \begin{aligned}
    &\dfrac{1}{1-\rho}\ttheta, & \text{for } \rho < 1, \\
    &\dfrac{1}{\rho(\rho-1)}\ttheta, & \text{for } \rho > 1.
    \end{aligned}
\right.\nonumber
\end{equation}

\bigskip

\noindent\textbf{\underline{Limit of $D2$}}: Recall that \( \hatbsm = \bSig^{\frac{1}{2}}\bY\T\be/T \) represents the vector of column means of \( \bR \), adjusted by subtracting the mean vector \( \bmu \). We have 
\begin{align}
\hatbsm\T\hatbSig\pseudoinv\hatbsm
=& \left(\bY\T\be/T\right)\T\left(\frac{1}{T}\bY\T\bH\bY\right)\pseudoinv\left(\bY\T\be/T\right)
\nonumber\\
=& 
\left(\bY\T\be/T\right)\T\left(\frac{1}{T}\bY\T\bH\bY\right)\pseudoinv\left(\bY\T\be/T\right)
\nonumber\\
=& 
\dfrac{1}{T}\left(\sqrt{T}\barbs\right)\T\calS\pseudoinv\left(\sqrt{T}\barbs\right),
\nonumber
\end{align}
where $\barbs=\bY\T\be/T$. This framework parallels the quadratic form of $\calS$ in the proof for $D1$, where a fixed unit vector $\bnu$ is present on both the left and right sides of the equation, thereby creating a quadratic form. However, in this case, $\sqrt{T}\barbs$ is a vector of $N$ random variables, with each entry following normal distribution $\NormDis(0,1)$. In this case, we can consider the trace Lemma \ref{lemma:Trace_Lemma} in Lemma B.26 of \cite{bai2010spectral}, which is fundamental in establishing initial heuristics for key identities of random matrix theory. It utilizes the approximation $\frac{1}{p}\bx\T\bA\bx\simeq\frac{1}{p}\Tr(\bA)$ for $\bx$ with independent zero-mean unit-variance entries and independent of $\bA$. Returning to our case, it is clear that \(\barbs\) is not independent of \(\calS\) (unless we assume $\by_t$ is multivariate Gaussian), thereby we can not apply this lemma directly. However, it is noteworthy that $T\times\barbs\T\calS\inv\barbs$ is the Hotelling's $T^2$ statistic. This aspect has been extensively investigated under large dimension scenarios by \cite{pan2011central}. By Theorem 2 in their work, under the same moment assumptions in Assumption \ref{ASSU2:DGP}, they suggested that $\barbs/\left\Vert\barbs\right\Vert$ can be viewed as a fixed unit vector when dealing with $\barbs\T\calS\inv\barbs/\left\Vert\barbs\right\Vert^2$ even if $\barbs$ is not independent of $\calS$. This enables us to find the limiting behaviour of $\barbs\T\calS\inv\barbs$ by using the same proof argument as in the proof for $D1$.

More broadly, \cite{pan2011central} have elucidated the asymptotic distributions of random quadratic forms, particularly those involving the sample mean and sample covariance. For a thorough understanding of their findings, readers are directed to Theorem 2 in their paper. The result related to our proof for D2, D3, D5, D6 (D5, D6 are defined in the proof of Theorem \ref{Thm:pl}) is as follows: given that $f(x)$ is analytic within an open interval, the expression
\begin{equation}
\sqrt{T} \left[\dfrac{\barbs\T f(\calS)\barbs}{\left\Vert\barbs\right\Vert^2}-\int f(s)dF_{MP,\rho_T}(s) \right] \convergeInLaw\NormDis\left(0,\mathbb{V}\right)
\end{equation}
holds, where $\mathbb{V}=\dfrac{2}{\rho}\left[\int f^2(s)dF_{MP,\rho}(s) - \left(\int f(s)dF_{MP,\rho}(s)\right)^2\right]$.
Furthermore,
\begin{equation}
\sqrt{T}\left[\left\Vert\barbs\right\Vert^2-\rho_T\right]\convergeInLaw\NormDis(0,2\rho).
\end{equation}
This implies that $\left\Vert\barbs\right\Vert^2$ converges towards $\rho$ in probability, irrespective of whether $\rho$ is larger than 1 or not. And now it becomes clear that the only difference between the limiting behaviors of $\hatbsm\T\hatbSig\pseudoinv\hatbsm$ and $\bmu\T\hatbSig\pseudoinv\bmu/\bmu\T\bI_N\inv\bmu$ lies in the multiplication factor of $\left\Vert\barbs\right\Vert^2$. Thus it is easy to derive from results of $D1$ that $\hatbsm\T\hatbSig\pseudoinv\hatbsm
\rightarrow {\rho}/{(1-\rho)}$ when $\rho<1$, and $\hatbsm\T\hatbSig\pseudoinv\hatbsm
\rightarrow {1}/{(\rho-1)}$ when $\rho>1$.
To sum up,
\begin{equation}
\hatbsm\T\hatbSig\pseudoinv\hatbsm\convergeInP
\left\{
    \begin{aligned}
    &\dfrac{\rho}{1-\rho}, & \text{for } \rho < 1, \\
    &\dfrac{1}{\rho-1}, & \text{for } \rho > 1.
    \end{aligned}
\right.\nonumber
\end{equation}

\bigskip

\noindent\textbf{\underline{Limit of $D3$}}: The challenge in determining the limit of $\bmu\T\hatbSig\pseudoinv\hatbsm$ arise from the fact that the quadratic form involves $\hatbSig\pseudoinv$ being multiplied by distinct vectors on its left and right sides. Therefore, we examine the square of $D3$, $\hatbsm\T\hatbSig\pseudoinv\bmu\bmu\T\hatbSig\pseudoinv\hatbsm$, where the vectors on both the left and right sides of the quadratic form are identical. We have 
\begin{align}
\hatbsm\T\hatbSig\pseudoinv\bmu\bmu\T\hatbSig\pseudoinv\hatbsm
=& \left(\dfrac{\bY\T\be}{T}\right)\T\left(\frac{1}{T}\bY\T\bH\bY\right)\pseudoinv\bmu\bmu\T\left(\frac{1}{T}\bY\T\bH\bY\right)\pseudoinv\left(\dfrac{\bY\T\be}{T}\right)
\nonumber\\
=& 
\dfrac{1}{T}\left(\sqrt{T}\barbs\right)\T\calS\pseudoinv\bmu\bmu\T\calS\pseudoinv\left(\sqrt{T}\barbs\right).
\nonumber
\end{align}
By adopting this approach, we can apply the same reasoning used in $D2$, where $\sqrt{T}\barbs$ is considered a vector of $N$ random variables, each following a normal distribution $\NormDis(0,1)$. Consequently we have 
\[
\dfrac{T}{N}\hatbsm\T\hatbSig\pseudoinv\bmu\bmu\T\hatbSig\pseudoinv\hatbsm = \dfrac{1}{N}\left(\sqrt{T}\barbs\right)\T\calS\pseudoinv\bmu\bmu\T\calS\pseudoinv\left(\sqrt{T}\barbs\right),
\]
such that according to the trace Lemma \ref{lemma:Trace_Lemma},
\[
\dfrac{T}{N}\hatbsm\T\hatbSig\pseudoinv\bmu\bmu\T\hatbSig\pseudoinv\hatbsm - \dfrac{1}{N}\Tr\left(\calS\pseudoinv\bmu\bmu\T\calS\pseudoinv\right)=O_p\left(N^{-\frac{1}{2}}\right).
\]
Subsequently, $\Tr\left(\calS\pseudoinv\bmu\bmu\T\calS\pseudoinv\right)=\bmu\T[\calS\pseudoinv]^2\bmu$. Easily we can get
\[
\frac{\bmu\T[\calS\pseudoinv]^2\bmu}{\bmu\T\bI_N\inv\bmu}=\bnu\T
\left[\left(\frac{1}{T}\bY\T\bH\bY\right)\pseudoinv\right]^2\bnu,
\]
where $\bnu=\bmu/\left\Vert\bmu\right\Vert_2$ is still a vector of $\ell_2$ norm 1. By the same argument as the proof for $D1$ that the eigenmatrix of sample covariance matrix $\calS$ is Haar distributed, we claim that when $\rho<1$,
\[
\frac{1}{N}\left(\sqrt{N}\bnu\right)\T{\calS}^{-2} \left(\sqrt{N}\bnu\right)- \frac{1}{p}\sum_{i=1}^N \frac{1}{s_i^2}=O_p\left(N^{-\frac{1}{2}}\right).
\]
And 
\[
\frac{1}{N}\sum_{i=1}^N \frac{1}{s_i^2} = \int\frac{1}{s^2}dF_{\calS}(s)\rightarrow \int\frac{1}{s^2}dF_{MP,\rho}(s),
\]
where $F_{\calS}$ is the spectral measure of $T\inv\bY\T\bH\bY$ and $F_{MP,\rho}$ is the limiting spectral distribution of $\calS$, the Mar\v cenko Pastur's law. 

Thus, when $\rho<1$,
\[
\frac{\bmu\T[\calS\pseudoinv]^2\bmu}{\bmu\T\bI_N\inv\bmu}
\rightarrow \dfrac{1}{(1-\rho)^3}.
\]

When $\rho>1$, we claim that 
\[
\frac{1}{N}\left(\sqrt{N}\bnu\right)\T{\calS}^{-2} \left(\sqrt{N}\bnu\right)- \frac{1}{N}\sum_{i=1}^T \frac{1}{s_i^2}=O_p\left(N^{-\frac{1}{2}}\right).
\]
Again, let $t_i$, $i=1,...,T$ denote the eigenvalues of $\bH\bY\bY\T\bH/N$. We write $s_i=(N/T)t_i$, $i=1,...T$, then we have
\[
\frac{1}{N}\sum_{i=1}^T \frac{1}{s_i^2} =\left(\frac{T}{N}\right)^3\frac{1}{T}\sum_{i=1}^T \frac{1}{t_i^2}= \left(\frac{T}{N}\right)^3\int\frac{1}{t^2}dF_{\bH\bY\bY\T\bH/N}(t)\rightarrow \left(\dfrac{1}{\rho}\right)^3\int\frac{1}{t^2}dF_{MP,1/\rho}(t),
\]
where $F_{\bH\bY\bY\T\bH/N}$ is the spectral measure of $\bH\bY\bY\T\bH/N$ and $F_{MP,1/\rho}$ is the LSD. Now by the same arguments as above, we conclude that in probability,
\[
\frac{\bmu\T\bSig^{-\frac{1}{2}}[\calS\pseudoinv]^2\bSig^{-\frac{1}{2}}\bmu}{\bmu\T\bSig\inv\bmu}
\rightarrow \dfrac{1}{(\rho-1)^3}.
\]
To sum up,
\begin{equation}
\frac{\bmu\T[\calS\pseudoinv]^2\bmu}{\bmu\T\bI_N\inv\bmu} \convergeInP
\left\{
    \begin{aligned}
    &\dfrac{1}{(1-\rho)^3}, & \text{for } \rho < 1, \\
    &\dfrac{1}{(\rho-1)^3}, & \text{for } \rho > 1.
    \end{aligned}
\right.\nonumber
\end{equation}
Furthermore, in accordance with Assumption \ref{ASSU3:MSR},
\begin{equation}
\bmu\T[\calS\pseudoinv]^2\bmu \convergeInP
\left\{
    \begin{aligned}
    &\dfrac{1}{(1-\rho)^3}\ttheta, & \text{for } \rho < 1, \\
    &\dfrac{1}{(\rho-1)^3}\ttheta, & \text{for } \rho > 1,
    \end{aligned}
\right.\nonumber
\end{equation}
Therefore, we have
\begin{equation}
\left\{
\begin{aligned}
&\dfrac{T}{N}\cdot\left(\hatbsm\T\hatbSig\pseudoinv\bmu\right)^2-\dfrac{1}{N}\cdot\dfrac{1}{(1-\rho)^3}\ttheta=O_p(N^{-\frac{1}{2}}), & \text{for } \rho < 1, \\
&\dfrac{T}{N}\cdot\left(\hatbsm\T\hatbSig\pseudoinv\bmu\right)^2-\dfrac{1}{N}\cdot\dfrac{1}{(\rho-1)^3}\ttheta=O_p(N^{-\frac{1}{2}}), & \text{for } \rho > 1,
\end{aligned}
\right.\nonumber
\end{equation}
\begin{equation}
\Rightarrow
\left\{
\begin{aligned}
&\left(\hatbsm\T\hatbSig\pseudoinv\bmu\right)^2-\dfrac{1}{T}\cdot\dfrac{1}{(1-\rho)^3}\ttheta=O_p(T^{-\frac{1}{2}}), & \text{for } \rho < 1, \\
&\left(\hatbsm\T\hatbSig\pseudoinv\bmu\right)^2-\dfrac{1}{T}\cdot\dfrac{1}{(\rho-1)^3}\ttheta=O_p(T^{-\frac{1}{2}}), & \text{for } \rho > 1.
\end{aligned}
\right.\nonumber
\end{equation}
Given that $\ttheta$ and $\rho$ are constants and bounded, and under Assumption \ref{ASSU1:HD} which assumes $T\rightarrow\infty$, it follows that $\bmu\T\hatbSig\pseudoinv\hatbsm$ converges to 0 for all values of $\rho$.

\bigskip

\noindent\textbf{\underline{Limit of $D4$}}: The proof is similar to that for $D3$. Easily we can get
\begin{equation}
\bmu\T\hatbSig\pseudoinv\bI_N\hatbSig\pseudoinv\bmu=\bmu\T[\calS\pseudoinv]^2\bmu \convergeInP
\left\{
    \begin{aligned}
    &\dfrac{1}{(1-\rho)^3}\ttheta, & \text{for } \rho < 1, \\
    &\dfrac{1}{(\rho-1)^3}\ttheta, & \text{for } \rho > 1.
    \end{aligned}
\right.\nonumber
\end{equation}

\bigskip

\noindent\textbf{\underline{Limit of $D5$}}: Again, similar to the statistic $\hatbsm\T\hatbSig\pseudoinv\hatbsm$, the only difference between the limiting behaviors of $\hatbsm\T\hatbSig\pseudoinv\bI_N\hatbSig\pseudoinv\hatbsm$ and $\bmu\T\hatbSig\pseudoinv\bI_N\hatbSig\pseudoinv\bmu/\bmu\T\bI_N\inv\bmu$ lies in the multiplication factor of $\left\Vert\barbs\right\Vert^2$, and we can easily get the following results:
\begin{equation}
\hatbsm\T\hatbSig\pseudoinv\bI_N\hatbSig\pseudoinv\hatbsm\convergeInP
\left\{
    \begin{aligned}
    &\dfrac{\rho}{(1-\rho)^3}, & \text{for } \rho < 1, \\
    &\dfrac{\rho}{(\rho-1)^3}, & \text{for } \rho > 1.
    \end{aligned}
\right.\nonumber
\end{equation}

\bigskip

\noindent\textbf{\underline{Limit of $D6$}}:
Mirroring the approach for $D3$, $D6$ also presents an imbalance in its quadratic form. To address this, we analyze the square of $D6$, given by $\hatbsm\T\hatbSig\pseudoinv\bI_N\hatbSig\pseudoinv\bmu\bmu\T\hatbSig\pseudoinv\bI_N\hatbSig\pseudoinv\hatbsm$. This formulation simplifies to:
\begin{align}
\hatbsm\T\hatbSig\pseudoinv\bI_N\hatbSig\pseudoinv\bmu\bmu\T\hatbSig\pseudoinv\bI_N\hatbSig\pseudoinv\hatbsm
=& \left(\dfrac{\bY\T\be}{T}\right)\T[\calS\pseudoinv]^2\bmu\bmu\T[\calS\pseudoinv]^2\left(\dfrac{\bY\T\be}{T}\right)
\nonumber\\
=& 
\dfrac{1}{T}\left(\sqrt{T}\barbs\right)\T[\calS\pseudoinv]^2\bmu\bmu\T[\calS\pseudoinv]^2\left(\sqrt{T}\barbs\right).
\nonumber
\end{align}
Using the same rationale as in $D3$ and invoking the Trace Lemma \ref{lemma:Trace_Lemma}, we deduce that
\[
\dfrac{T}{N}\hatbsm\T\hatbSig\pseudoinv\bI_N\hatbSig\pseudoinv\bmu\bmu\T\hatbSig\pseudoinv\bI_N\hatbSig\pseudoinv\hatbsm - \dfrac{1}{N}\Tr\left([\calS\pseudoinv]^2\bmu\bmu\T[\calS\pseudoinv]^2\right)=O_p\left(N^{-\frac{1}{2}}\right).
\]
Furthermore, $\Tr\left([\calS\pseudoinv]^2\bmu\bmu\T[\calS\pseudoinv]^2\right) = \bmu\T[\calS\pseudoinv]^4\bmu$. Similar to $D4$, we examine the ratio of $\bmu\T[\calS\pseudoinv]^4\bmu$ relative to $\bmu\T\bI_N\inv\bmu$. We have 
\[
\frac{\bmu\T[\calS\pseudoinv]^4\bmu}{\bmu\T\bI_N\inv\bmu}=\bnu\T
[\calS\pseudoinv]^4\bnu,
\]
where $\bnu=\bmu/\left\Vert\bmu\right\Vert_2$ is a vector of $\ell_2$ norm 1. By the same argument as the proof for $D1$ that the eigenmatrix of sample covariance matrix $\calS$ is Haar distributed, we claim that when $\rho<1$,
\[
\dfrac{1}{N}\left(\sqrt{N}\bnu\right)\T[\calS\pseudoinv]^4 \left(\sqrt{N}\bnu\right)- \frac{1}{N}\sum_{i=1}^N \frac{1}{s_i^4}=O_p\left(N^{-\frac{1}{2}}\right),
\]
and 
\[
\frac{1}{N}\sum_{i=1}^N \frac{1}{s_i^4} = \int\frac{1}{s^4}dF_{\calS}(s)\rightarrow \int\frac{1}{s^4}dF_{MP,\rho}(s).
\]
Thus, when $\rho<1$,
\[
\frac{\bmu\T[\calS\pseudoinv]^4\bmu}{\bmu\T\bI_N\inv\bmu}
\rightarrow \int\frac{1}{s^4}dF_{MP,\rho}(s).
\]

When $\rho>1$, we claim that 
\[
\dfrac{1}{N}\left(\sqrt{N}\bnu\right)\T[\calS\pseudoinv]^4 \left(\sqrt{N}\bnu\right)- \frac{1}{N}\sum_{i=1}^T \frac{1}{s_i^4}=O_p\left(N^{-\frac{1}{2}}\right),
\]
Again, let $t_i$, $i=1,...,T$ denote the eigenvalues of $\bH\bY\bY\T\bH/N$. We write $s_i=(N/T)t_i$, $i=1,...T$, then we have
\[
\frac{1}{N}\sum_{i=1}^T \frac{1}{s_i^4} =\left(\frac{T}{N}\right)^5\frac{1}{T}\sum_{i=1}^T \frac{1}{t_i^4}= \left(\frac{T}{N}\right)^5\int\frac{1}{t^4}dF_{\bH\bY\bY\T\bH/N}(t)\rightarrow \left(\dfrac{1}{\rho}\right)^5\int\frac{1}{t^4}dF_{MP,1/\rho}(t).
\]
Now by the same arguments as above, we conclude that in probability,
\[
\frac{\bmu\T[\calS\pseudoinv]^4\bmu}{\bmu\T\bI_N\inv\bmu}
\rightarrow \dfrac{1}{\rho^5}\int\frac{1}{t^4}dF_{MP,1/\rho}(t).
\]
To sum up,
\begin{equation}
\frac{\bmu\T[\calS\pseudoinv]^4\bmu}{\bmu\T\bI_N\inv\bmu} \rightarrow
\left\{
    \begin{aligned}
    &\int\frac{1}{s^4}dF_{MP,\rho}(s), & \text{for } \rho < 1, \\
    &\dfrac{1}{\rho^5}\int\frac{1}{t^4}dF_{MP,1/\rho}(t), & \text{for } \rho > 1.
    \end{aligned}
\right.\nonumber
\end{equation}
Finally, we have 
\begin{equation}
\left\{
\begin{aligned}
&\dfrac{T}{N}\cdot\left(\hatbsm\T\hatbSig\pseudoinv\bI_N\hatbSig\pseudoinv\bmu\right)^2-\dfrac{1}{N}\cdot\int\frac{1}{s^4}dF_{MP,\rho}(s)\cdot\ttheta=O_p(N^{-\frac{1}{2}}), & \text{for } \rho < 1, \\
&\dfrac{T}{N}\cdot\left(\hatbsm\T\hatbSig\pseudoinv\bI_N\hatbSig\pseudoinv\bmu\right)^2-\dfrac{1}{N}\cdot\dfrac{1}{\rho^5}\int\frac{1}{t^4}dF_{MP,1/\rho}(t)\cdot\ttheta=O_p(N^{-\frac{1}{2}}), & \text{for } \rho > 1,
\end{aligned}
\right.\nonumber
\end{equation}
\begin{equation}
\Rightarrow
\left\{
\begin{aligned}
&\left(\hatbsm\T\hatbSig\pseudoinv\bI_N\hatbSig\pseudoinv\bmu\right)^2-\dfrac{1}{T}\cdot\int\frac{1}{s^4}dF_{MP,\rho}(s)\cdot\ttheta=O_p(T^{-\frac{1}{2}}), & \text{for } \rho < 1, \\
&\left(\hatbsm\T\hatbSig\pseudoinv\bI_N\hatbSig\pseudoinv\bmu\right)^2-\dfrac{1}{T}\cdot\dfrac{1}{\rho^5}\int\frac{1}{t^4}dF_{MP,1/\rho}(t)\cdot\ttheta=O_p(T^{-\frac{1}{2}}), & \text{for } \rho > 1.
\end{aligned}
\right.\nonumber
\end{equation}
Given that $\ttheta$, $\rho$, $\int\frac{1}{s^4}dF_{MP,\rho}(s)$ and $\int\frac{1}{t^4}dF_{MP,1/\rho}(t)$ are bounded constants, and under Assumption \ref{ASSU1:HD} which assumes $T\rightarrow\infty$, it follows that $\bmu\T\hatbSig\pseudoinv\bI_N\hatbSig\pseudoinv\hatbsm$ converges to 0 for all values of $\rho$.

\bigskip

Now it is easy to get the asymptotic limit of $\htheta_s$:
\begin{equation}
\htheta_s=D1+D2+2D3\convergeInP
\left\{
    \begin{aligned}
    &\dfrac{\ttheta+\rho}{1-\rho}, & \text{for } \rho < 1, \\
    &\dfrac{\ttheta+\rho}{\rho(\rho-1)}, & \text{for } \rho > 1.
    \end{aligned}
\right.\nonumber 
\end{equation}
By rearrangement, we have the resulst in Theorem \ref{Prop:htheta}. Also, We can express $L_{\bR}\left[\hatbomeastpseudo;{\bomeast}\right]$ in terms of $D1$ to $D6$. The formulation is given by:
\begin{align}
L_{\bR}\left[\hatbomeastpseudo;{\bomeast}\right]
&= \sigma^2 \left[ \dfrac{D4 + D5 + 2D6}{\htheta} - 2 \dfrac{D1 + D3}{\sqrt{\htheta}\sqrt{\theta}} \right. \nonumber \\
& \left. + \dfrac{\left(D1 + D3\right)^2}{\htheta} - 2 \dfrac{\left(D1 + D3\right)\sqrt{\theta}}{\sqrt{\htheta}} + 1 + \theta \right]. \nonumber
\end{align}
By substituting the above limiting results, it is easy to get the conclusion in Theorem \ref{Thm:pl}.
Similarly, the Sharpe ratio can be expressed using $D1$ to $D6$: 
\begin{align}
SR\left[\hatbomeastpseudo\right]=\dfrac{D1+D3}{\sqrt{D4+D5+2D6}}.\nonumber
\end{align}

\bigskip

Here finishes the proof for Theorem \ref{Prop:htheta}, \ref{Thm:sr}, \ref{Thm:pl}. 
\end{proof}

\bigskip
\subsection{Proofs for Proposition \ref{Prop:htheta_sfm}, Theorem \ref{Thm:sr_sfm} and Theorem \ref{Thm:pl_sfm}}
\begin{proof}[\textbf{Proof:}]
Based on the DGP in Assumption \ref{ASSU_B2:DGP}, we can formulate single-factor stock returns as:
\[
\br_t=\bb\mu_f + \bb\sigma_f\cdot z_t + \sigmae\cdot\by_t, \quad 1\leq t\leq T,
\]
or in matrix form:
\[
\bR=\mu_f\cdot\be\bb\T + \sigma_f\bz\bb\T + \sigmae\bY,
\]
where $z_t$ and $\by_t$ are independent,  $\bz$ is a $T\times 1$ vector of $i.i.d.$ random variables with $\mathbb{E}[z_{1}]=0$, $\mathbb{E}[z_{1}]^2=1$, $\mathbb{E}[z_{1}]^4< \infty$, $\bY$ is a $T\times N$ matrix of $i.i.d.$ random variables with $\mathbb{E}[y_{11}]=0$, $\mathbb{E}[y_{11}]^2=1$, $\mathbb{E}[y_{11}]^4< \infty$ and $\be$ is a vector of ones. The proof becomes more straightforward when we formulated in matrices.

Again, let $\hatbsm=\hatbmu-\bmu$ represent the centered sample mean vector. Then under single factor model,
\[
\hatbsm = \frac{\sigma_f\bb\bz\T\be}{T}+\frac{\sigmae\bY\T\be}{T}.
\]

Recall $\bH=\bI-\frac{\be\be\T}{T}$ and the difference between $1/T$ and $1/(T-1)$ becomes negligible when considering the limit as $T\rightarrow\infty$, we consider
\begin{align}
    \hatbSig &= \dfrac{1}{T}\left(\bR-\bar{\bR}\right)\T\left(\bR-\bar{\bR}\right)
    \nonumber\\ 
    &=\dfrac{1}{T}\left(\bH\bR\right)\T\left(\bH\bR\right)
    \nonumber\\ 
    &=\dfrac{1}{T}\left(\sigma_f\bH\bz\bb\T +\bH\bY\sigmae\right)\T\left(\sigma_f\bH\bz\bb\T +\bH\bY\sigmae\right)
    \nonumber\\ 
    &= \sigmae^2\left(\dfrac{1}{T}\bY\T\bH\bY\right)+\sigma_f\sigmae\left(\dfrac{1}{T}\bY\T\bH\bz\right)\bb\T\nonumber\\
    &\quad+\sigma_f\sigmae\bb\left(\dfrac{1}{T}\bY\T\bH\bz\right)\T+\sigma_f^2\bb\bb\T\left(\dfrac{1}{T}\bz\T\bH\bz\right).
    \nonumber
\end{align}

Let us infer $\frac{1}{T}\bz\T\bH\bz$ first. We know that  $\frac{1}{T}\bz\T\bz=\frac{1}{T}\left(z_1^2+\cdots+z_T^2\right)$, and it is easy to get $\mathbb{E}[\frac{1}{T}\bz\T\bz]=1$ and $\Var[\frac{1}{T}\bz\T\bz]=\frac{1}{T}\left(\mathbb{E}[z_1]^4-1\right)$. As we assume $\mathbb{E}[z_1]^4$ is bounded, we have $\frac{1}{T}\bz\T\bz=1 + O_p(T^{-\frac{1}{2}})$. Also $\frac{\bz\T\be}{\sqrt{T}}$ follows $\NormDis(0,1)$, $\left(\frac{\bz\T\be}{\sqrt{T}}\right)^2$ follows $\chi^2(1)$, we have $\left(\frac{\bz\T\be}{T}\right)^2=O_p(T^{-1})$. Thus, $\frac{1}{T}\bz\T\bH\bz=1 + O_p(T^{-\frac{1}{2}})$.

Denote $\calS=\frac{1}{T}\bY\T\bH\bY$, we consider
\begin{align}
    \hatbSig =\sigmae^2\calS+\sigma_f\sigmae\left(\dfrac{1}{T}\bY\T\bH\bz\right)\bb\T+\sigma_f\sigmae\bb\left(\dfrac{1}{T}\bY\T\bH\bz\right)\T+\sigma_f^2\bb\bb\T.
    \nonumber
\end{align}
In the following, we prove for the case $\rho<1$ and $\rho>1$ separately. Similar to the proof of Theorem \ref{Prop:htheta}, \ref{Thm:sr} and \ref{Thm:pl}, we will determine the limits of 6 terms. They are $E1=\bmu\T\hatbSig\inv\bmu$, $E2=\hatbsm\T\hatbSig\inv\hatbsm$, $E3=\bmu\T\hatbSig\inv\hatbsm$,
$E4=\bmu\T\hatbSig\inv\bSig\hatbSig\inv\bmu$, $E5=\hatbsm\T\hatbSig\inv\bSig\hatbSig\inv\hatbsm$,
$E6=\bmu\T\hatbSig\inv\bSig\hatbSig\inv\hatbsm$. 

\bigskip
\subsubsection{\texorpdfstring{Case $\rho<1$ }%
{Case rho<1 }}
Before proving, we introduce the Sherman–Morrison formula. For any invertible matrix $\bM$, vectors $\bsm$, $\bn$ and a scalar $q$,
\begin{equation}
\left(\bM+q\bsm\bn\T\right)\inv=\bM\inv-\frac{q\bM\inv\bsm\bn\T\bM\inv}{1+q\bn\T\bM\inv\bsm}.\nonumber
\end{equation}

In this sub-section, we consider
\begin{align}
    \hatbSig =\sigmae^2\calS+\left(\sigma_f\sigmae\dfrac{1}{T}\bY\T\bH\bz\right)\bb\T+\bb\left(\sigma_f\sigmae\dfrac{1}{T}\bY\T\bH\bz\right)\T+\sigma_f^2\bb\bb\T.
    \nonumber
\end{align}

Denote 
\begin{align}
    \bC &=\sigmae^2\calS,\nonumber\\
    \bB &=\bC+\left(\sigma_f\sigmae\frac{1}{T}\bY\T\bH\bz\right)\bb\T,\nonumber\\
    \bA &=\bB+\bb\left(\sigma_f\sigmae\dfrac{1}{T}\bY\T\bH\bz\right)\T,\nonumber
\end{align}
then using the above Sherman–Morrison formula, we have
\begin{align}
    \bC\inv &=\frac{1}{\sigmae^2}\calS\inv,\nonumber\\
    \bB\inv &=\bC\inv-\frac{\frac{\sigma_f}{\sigmae^3}\calS\inv\left(\frac{1}{T}\bY\T\bH\bz\right)\bb\T\calS\inv}{1+\frac{\sigma_f}{\sigmae}\bb\T\calS\inv\frac{1}{T}\bY\T\bH\bz},\nonumber\\
    \bA\inv &=\bB\inv-\frac{\sigma_f\sigmae\bB\inv\bb\left(\frac{1}{T}\bY\T\bH\bz\right)\T\bB\inv}{1+\sigma_f\sigmae\left(\frac{1}{T}\bY\T\bH\bz\right)\T\bB\inv\bb},\nonumber\\
    \hatbSig\inv&=\bA\inv-\frac{\sigma_f^2\bA\inv\bb\bb\T\bA\inv}{1+\sigma_f^2\bb\T\bA\inv\bb}.\nonumber
\end{align}

We organize the following into three parts: First, in \ref{B311}, we present the asymptotic results for $E1$-$E6$. Second, in \ref{B312}, we list the asymptotic results for the intermediate terms that appear in \ref{B311}. Third, in \ref{B313}, we provide the proofs for the asymptotic results of the intermediate terms. Also, recall that $\phi=\frac{\sigma_f^2\left\Vert\bb\right\Vert^2}{\sigmae^2}$.

\paragraph{\texorpdfstring{Asymptotic results of $E1$-$E6$}{Asymptotic results of E1-E6}}\label{B311}

\noindent\textbf{\underline{Limit of $E1$}}: We have 
\begin{align}
E1&=\bmu\T\hatbSig\inv\bmu\nonumber\\
&=\mu_f^2\bb\T\hatbSig\inv\bb\nonumber\\
&=\mu_f^2\gamma_{\hatSig1}\nonumber\\
&\convergeInP \frac{\mu_f^2\frac{1}{\sigmae^2}c}{1-\frac{\sigma_f^2}{\sigmae^2}\rho c+\frac{\sigma_f^2}{\sigmae^2} c}=\frac{\ttheta}{1-\rho}.\nonumber
\end{align}

\bigskip

\noindent\textbf{\underline{Limit of $E2$}}: We have 
\begin{align}
E2&=\hatbsm\T\hatbSig\inv\hatbsm\nonumber\\
&=\left(\frac{\sigma_f\bb\bz\T\be}{T}+\frac{\sigmae\bY\T\be}{T}\right)\T\hatbSig\inv\left(\frac{\sigma_f\bb\bz\T\be}{T}+\frac{\sigmae\bY\T\be}{T}\right)\nonumber\\
&=E21+E22+2E23,\nonumber
\end{align}
where 
\begin{align}
E21&=\sigma_f^2\left(\frac{\bz\T\be}{\sqrt{T}}\right)^2\frac{1}{T}\bb\T\hatbSig\inv\bb=\sigma_f^2\left(\frac{\bz\T\be}{\sqrt{T}}\right)^2\frac{1}{T}\gamma_{\hatSig1}=O_p(T^{-1}),\nonumber\\
E22&=\sigmae^2\ttalpha_{\hatSig1}\convergeInP \frac{\rho}{1-\rho},\nonumber\\
E23&=\frac{1}{\sqrt{T}}\sigma_f\sigmae\left(\frac{\bz\T\be}{\sqrt{T}}\right)\tbeta_{\hatSig1}=O_p(T^{-1}).\nonumber
\end{align}
Thus, $E2\convergeInP \frac{\rho}{1-\rho}$.

\bigskip

\noindent\textbf{\underline{Limit of $E3$}}: We have 
\begin{align}
E3&=\bmu\T\hatbSig\inv\hatbsm\nonumber\\
&=\mu_f\bb\T\hatbSig\inv\left(\frac{\sigma_f\bb\bz\T\be}{T}+\frac{\sigmae\bY\T\be}{T}\right)\nonumber\\
&=E31+E32,\nonumber
\end{align}
where 
\begin{align}
E31&=\frac{1}{\sqrt{T}}\mu_f\sigma_f\left(\frac{\bz\T\be}{\sqrt{T}}\right)\bb\T\hatbSig\inv\bb=\frac{1}{\sqrt{T}}\mu_f\sigma_f\left(\frac{\bz\T\be}{\sqrt{T}}\right)\gamma_{\hatSig1}=O_p(T^{-\frac{1}{2}}),\nonumber\\
E32&=\mu_f\sigmae\tbeta_{\hatSig1}=O_p(T^{-\frac{1}{2}}).\nonumber
\end{align}
Thus, $E3=O_p(T^{-\frac{1}{2}})$.

\bigskip

\noindent\textbf{\underline{Limit of $E4$}}: We have 
\begin{align}
E4&=\bmu\T\hatbSig\inv\bSig\hatbSig\inv\bmu\nonumber\\
&=\mu_f^2\bb\T\hatbSig\inv\left(\sigma_f^2\bb\bb\T+\sigmae^2\bI_N\right)\hatbSig\inv\bb\nonumber\\
&=\mu_f^2\sigma_f^2\left(\bb\T\hatbSig\inv\bb\right)^2+\mu_f^2\sigmae^2\left(\bb\T\hatbSig\inv\hatbSig\inv\bb\right)\nonumber\\
&=\mu_f^2\sigma_f^2\gamma_{\hatSig1}^2+\mu_f^2\sigmae^2\gamma_{\hatSig2}\nonumber\\
&\convergeInP \frac{\frac{\mu_f^2}{\sigmae^2}\tilde{b}^2\frac{1}{(1-\rho)^3}\left[1+\frac{\sigma_f^2}{\sigmae^2}\tilde{b}^2\right]}{\left(1-\frac{\sigma_f^2}{\sigmae^2}\rho c+\frac{\sigma_f^2}{\sigmae^2} c\right)^2}=\frac{\ttheta}{(1-\rho)^3}.\nonumber
\end{align}

\bigskip

\noindent\textbf{\underline{Limit of $E5$}}: We have 
\begin{align}
E5&=\hatbsm\T\hatbSig\inv\bSig\hatbSig\inv\hatbsm\nonumber\\
&=\left(\frac{\sigma_f\bb\bz\T\be}{T}+\frac{\sigmae\bY\T\be}{T}\right)\T\hatbSig\inv\left(\sigma_f^2\bb\bb\T+\sigmae^2\bI_N\right)\hatbSig\inv\left(\frac{\sigma_f\bb\bz\T\be}{T}+\frac{\sigmae\bY\T\be}{T}\right)\nonumber\\
&=E51+E52+2E53+E54+E55+2E56,\nonumber
\end{align}
where 
\begin{align}
E51&=\sigma_f^4\left(\frac{\bz\T\be}{\sqrt{T}}\right)^2\frac{1}{T}\left(\bb\T\hatbSig\inv\bb\right)^2=\sigma_f^4\left(\frac{\bz\T\be}{\sqrt{T}}\right)^2\frac{1}{T}\gamma_{\hatSig1}^2=O_p(T^{-1}),\nonumber\\
E52&=\sigma_f^2\sigmae^2\tbeta_{\hatSig1}^2=O_p(T^{-1}),\nonumber\\
E53&=\sigma_f^3\sigmae\left(\frac{\bz\T\be}{\sqrt{T}}\right)\frac{1}{\sqrt{T}}\gamma_{\hatSig1}\tbeta_{\hatSig1}=O_p(T^{-1}),\nonumber\\
E54&=\sigma_f^2\sigmae^2\left(\frac{\bz\T\be}{\sqrt{T}}\right)^2\frac{1}{T}\gamma_{\hatSig2}=O_p(T^{-1})\nonumber\\
E55&=\sigmae^4\ttalpha_{\hatSig2}\convergeInP \frac{\rho}{(1-\rho)^3},\nonumber\\
E56 &=\sigma_f\sigmae^3\left(\frac{\bz\T\be}{\sqrt{T}}\right)\frac{1}{\sqrt{T}}\tbeta_{\hatSig2}=O_p(T^{-1})
\end{align}
Thus, $E5\convergeInP \frac{\rho}{(1-\rho)^3}$.

\bigskip

\noindent\textbf{\underline{Limit of $E6$}}: We have 
\begin{align}
E6&=\bmu\T\hatbSig\inv\bSig\hatbSig\inv\hatbsm\nonumber\\
&=\mu_f\bb\T\hatbSig\inv\left(\sigma_f^2\bb\bb\T+\sigmae^2\bI_N\right)\hatbSig\inv\left(\frac{\sigma_f\bb\bz\T\be}{T}+\frac{\sigmae\bY\T\be}{T}\right)\nonumber\\
&=E61+E62+E63+E64,\nonumber
\end{align}
where 
\begin{align}
E61&=\mu_f\sigma_f^3\left(\frac{\bz\T\be}{\sqrt{T}}\right)\left(\bb\T\hatbSig\inv\bb\right)^2\frac{1}{\sqrt{T}}=\mu_f\sigma_f^3\left(\frac{\bz\T\be}{\sqrt{T}}\right)\gamma_{\hatSig1}^2\frac{1}{\sqrt{T}}=O_p(T^{-\frac{1}{2}}),\nonumber\\
E62&=\mu_f\sigmae\sigma_f^2\gamma_{\hatSig1}\tbeta_{\hatSig1}=O_p(T^{-\frac{1}{2}}),\nonumber\\
E63&=\mu_f\sigmae^2\sigma_f\gamma_{\hatSig2}\left(\frac{\bz\T\be}{\sqrt{T}}\right)\frac{1}{\sqrt{T}}=O_p(T^{-\frac{1}{2}}),\nonumber\\
E64&=\mu_f\sigmae^3\tbeta_{\hatSig2}=O_p(T^{-\frac{1}{2}}).\nonumber
\end{align}
Thus, $E6=O_p(T^{-\frac{1}{2}})$.

\paragraph{\texorpdfstring{Asymptotic results of intermediate terms used in \ref{B311}}{Asymptotic results of intermediate terms used in \ref{B311}}}\label{B312}

\begin{align}
c&=\frac{\tilde{b}^2}{1-\rho}\nonumber\\
d&=\frac{\tilde{b}^2}{(1-\rho)^3}\nonumber\\
\beta_{S1}&=\left(\frac{1}{T}\bY\T\bH\bz\right)\T\calS\inv\bb=O_p\left(T^{-\frac{1}{2}}\right)\nonumber\\
\beta_{S2}&=\left(\frac{1}{T}\bY\T\bH\bz\right)\T\calS^{-2}\bb=O_p\left(T^{-\frac{1}{2}}\right)\nonumber\\
\tbeta_{S1}&=\left(\frac{1}{T}\bY\T\be\right)\T\calS\inv\bb=O_p\left(T^{-\frac{1}{2}}\right)\nonumber\\
\tbeta_{S2}&=\left(\frac{1}{T}\bY\T\be\right)\T\calS^{-2}\bb=O_p\left(T^{-\frac{1}{2}}\right)\nonumber\\
\beta_{B1}&=\left(\frac{1}{T}\bY\T\bH\bz\right)\T\bB\inv\bb\convergeInP -\frac{\sigma_f}{\sigmae^3}\frac{\rho}{1-\rho}\tilde{b}^2\nonumber\\
\tbeta_{B1}&=\left(\frac{1}{T}\bY\T\be\right)\T\bB\inv\bb=O_p\left(T^{-\frac{1}{2}}\right)\nonumber\\
\beta_{B2}&=\left(\frac{1}{T}\bY\T\bH\bz\right)\T\bB^{-2}\bb\convergeInP -\frac{\sigma_f}{\sigmae^5}\frac{2\rho-\rho^2}{(1-\rho)^3}\tilde{b}^2\nonumber\\
\tbeta_{B2}&=\left(\frac{1}{T}\bY\T\be\right)\T\bB^{-2}\bb=O_p\left(T^{-\frac{1}{2}}\right)\nonumber\\
\tbeta_{A1}&=\left(\frac{1}{T}\bY\T\be\right)\T\bA\inv\bb=O_p\left(T^{-\frac{1}{2}}\right)\nonumber\\
\tbeta_{A2}&=\left(\frac{1}{T}\bY\T\be\right)\T\bA^{-2}\bb=O_p\left(T^{-\frac{1}{2}}\right)\nonumber\\
\tbeta_{\hatSig1}&=\left(\frac{1}{T}\bY\T\be\right)\T\hatbSig\inv\bb=O_p\left(T^{-\frac{1}{2}}\right)\nonumber\\
\tbeta_{\hatSig2}&=\left(\frac{1}{T}\bY\T\be\right)\T\hatbSig^{-2}\bb=O_p\left(T^{-\frac{1}{2}}\right)\nonumber\\
\alpha_{S1} &= \left(\frac{1}{T}\bY\T\bH\bz\right)\T\calS\inv\left(\frac{1}{T}\bY\T\bH\bz\right)\convergeInP \rho\nonumber\\
\talpha_{S1} &= \left(\frac{1}{T}\bY\T\be\right)\T\calS\inv\left(\frac{1}{T}\bY\T\bH\bz\right)=O_p\left(T^{-\frac{1}{2}}\right)\nonumber\\
\ttalpha_{S1} &= \left(\frac{1}{T}\bY\T\be\right)\T\calS\inv\left(\frac{1}{T}\bY\T\be\right)\convergeInP \frac{\rho}{1-\rho}\nonumber\\
\alpha_{S2} &= \left(\frac{1}{T}\bY\T\bH\bz\right)\T\calS^{-2}\left(\frac{1}{T}\bY\T\bH\bz\right)\convergeInP \frac{\rho}{1-\rho}\nonumber\\
\talpha_{S2} &= \left(\frac{1}{T}\bY\T\be\right)\T\calS^{-2}\left(\frac{1}{T}\bY\T\bH\bz\right)=O_p\left(T^{-\frac{1}{2}}\right)\nonumber\\
\ttalpha_{S2} &= \left(\frac{1}{T}\bY\T\be\right)\T\calS^{-2}\left(\frac{1}{T}\bY\T\be\right)\convergeInP \frac{\rho}{(1-\rho)^3}\nonumber\\
\talpha_{B1} &= \left(\frac{1}{T}\bY\T\bH\bz\right)\T\bB\inv\left(\frac{1}{T}\bY\T\be\right)=O_p\left(T^{-\frac{1}{2}}\right)\nonumber\\
\ttalpha_{B1} &= \left(\frac{1}{T}\bY\T\be\right)\T\bB\inv\left(\frac{1}{T}\bY\T\be\right)\convergeInP \frac{1}{\sigmae^2}\frac{\rho}{1-\rho}\nonumber\\
\talpha_{B2} &= \left(\frac{1}{T}\bY\T\bH\bz\right)\T\bB^{-2}\left(\frac{1}{T}\bY\T\be\right)=O_p\left(T^{-\frac{1}{2}}\right)\nonumber\\
\ttalpha_{B2} &= \left(\frac{1}{T}\bY\T\be\right)\T\bB^{-2}\left(\frac{1}{T}\bY\T\be\right)\convergeInP \frac{1}{\sigmae^4}\frac{\rho}{(1-\rho)^3}\nonumber\\
\ttalpha_{A1} &= \left(\frac{1}{T}\bY\T\be\right)\T\bA\inv\left(\frac{1}{T}\bY\T\be\right)\convergeInP \frac{1}{\sigmae^2}\frac{\rho}{1-\rho}\nonumber\\
\ttalpha_{A2} &= \left(\frac{1}{T}\bY\T\be\right)\T\bA^{-2}\left(\frac{1}{T}\bY\T\be\right)\convergeInP \frac{1}{\sigmae^4}\frac{\rho}{(1-\rho)^3}\nonumber\\
\ttalpha_{\hatSig1} &= \left(\frac{1}{T}\bY\T\be\right)\T\hatbSig\inv\left(\frac{1}{T}\bY\T\be\right)\convergeInP \frac{1}{\sigmae^2}\frac{\rho}{1-\rho}\nonumber\\
\ttalpha_{\hatSig2} &= \left(\frac{1}{T}\bY\T\be\right)\T\hatbSig^{-2}\left(\frac{1}{T}\bY\T\be\right)\convergeInP \frac{1}{\sigmae^4}\frac{\rho}{(1-\rho)^3}\nonumber\\
\gamma_{S1} &=\bb\T\calS\inv\bb\convergeInP \frac{1}{1-\rho}\tilde{b}^2\nonumber\\
\gamma_{S2}&=\bb\T\calS^{-2}\bb\convergeInP\frac{1}{(1-\rho)^3}\tilde{b}^2\nonumber\\
\gamma_{B1}&=\bb\T\bB\inv\bb\convergeInP \frac{1}{\sigmae^2}\frac{1}{1-\rho}\tilde{b}^2\nonumber\\
\gamma_{B2}&=\bb\T\bB^{-2}\bb\convergeInP \frac{1}{\sigmae^4}\frac{1}{(1-\rho)^3}\tilde{b}^2\nonumber\nonumber\\
\gamma_{A1}&=\bb\T\bA\inv\bb\convergeInP \frac{\frac{1}{\sigmae^2}c}{1-\frac{\sigma_f^2}{\sigmae^2}\rho c}\nonumber\\
\gamma_{A2}&=\bb\T\bA^{-2}\bb\convergeInP \frac{\frac{1}{\sigmae^4}d\left[1+\frac{\sigma_f^2}{\sigmae^2}\rho(1-\rho)^3d\right]}{\left(1-\frac{\sigma_f^2}{\sigmae^2}\rho c\right)^2}\nonumber\\
\gamma_{\hatSig1}&=\bb\T\hatbSig\inv\bb\convergeInP \frac{\frac{1}{\sigmae^2}c}{1-\frac{\sigma_f^2}{\sigmae^2}\rho c+\frac{\sigma_f^2}{\sigmae^2} c}\nonumber\\
\gamma_{\hatSig2}&=\bb\T\hatbSig^{-2}\bb\convergeInP \frac{\frac{1}{\sigmae^4}d\left[1+\frac{\sigma_f^2}{\sigmae^2}\rho(1-\rho)^3d\right]}{\left(1-\frac{\sigma_f^2}{\sigmae^2}\rho c+\frac{\sigma_f^2}{\sigmae^2} c\right)^2}\nonumber
\end{align}

\paragraph{\texorpdfstring{Proofs for \ref{B312}}{Proofs for \ref{B312}}}\label{B313}

\noindent\textbf{\underline{Limit of $\gamma_{S1}$}}: By the same justification in $D1$, we have
\begin{equation}
\gamma_{S1}=\bb\T\calS\inv\bb \convergeInP \frac{\tilde{b}^2}{1-\rho}=c.\nonumber
\end{equation}

\bigskip
\noindent\textbf{\underline{Limit of $\beta_{S1}$}}: Consider the square of $\beta_{S1}$, we have
\begin{equation}
    T\beta_{S1}^2=\frac{1}{T}\bz\T\bH\bY\calS\inv\bb\bb\T\calS\inv\bY\T\bH\bz\nonumber.
\end{equation}
Based on trace Lemma \ref{lemma:Trace_Lemma}, we have 
\begin{align}
T\beta_{S1}^2&-\frac{1}{T}\bb\T\calS\inv\bY\T\bH\bY\calS\inv\bb=O_p(T^{-\frac{1}{2}}),\nonumber\\
T\beta_{S1}^2&-\bb\T\calS\inv\bb=O_p(T^{-\frac{1}{2}}).\nonumber
\end{align}
Thus, $\beta_{S1}=O_p(T^{-\frac{1}{2}})$.

\bigskip
\noindent\textbf{\underline{Limit of $\alpha_{S1}$}}: Based on trace Lemma \ref{lemma:Trace_Lemma}, we have
\begin{align}
    \alpha_{S1}&-\frac{1}{T}\Tr\left[\frac{1}{T}\bH\bY\calS\inv\bY\T\bH\right]=O_p(T^{-\frac{1}{2}})\nonumber\\
    \alpha_{S1}&-\frac{1}{T}\Tr\left[\calS\inv\frac{1}{T}\bY\T\bH\bY\right]=O_p(T^{-\frac{1}{2}})\nonumber\\
    \alpha_{S1}&-\frac{1}{T}\Tr\left[\bI_N\right]=O_p(T^{-\frac{1}{2}}).\nonumber  
\end{align}
Thus, $\alpha_{S1}\convergeInP\rho$.

\bigskip
\noindent\textbf{\underline{Limit of $\beta_{B1}$}}: 
\begin{align}
\beta_{B1}&=\left[\frac{1}{T}\bY\T\bH\bz\right]\T\bB\inv\bb\nonumber\\
&=\left[\frac{1}{T}\bY\T\bH\bz\right]\T\left[\bC\inv-\frac{\frac{\sigma_f}{\sigmae^3}\calS\inv\left(\frac{1}{T}\bY\T\bH\bz\right)\bb\T\calS\inv}{1+\frac{\sigma_f}{\sigmae}\bb\T\calS\inv\frac{1}{T}\bY\T\bH\bz}\right]\bb\nonumber\\
&=\frac{1}{\sigmae^2}\beta_{S1}-\frac{\frac{\sigma_f}{\sigmae^3}\alpha_{S1}\gamma_{S1}}{1+\frac{\sigma_f}{\sigmae}\beta_{S1}}\convergeInP-\frac{\sigma_f}{\sigmae^3}\rho c.\nonumber
\end{align}

\bigskip
\noindent\textbf{\underline{Limit of $\gamma_{B1}$}}: 
\begin{align}
\gamma_{B1}&=\bb\T\bB\inv\bb\nonumber\\
&=\bb\T\left[\bC\inv-\frac{\frac{\sigma_f}{\sigmae^3}\calS\inv\left(\frac{1}{T}\bY\T\bH\bz\right)\bb\T\calS\inv}{1+\frac{\sigma_f}{\sigmae}\bb\T\calS\inv\frac{1}{T}\bY\T\bH\bz}\right]\bb\nonumber\\
&=\frac{1}{\sigmae^2}\gamma_{S1}-\frac{\frac{\sigma_f}{\sigmae^3}\beta_{S1}\gamma_{S1}}{1+\frac{\sigma_f}{\sigmae}\beta_{S1}}\convergeInP\frac{1}{\sigmae^2} c.\nonumber
\end{align}

\bigskip
\noindent\textbf{\underline{Limit of $\gamma_{A1}$}}: 
\begin{align}
\gamma_{A1}&=\bb\T\bA\inv\bb\nonumber\\
&=\bb\T\left[\bB\inv-\frac{\sigma_f\sigmae\bB\inv\bb\left(\frac{1}{T}\bY\T\bH\bz\right)\T\bB\inv}{1+\sigma_f\sigmae\left(\frac{1}{T}\bY\T\bH\bz\right)\T\bB\inv\bb}\right]\bb\nonumber\\
&=\frac{\gamma_{B1}}{1+\sigma_f\sigmae\beta_{B1}}\convergeInP \frac{c}{\sigmae^2-\sigma_f^2\rho c}.\nonumber
\end{align}

\bigskip
\noindent\textbf{\underline{Limit of $\gamma_{\hatSig1}$}}:
\begin{align}
\gamma_{\hatSig1}&=\bb\T\hatbSig\inv\bb\nonumber\\
&=\bb\T\left[ \bA\inv-\frac{\sigma_f^2\bA\inv\bb\bb\T\bA\inv}{1+\sigma_f^2\bb\T\bA\inv\bb}\right]\bb\nonumber\\
&=\frac{\gamma_{A1}}{1+\sigma_f^2\gamma_{A1}}\convergeInP \frac{c}{\sigmae^2-\sigma_f^2\rho c+\sigma_f^2 c}.\nonumber
\end{align}

\bigskip
\noindent\textbf{\underline{Limit of $\tbeta_{S1}$}}: By the same justification in $D3$, we have
\begin{align}
\tbeta_{S1}=\left(\frac{1}{T}\bY\T\be\right)\T\calS\inv\bb=O_p(T^{-\frac{1}{2}}).\nonumber
\end{align}

\bigskip
\noindent\textbf{\underline{Limit of $\ttalpha_{S1}$}}: By the same justification in $D2$, we have
\begin{equation}
\ttalpha_{S1}=\left(\frac{1}{T}\bY\T\be\right)\T\calS\inv\left(\frac{1}{T}\bY\T\be\right)\convergeInP \frac{\rho}{1-\rho}.\nonumber
\end{equation}

\bigskip
\noindent\textbf{\underline{Limit of $\talpha_{S1}$}}: Consider the square of $\talpha_{S1}$, we have
\begin{equation}
    T\talpha_{S1}^2=\frac{1}{T}\bz\T\bH\bY\frac{1}{T}\calS\inv\bY\T\be\be\T\bY\calS\inv\frac{1}{T}\bY\T\bH\bz\nonumber.
\end{equation}
Based on trace Lemma \ref{lemma:Trace_Lemma}, we have 
\begin{align}
T\talpha_{S1}^2&-\frac{1}{T}\be\T\bY\calS\inv\frac{1}{T}\bY\T\bH\bY\frac{1}{T}\calS\inv\bY\T\be=O_p(T^{-\frac{1}{2}}),\nonumber\\
T\talpha_{S1}^2&-\ttalpha_{S1}=O_p(T^{-\frac{1}{2}}).\nonumber
\end{align}
Thus, $\talpha_{S1}=O_p(T^{-\frac{1}{2}})$.

\bigskip
\noindent\textbf{\underline{Limit of $\ttalpha_{B1}$}}: 
\begin{align}
\ttalpha_{B1}&=\left(\frac{1}{T}\bY\T\be\right)\T\bB\inv\left(\frac{1}{T}\bY\T\be\right)\nonumber\\
&=\left(\frac{1}{T}\bY\T\be\right)\T\left[\bC\inv-\frac{\frac{\sigma_f}{\sigmae^3}\calS\inv\left(\frac{1}{T}\bY\T\bH\bz\right)\bb\T\calS\inv}{1+\frac{\sigma_f}{\sigmae}\bb\T\calS\inv\frac{1}{T}\bY\T\bH\bz}\right]\inv\left(\frac{1}{T}\bY\T\be\right)\nonumber\\
&=\frac{1}{\sigmae^2}\ttalpha_{S1}-\frac{\frac{\sigma_f}{\sigmae^3}\talpha_{S1}\tbeta_{S1}}{1+\frac{\sigma_f}{\sigmae}\beta_{S1}}\convergeInP \frac{1}{\sigmae^2}\frac{\rho}{1-\rho}.\nonumber
\end{align}

\bigskip
\noindent\textbf{\underline{Limit of $\tbeta_{B1}$}}: 
\begin{align}
\tbeta_{B1}&=\left[\frac{1}{T}\bY\T\be\right]\T\bB\inv\bb\nonumber\\
&=\left[\frac{1}{T}\bY\T\be\right]\T\left[\bC\inv-\frac{\frac{\sigma_f}{\sigmae^3}\calS\inv\left(\frac{1}{T}\bY\T\bH\bz\right)\bb\T\calS\inv}{1+\frac{\sigma_f}{\sigmae}\bb\T\calS\inv\frac{1}{T}\bY\T\bH\bz}\right]\bb\nonumber\\
&=\frac{1}{\sigmae^2}\tbeta_{S1}-\frac{\frac{\sigma_f}{\sigmae^3}\talpha_{S1}\gamma_{S1}}{1+\frac{\sigma_f}{\sigmae}\beta_{S1}}=O_p(T^{-\frac{1}{2}})\nonumber
\end{align}

\bigskip
\noindent\textbf{\underline{Limit of $\talpha_{B1}$}}: 
\begin{align}
\talpha_{B1}&=\left(\frac{1}{T}\bY\T\bH\bz\right)\T\bB\inv\left(\frac{1}{T}\bY\T\be\right)\nonumber\\
&=\left(\frac{1}{T}\bY\T\bH\bz\right)\T\left[\bC\inv-\frac{\frac{\sigma_f}{\sigmae^3}\calS\inv\left(\frac{1}{T}\bY\T\bH\bz\right)\bb\T\calS\inv}{1+\frac{\sigma_f}{\sigmae}\bb\T\calS\inv\frac{1}{T}\bY\T\bH\bz}\right]\inv\left(\frac{1}{T}\bY\T\be\right)\nonumber\\
&=\frac{1}{\sigmae^2}\talpha_{S1}-\frac{\frac{\sigma_f}{\sigmae^3}\alpha_{S1}\tbeta_{S1}}{1+\frac{\sigma_f}{\sigmae}\beta_{S1}}=O_p(T^{-\frac{1}{2}}).\nonumber
\end{align}

\bigskip
\noindent\textbf{\underline{Limit of $\ttalpha_{A1}$}}: 
\begin{align}
\ttalpha_{A1}&=\left(\frac{1}{T}\bY\T\be\right)\T\bA\inv\left(\frac{1}{T}\bY\T\be\right)\nonumber\\
&=\left(\frac{1}{T}\bY\T\be\right)\T\left[\bB\inv-\frac{\sigma_f\sigmae\bB\inv\bb\left(\frac{1}{T}\bY\T\bH\bz\right)\T\bB\inv}{1+\sigma_f\sigmae\left(\frac{1}{T}\bY\T\bH\bz\right)\T\bB\inv\bb}\right]\inv\left(\frac{1}{T}\bY\T\be\right)\nonumber\\
&=\ttalpha_{B1}-\frac{{\sigma_f}{\sigmae}\tbeta_{B1}\talpha_{B1}}{1+{\sigma_f}{\sigmae}\beta_{B1}}\convergeInP\frac{1}{\sigmae^2}\frac{\rho}{1-\rho}.\nonumber
\end{align}

\bigskip
\noindent\textbf{\underline{Limit of $\tbeta_{A1}$}}: 
\begin{align}
\tbeta_{A1}&=\left(\frac{1}{T}\bY\T\be\right)\T\bA\inv\bb\nonumber\\
&=\left(\frac{1}{T}\bY\T\be\right)\T\left[\bB\inv-\frac{\sigma_f\sigmae\bB\inv\bb\left(\frac{1}{T}\bY\T\bH\bz\right)\T\bB\inv}{1+\sigma_f\sigmae\left(\frac{1}{T}\bY\T\bH\bz\right)\T\bB\inv\bb}\right]\inv\bb\nonumber\\
&=\tbeta_{B1}-\frac{{\sigma_f}{\sigmae}\tbeta_{B1}\beta_{B1}}{1+{\sigma_f}{\sigmae}\beta_{B1}}=O_p(T^{-\frac{1}{2}}).\nonumber
\end{align}

\bigskip
\noindent\textbf{\underline{Limit of $\ttalpha_{\hatSig1}$}}: 
\begin{align}
\ttalpha_{\hatSig1}&=\left(\frac{1}{T}\bY\T\be\right)\T\hatbSig\inv\left(\frac{1}{T}\bY\T\be\right)\nonumber\\
&=\left(\frac{1}{T}\bY\T\be\right)\T\left[\bA\inv-\frac{\sigma_f^2\bA\inv\bb\bb\T\bA\inv}{1+\sigma_f^2\bb\T\bA\inv\bb}\right]\inv\left(\frac{1}{T}\bY\T\be\right)\nonumber\\
&=\ttalpha_{A1}-\frac{{\sigma_f^2}\tbeta_{A1}\tbeta_{A1}}{1+{\sigma_f^2}\gamma_{A1}}\convergeInP\frac{1}{\sigmae^2}\frac{\rho}{1-\rho}.\nonumber
\end{align}

\bigskip
\noindent\textbf{\underline{Limit of $\tbeta_{\hatSig1}$}}: 
\begin{align}
\tbeta_{\hatSig1}&=\left(\frac{1}{T}\bY\T\be\right)\T\hatbSig\inv\bb\nonumber\\
&=\left(\frac{1}{T}\bY\T\be\right)\T\left[\bA\inv-\frac{\sigma_f^2\bA\inv\bb\bb\T\bA\inv}{1+\sigma_f^2\bb\T\bA\inv\bb}\right]\inv\bb\nonumber\\
&=\frac{\tbeta_{A1}}{1+{\sigma_f^2}\gamma_{A1}}=O_p(T^{-\frac{1}{2}}).\nonumber
\end{align}

\bigskip
\noindent\textbf{\underline{Limit of $\gamma_{S2}$}}: By the same justification in $D4$, we have
\begin{equation}
\bb\T\calS^{-2}\bb \convergeInP \dfrac{\tilde{b}^2}{(1-\rho)^3}=d.\nonumber
\end{equation}

\bigskip
\noindent\textbf{\underline{Limit of $\beta_{S2}$}}: Consider the square of $\beta_{S2}$, we have
\begin{equation}
    T\beta_{S2}^2=\frac{1}{T}\bz\T\bH\bY\calS^{-2}\bb\bb\T\calS^{-2}\bY\T\bH\bz\nonumber.
\end{equation}
Based on trace Lemma \ref{lemma:Trace_Lemma}, we have 
\begin{align}
T\beta_{S2}^2&-\frac{1}{T}\bb\T\calS^{-2}\bY\T\bH\bY\calS^{-2}\bb=O_p(T^{-\frac{1}{2}}),\nonumber\\
T\beta_{S2}^2&-\bb\T\calS^{-3}\bb=O_p(T^{-\frac{1}{2}}),\nonumber\\
T\beta_{S2}^2&=\dfrac{\tilde{b}^2\left(1+\rho\right)}{(1-\rho)^5}+O_p(T^{-\frac{1}{2}}).\nonumber
\end{align}
Thus, $\beta_{S2}=O_p(T^{-\frac{1}{2}})$.

\bigskip
\noindent\textbf{\underline{Limit of $\alpha_{S2}$}}: Based on trace Lemma \ref{lemma:Trace_Lemma} and the fact that the eigenmatrix of $\calS$ is uniformly (i.e., Haar) distributed, we have
\begin{align}
    \alpha_{S2}&-\frac{1}{T}\Tr\left[\frac{1}{T}\bH\bY\calS^{-2}\bY\T\bH\right]=O_p(T^{-\frac{1}{2}})\nonumber\\
    \alpha_{S2}&-\frac{1}{T}\Tr\left[\calS\inv\right]=O_p(T^{-\frac{1}{2}}).\nonumber
\end{align}
Thus, $\alpha_{S2}\convergeInP\frac{\rho}{1-\rho}$.

\bigskip
\noindent\textbf{\underline{Limit of $\beta_{B2}$}}: 
\begin{align}
\beta_{B2}&=\left(\frac{1}{T}\bY\T\bH\bz\right)\T\bB\inv\bB\inv\bb\nonumber\\
&=\left(\frac{1}{T}\bY\T\bH\bz\right)\T\left[\bC\inv-\frac{\frac{\sigma_f}{\sigmae^3}\calS\inv\left(\frac{1}{T}\bY\T\bH\bz\right)\bb\T\calS\inv}{1+\frac{\sigma_f}{\sigmae}\bb\T\calS\inv\frac{1}{T}\bY\T\bH\bz}\right]\nonumber\\
&\quad\left[\bC\inv-\frac{\frac{\sigma_f}{\sigmae^3}\calS\inv\left(\frac{1}{T}\bY\T\bH\bz\right)\bb\T\calS\inv}{1+\frac{\sigma_f}{\sigmae}\bb\T\calS\inv\frac{1}{T}\bY\T\bH\bz}\right]\bb\nonumber\\
&=\frac{1}{\sigmae^4}\beta_{S2}-\frac{\frac{\sigma_f}{\sigmae^5}\alpha_{S2}\gamma_{S1}}{1+\frac{\sigma_f}{\sigmae}\beta_{S1}}-\frac{\frac{\sigma_f}{\sigmae^5}\alpha_{S1}\gamma_{S2}}{1+\frac{\sigma_f}{\sigmae}\beta_{S1}}+\frac{\frac{\sigma_f^2}{\sigmae^6}\alpha_{S1}\beta_{S2}\gamma_{S1}}{(1+\frac{\sigma_f}{\sigmae}\beta_{S1})^2}\nonumber\\
&\convergeInP-\frac{\sigma_f}{\sigmae^5}\frac{\rho}{1-\rho}c-\frac{\sigma_f}{\sigmae^5}\rho d = -\frac{\sigma_f}{\sigmae^5}\tilde{b}^2\frac{2\rho-\rho^2}{(1-\rho)^3}.\nonumber
\end{align}

\bigskip
\noindent\textbf{\underline{Limit of $\gamma_{B2}$}}:
\begin{align}
\gamma_{B2}&=\bb\T\bB\inv\bB\inv\bb\nonumber\\
&=\frac{1}{\sigmae^4}\gamma_{S2}-\frac{\frac{\sigma_f}{\sigmae^5}\beta_{S2}\gamma_{S1}}{1+\frac{\sigma_f}{\sigmae}\beta_{S1}}-\frac{\frac{\sigma_f}{\sigmae^5}\beta_{S1}\gamma_{S2}}{1+\frac{\sigma_f}{\sigmae}\beta_{S1}}+\frac{\frac{\sigma_f^2}{\sigmae^6}\beta_{S1}\beta_{S2}\gamma_{S1}}{(1+\frac{\sigma_f}{\sigmae}\beta_{S1})^2}\nonumber\\
&\convergeInP \frac{1}{\sigmae^4}d=\frac{1}{\sigmae^4}\tilde{b}^2\frac{1}{(1-\rho)^3}.\nonumber
\end{align}

\bigskip
\noindent\textbf{\underline{Limit of $\gamma_{A2}$}}: 
\begin{align}
\gamma_{A2}&=\bb\T\bA\inv\bA\inv\bb\nonumber\\
&=\bb\T\left[\bB\inv-\frac{\sigma_f\sigmae\bB\inv\bb\left(\frac{1}{T}\bY\T\bH\bz\right)\T\bB\inv}{1+\sigma_f\sigmae\beta_{B1}}\right]\nonumber\\
&\quad\left[\bB\inv-\frac{\sigma_f\sigmae\bB\inv\bb\left(\frac{1}{T}\bY\T\bH\bz\right)\T\bB\inv}{1+\sigma_f\sigmae\beta_{B1}}\right]\bb\nonumber\\
&=\gamma_{B2}-\frac{{\sigma_f}{\sigmae}\gamma_{B2}\beta_{B1}}{1+{\sigma_f}{\sigmae}\beta_{B1}}-\frac{{\sigma_f}{\sigmae}\gamma_{B1}\beta_{B2}}{1+{\sigma_f}{\sigmae}\beta_{B1}}+\frac{{\sigma_f^2}{\sigmae^2}\gamma_{B1}\beta_{B2}\beta_{B1}}{(1+{\sigma_f}{\sigmae}\beta_{B1})^2}\nonumber\\
& = \frac{\gamma_{B2}+\sigma_f\sigmae\gamma_{B2}\beta_{B1}-\sigma_f\sigmae\beta_{B2}\gamma_{B1}}{(1+{\sigma_f}{\sigmae}\beta_{B1})^2}\nonumber\\
&\convergeInP \frac{\frac{1}{\sigmae^4}d\left[1+\frac{\sigma_f^2}{\sigmae^2}\rho(1-\rho)^3d\right]}{\left[1-\frac{\sigma_f^2}{\sigmae^2}\rho c\right]^2}.\nonumber
\end{align}

\bigskip
\noindent\textbf{\underline{Limit of $\gamma_{\hatSig2}$}}:
\begin{align}
\gamma_{\hatSig2}&=\bb\T\hatbSig\inv\hatbSig\inv\bb\nonumber\\
&=\bb\T\left[\bA\inv-\frac{\sigma_f^2\bA\inv\bb\bb\T\bA\inv}{1+\sigma_f^2\gamma_{A1}}\right]\left[\bA\inv-\frac{\sigma_f^2\bA\inv\bb\bb\T\bA\inv}{1+\sigma_f^2\gamma_{A1}}\right]\bb\nonumber\\
& = \frac{\gamma_{A2}}{(1+\sigma_f^2\gamma_{A1})^2}\nonumber\\
&\convergeInP \frac{\frac{1}{\sigmae^4}d\left[1+\frac{\sigma_f^2}{\sigmae^2}\rho(1-\rho)^3d\right]}{\left[1-\frac{\sigma_f^2}{\sigmae^2}\rho c+\frac{\sigma_f^2}{\sigmae^2} c\right]^2}.\nonumber
\end{align}

\bigskip
\noindent\textbf{\underline{Limit of $\talpha_{S2}$}}: Consider the square of $\talpha_{S2}$, we have
\begin{equation}
    T\talpha_{S2}^2=\frac{1}{T}\bz\T\bH\bY\frac{1}{T}\calS^{-2}\bY\T\be\be\T\bY\calS^{-2}\frac{1}{T}\bY\T\bH\bz\nonumber.
\end{equation}
Based on trace Lemma \ref{lemma:Trace_Lemma}, we have 
\begin{align}
T\talpha_{S2}^2&-\left(\frac{1}{T}\bY\T\be\right)\T\calS^{-3}\left(\frac{1}{T}\bY\T\be\right)=O_p(T^{-\frac{1}{2}}),\nonumber\\
T\talpha_{S2}^2&\convergeInP\frac{(1+\rho)\rho}{(1-\rho)^5}.\nonumber
\end{align}
Thus, $\talpha_{S2}=O_p(T^{-\frac{1}{2}})$.

\bigskip
\noindent\textbf{\underline{Limit of $\ttalpha_{S2}$}}: By the same justification in $D5$, we have
\begin{equation}
\ttalpha_{S2}=\left(\frac{1}{T}\bY\T\be\right)\T\calS^{-2}\left(\frac{1}{T}\bY\T\be\right)\convergeInP \frac{\rho}{(1-\rho)^3}.\nonumber
\end{equation}

\bigskip
\noindent\textbf{\underline{Limit of $\tbeta_{S2}$}}: By the same justification in $D6$, we have
\begin{align}
\tbeta_{S2}=\left(\frac{1}{T}\bY\T\be\right)\T\calS^{-2}\bb=O_p(T^{-\frac{1}{2}}).\nonumber
\end{align}

\bigskip
\noindent\textbf{\underline{Limit of $\ttalpha_{B2}$}}:
\begin{align}
\ttalpha_{B2}&=\left(\frac{1}{T}\bY\T\be\right)\T\bB\inv\bB\inv\left(\frac{1}{T}\bY\T\be\right)\nonumber\\
&=\left(\frac{1}{T}\bY\T\be\right)\T\left[\bC\inv-\frac{\frac{\sigma_f}{\sigmae^3}\calS\inv\left(\frac{1}{T}\bY\T\bH\bz\right)\bb\T\calS\inv}{1+\frac{\sigma_f}{\sigmae}\bb\T\calS\inv\frac{1}{T}\bY\T\bH\bz}\right]\nonumber\\
&\quad\left[\bC\inv-\frac{\frac{\sigma_f}{\sigmae^3}\calS\inv\left(\frac{1}{T}\bY\T\bH\bz\right)\bb\T\calS\inv}{1+\frac{\sigma_f}{\sigmae}\bb\T\calS\inv\frac{1}{T}\bY\T\bH\bz}\right]\left(\frac{1}{T}\bY\T\be\right)\nonumber\\
&=\frac{1}{\sigmae^4}\ttalpha_{S2}+O_p(T^{-\frac{1}{2}})\nonumber\\
&\convergeInP\frac{1}{\sigmae^4}\frac{\rho}{(1-\rho)^3}.\nonumber
\end{align}

\bigskip
\noindent\textbf{\underline{Limit of $\tbeta_{B2}$}}: 
\begin{align}
\tbeta_{B2}&=\left(\frac{1}{T}\bY\T\be\right)\T\bB\inv\bB\inv\bb\nonumber\\
&=\left(\frac{1}{T}\bY\T\be\right)\T\left[\bC\inv-\frac{\frac{\sigma_f}{\sigmae^3}\calS\inv\left(\frac{1}{T}\bY\T\bH\bz\right)\bb\T\calS\inv}{1+\frac{\sigma_f}{\sigmae}\bb\T\calS\inv\frac{1}{T}\bY\T\bH\bz}\right]\nonumber\\
&\quad\left[\bC\inv-\frac{\frac{\sigma_f}{\sigmae^3}\calS\inv\left(\frac{1}{T}\bY\T\bH\bz\right)\bb\T\calS\inv}{1+\frac{\sigma_f}{\sigmae}\bb\T\calS\inv\frac{1}{T}\bY\T\bH\bz}\right]\bb\nonumber\\
&=\frac{1}{\sigmae^4}\tbeta_{S2}-\frac{\frac{\sigma_f}{\sigmae^5}\talpha_{S2}\gamma_{S1}}{1+\frac{\sigma_f}{\sigmae}\beta_{S1}}-\frac{\frac{\sigma_f}{\sigmae^5}\talpha_{S1}\gamma_{S2}}{1+\frac{\sigma_f}{\sigmae}\beta_{S1}}+\frac{\frac{\sigma_f^2}{\sigmae^6}\talpha_{S1}\beta_{S2}\gamma_{S1}}{(1+\frac{\sigma_f}{\sigmae}\beta_{S1})^2}\nonumber\\
& =O_p(T^{-\frac{1}{2}}).\nonumber
\end{align}

\bigskip
\noindent\textbf{\underline{Limit of $\talpha_{B2}$}}: 
\begin{align}
\talpha_{B2}&=\left(\frac{1}{T}\bY\T\bH\bz\right)\T\bB\inv\bB\inv\left(\frac{1}{T}\bY\T\be\right)\nonumber\\
&=\left(\frac{1}{T}\bY\T\bH\bz\right)\T\left[\bC\inv-\frac{\frac{\sigma_f}{\sigmae^3}\calS\inv\left(\frac{1}{T}\bY\T\bH\bz\right)\bb\T\calS\inv}{1+\frac{\sigma_f}{\sigmae}\bb\T\calS\inv\frac{1}{T}\bY\T\bH\bz}\right]\nonumber\\
&\quad\left[\bC\inv-\frac{\frac{\sigma_f}{\sigmae^3}\calS\inv\left(\frac{1}{T}\bY\T\bH\bz\right)\bb\T\calS\inv}{1+\frac{\sigma_f}{\sigmae}\bb\T\calS\inv\frac{1}{T}\bY\T\bH\bz}\right]\left(\frac{1}{T}\bY\T\be\right)\nonumber\\
&=\frac{1}{\sigmae^4}\talpha_{S2}-\frac{\frac{\sigma_f}{\sigmae^5}\alpha_{S2}\tbeta_{S1}}{1+\frac{\sigma_f}{\sigmae}\beta_{S1}}-\frac{\frac{\sigma_f}{\sigmae^5}\alpha_{S1}\tbeta_{S2}}{1+\frac{\sigma_f}{\sigmae}\beta_{S1}}+\frac{\frac{\sigma_f^2}{\sigmae^6}\alpha_{S1}\beta_{S2}\tbeta_{S1}}{(1+\frac{\sigma_f}{\sigmae}\beta_{S1})^2}\nonumber\\
& =O_p(T^{-\frac{1}{2}}).\nonumber
\end{align}

\bigskip
\noindent\textbf{\underline{Limit of $\ttalpha_{A2}$}}: 
\begin{align}
\ttalpha_{A2}&=\left(\frac{1}{T}\bY\T\be\right)\T\bA\inv\bA\inv\left(\frac{1}{T}\bY\T\be\right)\nonumber\\
&=\left(\frac{1}{T}\bY\T\be\right)\T\left[\bB\inv-\frac{\sigma_f\sigmae\bB\inv\bb\left(\frac{1}{T}\bY\T\bH\bz\right)\T\bB\inv}{1+\sigma_f\sigmae\beta_{B1}}\right]\nonumber\\
&\quad\left[\bB\inv-\frac{\sigma_f\sigmae\bB\inv\bb\left(\frac{1}{T}\bY\T\bH\bz\right)\T\bB\inv}{1+\sigma_f\sigmae\beta_{B1}}\right]\left(\frac{1}{T}\bY\T\be\right)\nonumber\\
&=\ttalpha_{B2}-\frac{{\sigma_f}{\sigmae}\tbeta_{B2}\talpha_{B1}}{1+{\sigma_f}{\sigmae}\beta_{B1}}-\frac{{\sigma_f}{\sigmae}\tbeta_{B1}\talpha_{B2}}{1+{\sigma_f}{\sigmae}\beta_{B1}}+\frac{{\sigma_f^2}{\sigmae^2}\tbeta_{B1}\beta_{B2}\talpha_{B1}}{(1+{\sigma_f}{\sigmae}\beta_{B1})^2}\nonumber\\
&\convergeInP \frac{1}{\sigmae^4}\frac{\rho}{(1-\rho)^3}.\nonumber
\end{align}

\bigskip
\noindent\textbf{\underline{Limit of $\tbeta_{A2}$}}: 
\begin{align}
\tbeta_{A2}&=\left(\frac{1}{T}\bY\T\be\right)\T\bA\inv\bA\inv\bb\nonumber\\
&=\left(\frac{1}{T}\bY\T\be\right)\T\left[\bB\inv-\frac{\sigma_f\sigmae\bB\inv\bb\left(\frac{1}{T}\bY\T\bH\bz\right)\T\bB\inv}{1+\sigma_f\sigmae\beta_{B1}}\right]\nonumber\\
&\quad\left[\bB\inv-\frac{\sigma_f\sigmae\bB\inv\bb\left(\frac{1}{T}\bY\T\bH\bz\right)\T\bB\inv}{1+\sigma_f\sigmae\beta_{B1}}\right]\bb\nonumber\\
&=\tbeta_{B2}-\frac{{\sigma_f}{\sigmae}\tbeta_{B2}\beta_{B1}}{1+{\sigma_f}{\sigmae}\beta_{B1}}-\frac{{\sigma_f}{\sigmae}\tbeta_{B1}\beta_{B2}}{1+{\sigma_f}{\sigmae}\beta_{B1}}+\frac{{\sigma_f^2}{\sigmae^2}\tbeta_{B1}\beta_{B2}\beta_{B1}}{(1+{\sigma_f}{\sigmae}\beta_{B1})^2}\nonumber\\
&=\frac{\tbeta_{B2}+\sigma_f\sigmae\beta_{B1}\tbeta_{B2}-\sigma_f\sigmae\tbeta_{B1}\beta_{B2}}{(1+{\sigma_f}{\sigmae}\beta_{B1})^2}=O_p(T^{-\frac{1}{2}}).\nonumber
\end{align}

\bigskip
\noindent\textbf{\underline{Limit of $\ttalpha_{\hatSig2}$}}: 
\begin{align}
\ttalpha_{\hatSig2}&=\left(\frac{1}{T}\bY\T\be\right)\T\hatbSig\inv\hatbSig\inv\left(\frac{1}{T}\bY\T\be\right)\nonumber\\
&=\left(\frac{1}{T}\bY\T\be\right)\T\left[\bA\inv-\frac{\sigma_f^2\bA\inv\bb\bb\T\bA\inv}{1+\sigma_f^2\gamma_{A1}}\right]\left[\bA\inv-\frac{\sigma_f^2\bA\inv\bb\bb\T\bA\inv}{1+\sigma_f^2\gamma_{A1}}\right]\left(\frac{1}{T}\bY\T\be\right)\nonumber\\
&=\ttalpha_{A2}-\frac{\sigma_f^2\tbeta_{A2}\tbeta_{A1}}{1+\sigma_f^2\gamma_{A1}}-\frac{\sigma_f^2\tbeta_{A1}\tbeta_{A2}}{1+\sigma_f^2\gamma_{A1}}+\frac{{\sigma_f^4}\tbeta_{A1}\gamma_{A2}\tbeta_{A1}}{(1+\sigma_f^2\gamma_{A1})^2}\nonumber\\
&\convergeInP \frac{1}{\sigmae^4}\frac{\rho}{(1-\rho)^3}.\nonumber
\end{align}

\bigskip
\noindent\textbf{\underline{Limit of $\tbeta_{\hatSig2}$}}:
\begin{align}
\tbeta_{\hatSig2}&=\left(\frac{1}{T}\bY\T\be\right)\T\hatbSig\inv\hatbSig\inv\bb\nonumber\\
&=\left(\frac{1}{T}\bY\T\be\right)\T\left[\bA\inv-\frac{\sigma_f^2\bA\inv\bb\bb\T\bA\inv}{1+\sigma_f^2\gamma_{A1}}\right]\left[\bA\inv-\frac{\sigma_f^2\bA\inv\bb\bb\T\bA\inv}{1+\sigma_f^2\gamma_{A1}}\right]\bb\nonumber\\
&=\tbeta_{A2}-\frac{\sigma_f^2\gamma_{A2}\tbeta_{A1}}{1+\sigma_f^2\gamma_{A1}}-\frac{\sigma_f^2\gamma_{A1}\tbeta_{A2}}{1+\sigma_f^2\gamma_{A1}}+\frac{{\sigma_f^4}\gamma_{A1}\gamma_{A2}\tbeta_{A1}}{(1+\sigma_f^2\gamma_{A1})^2}\nonumber\\
&=O_p(T^{-\frac{1}{2}}).\nonumber
\end{align}

\bigskip

\subsubsection{\texorpdfstring{Case $\rho>1$ }%
{Case rho>1 }}
We first introduce formulas that may be used to obtain the pseudoinverse of a matrix $\bM$ modified by a rank-1 matrix. These formulas were developed in \cite{meyer1973generalized}. For a matrix $\bM$, vectors $\bsm$, $\bn$, we adopt the following notation:
\begin{itemize}
\renewcommand\labelitemi{}
\item $\left\Vert\cdot\right\Vert$ - the Euclidean norm,
\item $R(\cdot)$ - the range or column space,
\item $\bk$ - the column $\bM\pseudoinv \bsm$,
\item $\bh$ - the row $\bn\T\bM\pseudoinv $,
\item $\bu$ - the column $\left(\bI_N-\bM\bM\pseudoinv\right)\bsm$,
\item $\bv$ - the row $\bn\T\left(\bI_N-\bM\pseudoinv\bM\right)$,
\item $\beta$ - the scalar $1+\bn\T\bM\pseudoinv\bsm$.
\end{itemize}
For a nonzero vector $\bx$, its pseudoinverse is given by 
\begin{equation}
    \bx\pseudoinv=\frac{\bx\T}{\left\Vert\bx\right\Vert^2}.\nonumber
\end{equation}
In order to consider the structure of the matrix $\left(\bM+\bsm\bn\T\right)\pseudoinv$, we list three cases that will be used in the following proofs:
\begin{enumerate}
\item $\bsm\not\in R(\bM)$ and $\bn\not\in R(\bM\T)$:
\begin{equation}
\left(\bM+\bsm\bn\T\right)\pseudoinv=\bM\pseudoinv-\bk\bu\pseudoinv-\bv\pseudoinv\bh+\beta\bv\pseudoinv\bu\pseudoinv.\nonumber
\end{equation}
\item $\bsm\in R(\bM)$ and $\bn$ arbitrary and $\beta\neq 0$:
\begin{align}
\bp&=-\frac{\left\Vert\bk\right\Vert^2}{\beta}\bv\T-\bk,\nonumber\\
\bq\T&=-\frac{\left\Vert\bv\right\Vert^2}{\beta}\bk\T\bM\pseudoinv-\bh,\nonumber\\
\sigma&=\left\Vert\bk\right\Vert^2\left\Vert\bv\right\Vert^2+\left\vert\beta\right\vert^2,\nonumber\\
\left(\bM+\bsm\bn\T\right)\pseudoinv&=\bM\pseudoinv+\frac{1}{\beta}\bv\T\bk\T\bM\pseudoinv-\frac{\beta}{\sigma}\bp\bq\T.\nonumber
\end{align}
\item $\bsm\in R(\bM)$ and $\bn\in R(\bM\T)$ and $\beta=0$:
\begin{equation}
\left(\bM+\bsm\bn\T\right)\pseudoinv=\bM\pseudoinv-\bk\bk\pseudoinv\bM\pseudoinv-\bM\pseudoinv\bh\pseudoinv\bh+(\bk\pseudoinv\bM\pseudoinv\bh\pseudoinv)\bk\bh.\nonumber
\end{equation}
\end{enumerate}
We still consider
\begin{align}
    \hatbSig =\sigmae^2\calS+\left(\sigma_f\sigmae\dfrac{1}{T}\bY\T\bH\bz\right)\bb\T+\bb\left(\sigma_f\sigmae\dfrac{1}{T}\bY\T\bH\bz\right)\T+\sigma_f^2\bb\bb\T.
    \nonumber
\end{align}
Denote 
\begin{align}
    \bC &=\sigmae^2\calS,\nonumber\\
    \bB &=\bC+\left(\sigma_f\sigmae\frac{1}{T}\bY\T\bH\bz\right)\bb\T,\nonumber\\
    \bA &=\bB+\bb\left(\sigma_f\sigmae\dfrac{1}{T}\bY\T\bH\bz\right)\T,\nonumber
\end{align}
it is easy to verify that we need formula in case 2 to calculate $\bB\pseudoinv$, formula in case 1 to calculate $\bA\pseudoinv$ and formula in case 3 to calculate $\hatbSig\pseudoinv$. In the following, we present pseudoinverse of matrices $\bB,\bA$ and $\hatbSig$:

\bigskip

\noindent\textbf{\underline{ $\bB\pseudoinv$}}: 
\begin{align}
    \beta_{\bB}&=1+\frac{\sigma_f}{\sigmae}\bb\T\calS\pseudoinv\left(\frac{1}{T}\bY\T\bH\bz\right)=1+\frac{\sigma_f}{\sigmae}\ubeta_{S1}=1+O_p(T^{-\frac{1}{2}})\nonumber\\
    \bk_{\bB} &= \frac{\sigma_f}{\sigmae}\calS\pseudoinv\left(\frac{1}{T}\bY\T\bH\bz\right)\nonumber\\
    \bk_{\bB}\T &= \frac{\sigma_f}{\sigmae}\left(\frac{1}{T}\bY\T\bH\bz\right)\T\calS\pseudoinv\nonumber\\
    \left\Vert\bk_{\bB}\right\Vert^2&=\frac{\sigma_f^2}{\sigmae^2}\left(\frac{1}{T}\bY\T\bH\bz\right)\T\left[\calS\pseudoinv\right]^2\left(\frac{1}{T}\bY\T\bH\bz\right)=\frac{\sigma_f^2}{\sigmae^2}\ualpha_{S2}\convergeInP\frac{\sigma_f^2}{\sigmae^2}\frac{1}{\rho-1}\nonumber\\
    \bh_{\bB}&=\frac{1}{\sigmae^2}\bb\T\calS\pseudoinv\nonumber\\
    \bv_{\bB}\T&=\left(\bI_N-\calS\pseudoinv\calS\right)\bb\nonumber\\
    \left\Vert\bv_{\bB}\right\Vert^2&=\bb\T\left(\bI_N-\calS\pseudoinv\calS\right)\bb=\ugamma_{IS}\convergeInP\tilde{b}^2\frac{\rho-1}{\rho}\nonumber\\
    \sigma_{\bB}&=\left\Vert\bk_{\bB}\right\Vert^2\left\Vert\bv_{\bB}\right\Vert^2+\beta_{\bB}^2\convergeInP \tsigma_{\bB} =\frac{\sigma_f^2}{\sigmae^2}\frac{\tilde{b}^2}{\rho}+1 \nonumber\\
    \bp_{\bB}&=-\frac{\left\Vert\bk_{\bB}\right\Vert^2}{\beta_{\bB}}\bv_{\bB}\T-\bk_{\bB}\nonumber\\
    \bq_{\bB}\T&=-\frac{\left\Vert\bv_{\bB}\right\Vert^2}{\sigmae^2\beta_{\bB}}\bk_{\bB}\T\calS\pseudoinv-\bh_{\bB}\nonumber\\
    \bB\pseudoinv&=\frac{1}{\sigmae^2}\calS\pseudoinv+\frac{1}{\sigmae^2}\frac{1}{\beta_{\bB}}\bv_{\bB}\T\bk_{\bB}\T\calS\pseudoinv-\frac{\beta_{\bB}}{\sigma_{\bB}}\bp_{\bB}\bq_{\bB}\T\nonumber
\end{align}

\noindent\textbf{\underline{ $\bA\pseudoinv$}}: 
\begin{align}
    \beta_{\bA}&=1+\sigma_f\sigmae\left(\frac{1}{T}\bY\T\bH\bz\right)\T\bB\pseudoinv\bb\convergeInP1-\frac{\sigma_f^2}{\sigmae^2}\frac{1}{\tsigma_{\bB}}\frac{\tilde{b}^2}{\rho(\rho-1)}\nonumber\\
    \bk_{\bA} &= \bB\pseudoinv\bb\nonumber\\
    \bh_{\bA}&=\sigma_f\sigmae\left(\frac{1}{T}\bY\T\bH\bz\right)\T\bB\pseudoinv\nonumber\\
    \bu_{\bA}&=\left(\bI_N-\bB\bB\pseudoinv\right)\bb\nonumber\\
    \left\Vert\bu_{\bA}\right\Vert^2&=\bb\T\left(\bI_N-\bB\bB\pseudoinv\right)\bb=\ugamma_{IB}\convergeInP\tilde{b}^2\frac{\rho-1}{\rho}\nonumber\\
    \bu_{\bA}\pseudoinv&=\frac{\bu_{\bA}\T}{\left\Vert\bu_{\bA}\right\Vert^2}=\frac{\bb\T\left(\bI_N-\bB\bB\pseudoinv\right)}{\bb\T\left(\bI_N-\bB\bB\pseudoinv\right)\bb}=\frac{\bb\T\left(\bI_N-\bB\bB\pseudoinv\right)}{\ugamma_{IB}}\nonumber\\
    \bv_{\bA}&=\sigma_f\sigmae\left(\frac{1}{T}\bY\T\bH\bz\right)\T\left(\bI_N-\bB\pseudoinv\bB\right)\nonumber\\
    \left\Vert\bv_{\bA}\right\Vert^2&=\sigma_f^2\sigmae^2\left(\frac{1}{T}\bY\T\bH\bz\right)\T\left(\bI_N-\bB\pseudoinv\bB\right)\left(\frac{1}{T}\bY\T\bH\bz\right)=\sigma_f^2\sigmae^2\ualpha_{IB}\convergeInP\sigma_f^4\frac{1}{\tsigma_{\bB}}\tilde{b}^2\frac{\rho-1}{\rho}\nonumber\\
    \bv_{\bA}\pseudoinv&=\frac{\bv_{\bA}\T}{\left\Vert\bv_{\bA}\right\Vert^2}=\frac{\sigma_f\sigmae\left(\bI_N-\bB\pseudoinv\bB\right)\left(\frac{1}{T}\bY\T\bH\bz\right)}{\left\Vert\bv_{\bA}\right\Vert^2}\nonumber\\
    \bA\pseudoinv&=\bB\pseudoinv-\bk_{\bA}\bu_{\bA}\pseudoinv-\bv_{\bA}\pseudoinv\bh_{\bA}+\beta_{\bA}\bv_{\bA}\pseudoinv\bu_{\bA}\pseudoinv\nonumber
\end{align}

\noindent\textbf{\underline{ $\hatbSig\pseudoinv$}}: 
\begin{align}
    \beta_{\hatbSig}&=1+\sigma_f^2\bb\T\bA\pseudoinv\bb=1+\sigma_f^2\ugamma_{A1}\convergeInP 0\nonumber\\
    \bk_{\hatbSig} &= \sigma_f^2\bA\pseudoinv\bb\nonumber\\
    \bk_{\hatbSig}\pseudoinv &= \frac{\bb\T\bA\pseudoinv}{\sigma_f^2\bb\T\bA\pseudoinv\bA\pseudoinv\bb}=\frac{\bb\T\bA\pseudoinv}{\sigma_f^2\ugamma_{A2}}\nonumber\\
    \bh_{\hatbSig}&=\bb\T\bA\pseudoinv\nonumber\\
    \bh_{\hatbSig}\pseudoinv&=\frac{\bA\pseudoinv\bb}{\bb\T\bA\pseudoinv\bA\pseudoinv\bb}=\frac{\bA\pseudoinv\bb}{\ugamma_{A2}}\nonumber\\
    \hatbSig\pseudoinv&=\bA\pseudoinv-\frac{\bA\pseudoinv\bb\bb\T\bA\pseudoinv\bA\pseudoinv}{\ugamma_{A2}}-\frac{\bA\pseudoinv\bA\pseudoinv\bb\bb\T\bA\pseudoinv}{\ugamma_{A2}}+\frac{\bA\pseudoinv\bb\bb\T\bA\pseudoinv\ugamma_{A3}}{(\ugamma_{A2})^2}\nonumber
\end{align}
Before presenting the derivations for $E1$-$E6$, it is important to note that further expanding $\hatbSig\pseudoinv$ is already complex due to its nested nature. For example, we need to calculate the quadratic and cubic powers of $\bA\pseudoinv$, and when considering $E4$-$E6$, we will even have to consider $\left[\bA\pseudoinv\right]^4$. Simplifying these calculations first would be beneficial for subsequent proofs. Therefore, the following \ref{B321} and \ref{B322} will present necessary results, especially for $\bb\T\bA\pseudoinv\bb$, $\bb\T \left[\bA\pseudoinv\right]^2 \bb$, $\bb\T \left[\bA\pseudoinv\right]^3 \bb$, and $\bb\T \left[\bA\pseudoinv\right]^4 \bb$. \ref{B321} provides some intermediate results, while \ref{B322} contains the proofs for the results in \ref{B321}.

\paragraph{\texorpdfstring{Asymptotic results of terms in expanding $\hatbSig\pseudoinv$}{Asymptotic results of terms in expanding hatbSigpseudoinv}} \label{B321}

(Please note that we use underlining to denote terms that appear in the $\rho>1$ case.)
\begin{align}
\tsigma_{\bB}&=\frac{\sigma_f^2}{\sigmae^2}\frac{\tilde{b}^2}{\rho}+1\nonumber\\
\tbeta_{\bA}&=
1-\frac{\sigma_f^2}{\sigmae^2}\frac{1}{\tsigma_{\bB}}\frac{\tilde{b}^2}{\rho(\rho-1)}\nonumber\\
v_{\bA}^2&=\sigma_f^4\frac{1}{\tsigma_{\bB}}\tilde{b}^2\frac{\rho-1}{\rho}\nonumber\\
u_{\bA}^2&=\tilde{b}^2\frac{\rho-1}{\rho}\nonumber\\
\ubeta_{0}&=\left(\frac{1}{T}\bY\T\bH\bz\right)\T\bb=O_p\left(T^{-\frac{1}{2}}\right)\nonumber\\
\ubeta_{S11}&=\left(\frac{1}{T}\bY\T\bH\bz\right)\T\calS\pseudoinv\calS\bb=O_p\left(T^{-\frac{1}{2}}\right)\nonumber\\
\ubeta_{IS}&=\left(\frac{1}{T}\bY\T\bH\bz\right)\T\left(\bI_N-\calS\pseudoinv\calS\right)\bb=O_p\left(T^{-\frac{1}{2}}\right)\nonumber\\
\ubeta_{S1}&=\left(\frac{1}{T}\bY\T\bH\bz\right)\T\calS\pseudoinv\bb=O_p\left(T^{-\frac{1}{2}}\right)\nonumber\\
\ubeta_{S2}&=\left(\frac{1}{T}\bY\T\bH\bz\right)\T\left[\calS\pseudoinv\right]^2\bb=O_p\left(T^{-\frac{1}{2}}\right)\nonumber\\
\ubeta_{B1}&=\bb\T\bB\pseudoinv\left(\frac{1}{T}\bY\T\bH\bz\right)\convergeInP \frac{1}{\tsigma_{\bB}}\frac{\sigma_f}{\sigmae^3}\tilde{b}^2\frac{1}{\rho} \nonumber\\
\ubeta_{B11}&=\bb\T\bB\pseudoinv\bB\left(\frac{1}{T}\bY\T\bH\bz\right)\convergeInP \frac{1}{\tsigma_{\bB}}\frac{\sigma_f}{\sigmae}\tilde{b}^2\frac{\rho-1}{\rho} \nonumber\\
\ubeta_{B11_2}&=\bb\T\bB\bB\pseudoinv\left(\frac{1}{T}\bY\T\bH\bz\right)=O_p(T^{-\frac{1}{2}}) \nonumber\\
\ubeta_{B12}&=\bb\T\bB\bB\pseudoinv\bB\pseudoinv\left(\frac{1}{T}\bY\T\bH\bz\right)=O_p(T^{-\frac{1}{2}}) \nonumber\\
\ubeta_{B121}&=\bb\T\bB\bB\pseudoinv\bB\pseudoinv\bB\left(\frac{1}{T}\bY\T\bH\bz\right)=O_p(T^{-\frac{1}{2}}) \nonumber\\
\ubeta_{B21}&=\bb\T\bB\pseudoinv\bB\pseudoinv\bB\left(\frac{1}{T}\bY\T\bH\bz\right)\convergeInP -\frac{1}{\tsigma_{\bB}^2}\left(\frac{\sigma_f^2}{\sigmae^2}\tilde{b}^2\frac{1}{\rho(\rho-1)}-1\right)\left(\frac{\sigma_f}{\sigmae^3}\tilde{b}^2\frac{1}{\rho}\right) \nonumber\\
\ubeta_{BIB}&=\bb\T\bB\pseudoinv\left(\bI_N-\bB\pseudoinv\bB\right)\left(\frac{1}{T}\bY\T\bH\bz\right)\convergeInP \frac{1}{\tsigma_{\bB}^2}\frac{1}{\sigma_f\sigmae}(\tsigma_{\bB}-1)^2\frac{\rho}{\rho-1} \nonumber\\
\ubeta_{IB}&=\bb\T\left(\bI_N-\bB\pseudoinv\bB\right)\left(\frac{1}{T}\bY\T\bH\bz\right)\convergeInP -\frac{1}{\tsigma_{\bB}}\frac{\sigma_f}{\sigmae}\tilde{b}^2\frac{\rho-1}{\rho} \nonumber\\
\ualpha_{0}&=\left(\frac{1}{T}\bY\T\bH\bz\right)\T\left(\frac{1}{T}\bY\T\bH\bz\right)\convergeInP \rho\nonumber\\
\ualpha_{S1}&=\left(\frac{1}{T}\bY\T\bH\bz\right)\T\calS\pseudoinv\left(\frac{1}{T}\bY\T\bH\bz\right)\convergeInP 1\nonumber\\
\ualpha_{S11}&=\left(\frac{1}{T}\bY\T\bH\bz\right)\T\calS\pseudoinv\calS\left(\frac{1}{T}\bY\T\bH\bz\right)\convergeInP \rho\nonumber\\
\ualpha_{S2}&=\left(\frac{1}{T}\bY\T\bH\bz\right)\T\left[\calS\pseudoinv\right]^2\left(\frac{1}{T}\bY\T\bH\bz\right)\convergeInP\frac{1}{\rho-1}\nonumber\\
\ualpha_{S3}&=\left(\frac{1}{T}\bY\T\bH\bz\right)\T\left[\calS\pseudoinv\right]^3\left(\frac{1}{T}\bY\T\bH\bz\right)\convergeInP\frac{\rho}{(\rho-1)^3}\nonumber\\
\ualpha_{S4}&=\left(\frac{1}{T}\bY\T\bH\bz\right)\T\left[\calS\pseudoinv\right]^4\left(\frac{1}{T}\bY\T\bH\bz\right)\convergeInP\frac{\rho(\rho+1)}{(\rho-1)^5}\nonumber\\
\ualpha_{B1}&=\left(\frac{1}{T}\bY\T\bH\bz\right)\T\bB\pseudoinv\left(\frac{1}{T}\bY\T\bH\bz\right)\convergeInP\frac{1}{\tsigma_{\bB}}\frac{1}{\sigmae^2}\nonumber\\
\ualpha_{B21}&=\left(\frac{1}{T}\bY\T\bH\bz\right)\T\bB\pseudoinv\bB\pseudoinv\bB\left(\frac{1}{T}\bY\T\bH\bz\right)\convergeInP\frac{1}{\tsigma_{\bB}}\frac{1}{\sigmae^2} -\frac{(\tsigma_{\bB}-1)(\rho-\tsigma_{\bB})}{\tsigma_{\bB}^2(\rho-1)\sigmae^2} \nonumber\\
\ualpha_{IB}&=\left(\frac{1}{T}\bY\T\bH\bz\right)\T\left(\bI_N-\bB\pseudoinv\bB\right)\left(\frac{1}{T}\bY\T\bH\bz\right)\convergeInP\frac{1}{\tsigma_{\bB}}\frac{\sigma_f^2}{\sigmae^2}\tilde{b}^2\frac{\rho-1}{\rho}\nonumber\\
\ualpha_{BIB}&=\left(\frac{1}{T}\bY\T\bH\bz\right)\T\left(\bI_N-\bB\pseudoinv\bB\right)\bB\pseudoinv\left(\frac{1}{T}\bY\T\bH\bz\right)\convergeInP\frac{(\tsigma_{\bB}-1)(\rho-\tsigma_{\bB})}{\tsigma_{\bB}^2(\rho-1)\sigmae^2}\nonumber\\
\ugamma_{S1}&=\bb\T\calS\pseudoinv\bb\convergeInP\tilde{b}^2\frac{1}{\rho(\rho-1)}\nonumber\\
\ugamma_{S2}&=\bb\T\left[\calS\pseudoinv\right]^2\bb\convergeInP\tilde{b}^2\frac{1}{(\rho-1)^3}\nonumber\\
\ugamma_{IS}&=\bb\T\left(\bI_N-\calS\pseudoinv\calS\right)\bb\convergeInP\tilde{b}^2\frac{\rho-1}{\rho}\nonumber\\
\ugamma_{B1}&=\bb\T\bB\pseudoinv\bb\convergeInP\frac{1}{\tsigma_{\bB}}\frac{1}{\sigmae^2}\frac{\tilde{b}^2}{\rho(\rho-1)}\nonumber\\
\ugamma_{B2}&=\bb\T\bB\pseudoinv\bB\pseudoinv\bb\convergeInP\frac{1}{\tsigma_{\bB}^2}(\tsigma_{\bB}-1)\frac{1}{\sigma_f^2\sigmae^2}\frac{\rho}{(\rho-1)^3}\nonumber\\
\ugamma_{B12}&=\bb\T\bB\bB\pseudoinv\bB\pseudoinv\bb\convergeInP(\tsigma_{\bB}-1)\frac{1}{\sigma_f^2}\frac{1}{\rho-1}\nonumber\\
\ugamma_{IB}&=\bb\T\left(\bI_N-\bB\bB\pseudoinv\right)\bb\convergeInP\tilde{b}^2\frac{\rho-1}{\rho}\nonumber\\
\ugamma_{A1}&=\bb\T\bA\pseudoinv\bb\convergeInP-\frac{1}{\sigma_f^2}\nonumber\\
\ugamma_{A2}&=\bb\T\bA\pseudoinv\bA\pseudoinv\bb\convergeInP \frac{\tsigma_{\bB}}{\sigma_f^4\tilde{b}^2\frac{\rho-1}{\rho}}\nonumber\\
\ugamma_{A3}&=\bb\T\bA\pseudoinv\bA\pseudoinv\bA\pseudoinv\bb\convergeInP \frac{1}{v_{\bA}^2}\frac{\sigma_f^2}{v_{\bA}^2}\frac{1}{\tsigma_{\bB}^2}\frac{\tsigma_{\bB}^2-2\rho\tsigma_{\bB}+\tsigma_{\bB}+\rho-1}{\rho-1}\nonumber\\
\ugamma_{A4}&=\bb\T\bA\pseudoinv\bA\pseudoinv\bA\pseudoinv\bA\pseudoinv\bb\convergeInP \frac{1}{v_{\bA}^2}\left[\frac{1}{v_{\bA}^2}\frac{\sigma_f^2}{\tsigma_{\bB}^2}\frac{1}{\rho-1}\right]^2\left[\tsigma_{\bB}(\tsigma_{\bB}-\rho+1)-(\rho\tsigma_{\bB}-\rho+1)\right]^2\nonumber\\
&+\frac{1}{v_{\bA}^2}\left[\frac{1}{v_{\bA}^2}\frac{\sigma_f^2}{\tsigma_{\bB}^2}\frac{1}{\rho-1}\right]^2\left[\tsigma_{\bB}-1\right]^2\left[(\tsigma_{\bB}+\rho-1)(\rho\tsigma_{\bB}-\rho+1)\right]\nonumber\\
\uzeta_{1}&=\left(\frac{1}{T}\bY\T\bH\bz\right)\T\bv_{\bB}\T=\ubeta_{IS}=O_p(T^{-\frac{1}{2}})\nonumber\\
\uzeta_{2}&=\left(\frac{1}{T}\bY\T\bH\bz\right)\T\bk_{\bB}\T=\frac{\sigma_f}{\sigmae}\ualpha_{S1}\convergeInP\frac{\sigma_f}{\sigmae}\nonumber\\
\uzeta_{3}&=\left(\frac{1}{T}\bY\T\bH\bz\right)\T\bp_{\bB}\convergeInP-\frac{\sigma_f}{\sigmae}\nonumber\\
\uzeta_{4}&=\bk_{\bB}\T\calS\pseudoinv\left(\frac{1}{T}\bY\T\bH\bz\right)\convergeInP\frac{\sigma_f}{\sigmae}\frac{1}{\rho-1} \nonumber\\
\uzeta_{5}&=\bk_{\bB}\T\calS\pseudoinv\calS\left(\frac{1}{T}\bY\T\bH\bz\right)\convergeInP\frac{\sigma_f}{\sigmae} \nonumber\\
\uzeta_{6}&=\bq_{\bB}\T\calS\left(\frac{1}{T}\bY\T\bH\bz\right)\convergeInP-\frac{\sigma_f}{\sigmae^3}\tilde{b}^2\frac{\rho-1}{\rho} \nonumber\\
\uzeta_{7}&=\bh_{\bB}\T\calS\left(\frac{1}{T}\bY\T\bH\bz\right)=O_p(T^{-\frac{1}{2}}) \nonumber\\
\uzeta_{8}&=\bq_{\bB}\T\left(\frac{1}{T}\bY\T\bH\bz\right)\convergeInP-\frac{\sigma_f}{\sigmae^3}\tilde{b}^2\frac{1}{\rho} \nonumber\\
\uzeta_{9}&=\bh_{\bB}\T\left(\frac{1}{T}\bY\T\bH\bz\right)=O_p(T^{-\frac{1}{2}}) \nonumber\\
\uzeta_{10}&=\bk_{\bB}\T\left[\calS\pseudoinv\right]^2\left(\frac{1}{T}\bY\T\bH\bz\right)\convergeInP\frac{\sigma_f}{\sigmae}\frac{\rho}{(\rho-1)^3} \nonumber\\
\uzeta_{11}&=\bq_{\bB}\T\calS\pseudoinv\calS\left(\frac{1}{T}\bY\T\bH\bz\right)\convergeInP -\frac{\sigma_f}{\sigmae^3}\tilde{b}^2\frac{1}{\rho} \nonumber\\
\uzeta_{12}&=\bh_{\bB}\T\calS\pseudoinv\calS\left(\frac{1}{T}\bY\T\bH\bz\right)=O_p(T^{-\frac{1}{2}}) \nonumber\\
\uzeta_{13}&=\bq_{\bB}\T\calS\pseudoinv\left(\frac{1}{T}\bY\T\bH\bz\right)\convergeInP -\frac{\sigma_f}{\sigmae^3}\tilde{b}^2\frac{1}{(\rho-1)^2} \nonumber\\
\uzeta_{14}&=\bh_{\bB}\T\calS\pseudoinv\left(\frac{1}{T}\bY\T\bH\bz\right)=O_p(T^{-\frac{1}{2}}) \nonumber\\
\uzeta_{15}&=\left(\frac{1}{T}\bY\T\bH\bz\right)\calS\pseudoinv\bp_{\bB}\convergeInP-\frac{\sigma_f}{\sigmae}\frac{1}{\rho-1} \nonumber\\
\uxi_{1}&=\bk_{\bB}\T\calS\pseudoinv\bb=O_p(T^{-\frac{1}{2}})\nonumber\\
\uxi_{2}&=\bh_{\bB}\bb\convergeInP\frac{\tilde{b}^2}{\sigmae^2}\frac{1}{\rho(\rho-1)}\nonumber\\
\uxi_{3}&=\bq_{\bB}\T\bb\convergeInP-\frac{\tilde{b}^2}{\sigmae^2}\frac{1}{\rho(\rho-1)}\nonumber\\
\uxi_{4}&=\bb\T\calS\bv_{\bB}\T=0 \nonumber\\
\uxi_{5}&=\bb\T\calS\bk_{\bB}\T=O_p(T^{-\frac{1}{2}}) \nonumber\\
\uxi_{6}&=\bb\T\calS\bp_{\bB}=O_p(T^{-\frac{1}{2}}) \nonumber\\
\uxi_{7}&=\bb\T\bv_{\bB}\T\convergeInP \tilde{b}^2\frac{\rho-1}{\rho} \nonumber\\
\uxi_{8}&=\bb\T\bk_{\bB}\T=O_p(T^{-\frac{1}{2}}) \nonumber\\
\uxi_{9}&=\bb\T\bp_{\bB}\convergeInP-\frac{\sigma_f^2}{\sigmae^2}\tilde{b}^2\frac{1}{\rho} \nonumber\\
\uxi_{10}&=\bu_{\bA}\pseudoinv\bb\convergeInP 1 \nonumber\\
\uxi_{11}&=\bh_{\bA}\bb=\beta_{\bA}-1 \nonumber\\
\uxi_{12}&=\bb\T\bv_{\bA}\pseudoinv\convergeInP-\frac{1}{\sigma_f^2} \nonumber\\
\uxi_{13}&=\bk_{\bB}\T\calS\pseudoinv\bv_{\bB}\T=0 \nonumber\\
\uxi_{14}&=\bk_{\bB}\T\calS\pseudoinv\bp_{\bB}\T\convergeInP - \frac{\sigma_f^2}{\sigmae^2}\frac{\rho}{(\rho-1)^3}\nonumber\\
\uxi_{15}&=\bq_{\bB}\T\bv_{\bB}\T=0\nonumber\\
\uxi_{16}&=\bq_{\bB}\T\bp_{\bB}\convergeInP  \frac{\sigma_f^2}{\sigmae^4}\tilde{b}^2\frac{1}{(\rho-1)^2}\nonumber\\
\uxi_{17}&=\bq_{\bB}\T\bk_{\bB}\convergeInP - \frac{\sigma_f^2}{\sigmae^4}\tilde{b}^2\frac{1}{(\rho-1)^2}\nonumber\\
\uxi_{18}&=\bk_{\bB}\T\calS\pseudoinv\bk_{\bB}\convergeInP \frac{\sigma_f^2}{\sigmae^2}\frac{\rho}{(\rho-1)^3}\nonumber\\
\uxi_{19}&=\bh_{\bB}\bk_{\bB}=O_p(T^{-\frac{1}{2}})\nonumber\\
\uxi_{20}&=\bb\T\bB\pseudoinv\bv_{\bA}\pseudoinv\convergeInP \frac{1}{v_{\bA}^2}\frac{(\tsigma_{\bB}-1)^2\rho}{\tsigma_{\bB}^2(\rho-1)}\nonumber\\
\uxi_{21}&=\bu_{\bA}\pseudoinv\bv_{\bA}\pseudoinv\convergeInP -\sigma_f^2\frac{1}{\tsigma_{\bB}v_{\bA}^2}\nonumber\\
\uxi_{22}&=\bh_{\bA}\bv_{\bA}\pseudoinv\convergeInP \frac{1}{v_{\bA}^2}\sigma_f^2\frac{(\tsigma_{\bB}-1)(\rho-\tsigma_{\bB})}{\tsigma_{\bB}^2(\rho-1)}\nonumber\\
\uxi_{23}&=\bu_{\bA}\pseudoinv\bk_{\bA}\convergeInP -\frac{1}{\tsigma_{\bB}}(\tsigma_{\bB}-1)\frac{1}{\sigmae^2}\frac{1}{(\rho-1)^2}\nonumber
\end{align}

Here we present some useful results that are repeatedly used. We use $A \Rightarrow B$ to denote that considering $A$ is equivalent to considering $B$ asymptotically.

\begin{enumerate}
\item $\bb\T\bB\bB\pseudoinv \Rightarrow\bb\T\calS\calS\pseudoinv$
\item $\bb\T\bB\pseudoinv\bB\pseudoinv\Rightarrow\frac{1}{\sigmae^4}\bb\T\calS\pseudoinv\calS\pseudoinv+\frac{1}{\sigmae^4}\uxi_7\bk_{\bB}\T\calS\pseudoinv\calS\pseudoinv-\frac{1}{\tsigma_{\bB}}\frac{1}{\sigmae^2}\uxi_7\uxi_{14}\bq_{\bB}\T-\frac{1}{\tsigma_{\bB}}\frac{1}{\sigmae^2}\uxi_9\bq_{\bB}\T\calS\pseudoinv+\frac{1}{\tsigma_{\bB}^2}\uxi_9\uxi_{16}\bq_{\bB}\T$
\item $\left(\frac{1}{T}\bY\T\bH\bz\right)\T\bB\pseudoinv\bB\pseudoinv\Rightarrow\frac{1}{\sigmae^4}\left(\frac{1}{T}\bY\T\bH\bz\right)\T\left[\calS\pseudoinv\right]^2+\frac{1}{\tsigma_{\bB}}\frac{\sigma_f}{\sigmae^3}\frac{1}{\rho-1}\bq_{\bB}\T+\frac{1}{\tsigma_{\bB}}\frac{\sigma_f}{\sigmae^3}\frac{1}{\rho-1}\bq_{\bB}\T\calS\pseudoinv-\frac{1}{\tsigma_{\bB}^2}(\tsigma_{\bB}-1)\frac{\sigma_f}{\sigmae^3}\frac{\rho}{(\rho-1)^2}\bq_{\bB}\T$
\item $\bB\pseudoinv\bB\left(\frac{1}{T}\bY\T\bH\bz\right)\Rightarrow\calS\pseudoinv\calS\left(\frac{1}{T}\bY\T\bH\bz\right)+\bv_{\bB}\T\uzeta_5-\frac{\sigmae^2}{\tsigma_{\bB}}\bp_{\bB}\uzeta_6$
\item $\bB\pseudoinv\bB\pseudoinv\bB\left(\frac{1}{T}\bY\T\bH\bz\right)\Rightarrow\frac{1}{\sigmae^2}\calS\pseudoinv\left(\frac{1}{T}\bY\T\bH\bz\right)+\frac{1}{\sigmae^2}\calS\pseudoinv\bv_{\bB}\T\uzeta_5-\frac{1}{\tsigma_{\bB}}\calS\pseudoinv\bp_{\bB}\uzeta_6+\frac{1}{\sigmae^2}\bv_{\bB}\T\uzeta_4-\frac{1}{\tsigma_{\bB}}\bp_{\bB}\uzeta_{11}+\frac{\sigmae^2}{\tsigma_{\bB}^2}\bp_{\bB}\uxi_{16}\uzeta_6$
\item $\bb\T\bA\pseudoinv\Rightarrow\bb\T\bB\pseudoinv-\frac{1}{\sigma_f^2}\bu_{\bA}\pseudoinv+\frac{1}{\sigma_f^2}\bh_{\bA}$
\item $\bb\T\bA\pseudoinv\bA\pseudoinv\Rightarrow-\frac{1}{v_{\bA}^2}\bh_{\bA}+\tbeta_{\bA}\frac{1}{v_{\bA}^2}\bu_{\bA}\pseudoinv$
\item 
$\bA\pseudoinv\bA\pseudoinv\bb\Rightarrow\bB\pseudoinv\bv_{\bA}\pseudoinv-\bB\pseudoinv\bb\uxi_{21}-\bv_{\bA}\pseudoinv\uxi_{22}+\tbeta_{\bA}\bv_{\bA}\pseudoinv\uxi_{21}$
\item 
$\hatbSig\pseudoinv\hatbSig\pseudoinv=\left[\bA\pseudoinv\right]^2-\frac{1}{\ugamma_{A2}}\left(\left[\bA\pseudoinv\right]^2\bb\bb\T\left[\bA\pseudoinv\right]^2+\left[\bA\pseudoinv\right]^3\bb\bb\T\bA\pseudoinv+\bA\pseudoinv\bb\bb\T\left[\bA\pseudoinv\right]^3\right)\\+\frac{\ugamma_{A3}}{\ugamma_{A2}^2}\left(\left[\bA\pseudoinv\right]^2\bb\bb\T\bA\pseudoinv+\bA\pseudoinv\bb\bb\T\left[\bA\pseudoinv\right]^2\right)+\frac{\ugamma_{A4}}{\ugamma_{A2}^2}\bA\pseudoinv\bb\bb\T\bA\pseudoinv-\frac{\ugamma_{A3}^2}{\ugamma_{A2}^3}\bA\pseudoinv\bb\bb\T\bA\pseudoinv$
\end{enumerate}

\paragraph{\texorpdfstring{Proofs for \ref{B321}}{Proofs for \ref{B321}}} \label{B322}

We omit tedious and lengthy proofs for most of the terms but retain the most important and fundamental ones. Specifically, we omit the proofs for the $\uzeta$ and $\uxi$ terms, as they are transformations of other terms. Additionally, we omit most of the $\ualpha$, $\ubeta$ and $\ugamma$ terms involving $\bB$, $\bA$, and $\hatbSig$, since the $\bB$-related terms are entirely based on terms in $\calS$, the $\bA$-related terms are based on $\bB$, and the $\hatbSig$-related terms are based on $\bA$.

\bigskip
\noindent\textbf{\underline{Limit of $\ubeta_{0}$}}: Consider the square of $\ubeta_{0}$, we have
\begin{equation}
    T\ubeta_{0}^2=\frac{1}{T}\bz\T\bH\bY\bb\bb\T\bY\T\bH\bz\nonumber.
\end{equation}
Based on trace Lemma \ref{lemma:Trace_Lemma}, we have 
\begin{align}
T\ubeta_{0}^2&-\frac{1}{T}\bb\T\bY\T\bH\bY\bb=O_p(T^{-\frac{1}{2}}),\nonumber\\
T\ubeta_{0}^2&-\bb\T\calS\bb=O_p(T^{-\frac{1}{2}}).\nonumber
\end{align}
Thus, $\ubeta_{0}=O_p(T^{-\frac{1}{2}})$.

\bigskip
\noindent\textbf{\underline{Limit of $\ubeta_{S11}$}}: Consider the square of $\ubeta_{S11}$, we have
\begin{equation}
    T\ubeta_{S11}^2=\frac{1}{T}\bz\T\bH\bY\calS\pseudoinv\calS\bb\bb\T\calS\calS\pseudoinv\bY\T\bH\bz\nonumber.
\end{equation}
Based on trace Lemma \ref{lemma:Trace_Lemma}, we have 
\begin{align}
T\ubeta_{S11}^2&-\frac{1}{T}\bb\T\calS\calS\pseudoinv\bY\T\bH\bY\calS\pseudoinv\calS\bb=O_p(T^{-\frac{1}{2}}),\nonumber\\
T\ubeta_{S11}^2&-\bb\T\calS\bb=O_p(T^{-\frac{1}{2}}).\nonumber
\end{align}
Thus, $\ubeta_{S11}=O_p(T^{-\frac{1}{2}})$.

\bigskip
\noindent\textbf{\underline{Limit of $\ubeta_{IS}$}}: Based on the above two results, we have $\ubeta_{IS}=O_p\left(T^{-\frac{1}{2}}\right)$.

\bigskip
\noindent\textbf{\underline{Limit of $\ubeta_{S1}$}}: Consider the square of $\ubeta_{S1}$, we have
\begin{equation}
    T\ubeta_{S1}^2=\frac{1}{T}\bz\T\bH\bY\calS\pseudoinv\bb\bb\T\calS\pseudoinv\bY\T\bH\bz\nonumber.
\end{equation}
Based on trace Lemma \ref{lemma:Trace_Lemma}, we have 
\begin{align}
T\ubeta_{S1}^2&-\frac{1}{T}\bb\T\calS\pseudoinv\bY\T\bH\bY\calS\pseudoinv\bb=O_p(T^{-\frac{1}{2}}),\nonumber\\
T\ubeta_{S1}^2&-\bb\T\calS\pseudoinv\bb=O_p(T^{-\frac{1}{2}}).\nonumber
\end{align}
Thus, $\ubeta_{S1}=O_p(T^{-\frac{1}{2}})$.

\bigskip
\noindent\textbf{\underline{Limit of $\ubeta_{S2}$}}: Similar to the proof for $\ubeta_{S1}$, we have $\ubeta_{S2}=O_p(T^{-\frac{1}{2}})$.

\bigskip
\noindent\textbf{\underline{Limit of $\ubeta_{B11}$}}:
\begin{align}
    \ubeta_{B11}&=\bb\T\bB\pseudoinv\bB\left(\frac{1}{T}\bY\T\bH\bz\right)\nonumber\\
    &\Rightarrow\bb\T\calS\pseudoinv\calS\left(\frac{1}{T}\bY\T\bH\bz\right)+\bb\T\bv_{\bB}\T\uzeta_5-\frac{\sigmae^2}{\tsigma_{\bB}}\bb\T\bp_{\bB}\uzeta_6\nonumber\\
    &=\ubeta_{S11}+\uxi_{7}\uzeta_5-\frac{\sigmae^2}{\tsigma_{\bB}}\uxi_9\uzeta_6\nonumber\\
    &\convergeInP\frac{\sigma_f}{\sigmae}\frac{1}{\tsigma_{\bB}}\tilde{b}^2\frac{\rho-1}{\rho}.\nonumber
\end{align}

\bigskip
\noindent\textbf{\underline{Limit of $\ualpha_{0}$}}: Based on trace Lemma \ref{lemma:Trace_Lemma} we have
\begin{align}
    \ualpha_{0}&-\frac{1}{T}\Tr\left[\frac{1}{T}\bH\bY\bY\T\bH\right]=O_p(T^{-\frac{1}{2}})\nonumber\\
    \ualpha_{0}&-\frac{1}{T}\Tr\left[\calS\right]=O_p(T^{-\frac{1}{2}})\nonumber\\
    \ualpha_{0}&-\rho=O_p(T^{-\frac{1}{2}})\nonumber.
\end{align}
Thus, $\ualpha_{0}\convergeInP \rho$.

\bigskip
\noindent\textbf{\underline{Limit of $\ualpha_{S1}$}}: Based on trace Lemma \ref{lemma:Trace_Lemma} we have
\begin{align}
    \ualpha_{S1}&-\frac{1}{T}\Tr\left[\frac{1}{T}\bH\bY\calS\pseudoinv\bY\T\bH\right]=O_p(T^{-\frac{1}{2}})\nonumber\\
    \ualpha_{S1}&-\frac{1}{T}\Tr\left[\calS\pseudoinv\calS\right]=O_p(T^{-\frac{1}{2}})\nonumber\\
    \ualpha_{S1}&-1=O_p(T^{-\frac{1}{2}})\nonumber.
\end{align}
Thus, $\ualpha_{S1}\convergeInP 1$.

\bigskip
\noindent\textbf{\underline{Limit of $\ualpha_{S11}$}}: Based on trace Lemma \ref{lemma:Trace_Lemma} we have
\begin{align}
    \ualpha_{S11}&-\frac{1}{T}\Tr\left[\frac{1}{T}\bH\bY\calS\pseudoinv\calS\bY\T\bH\right]=O_p(T^{-\frac{1}{2}})\nonumber\\
    \ualpha_{S11}&-\frac{1}{T}\Tr\left[\calS\right]=O_p(T^{-\frac{1}{2}})\nonumber\\
    \ualpha_{S1}&-\rho=O_p(T^{-\frac{1}{2}})\nonumber.
\end{align}
Thus, $\ualpha_{S1}\convergeInP \rho$.

\bigskip
\noindent\textbf{\underline{Limit of $\ualpha_{S2}$}}: Based on trace Lemma \ref{lemma:Trace_Lemma} we have
\begin{align}
    \ualpha_{S2}&-\frac{1}{T}\Tr\left[\frac{1}{T}\bH\bY\left[\calS\pseudoinv\right]^2\bY\T\bH\right]=O_p(T^{-\frac{1}{2}})\nonumber\\
    \ualpha_{S2}&-\frac{1}{T}\Tr\left[\calS\pseudoinv\right]=O_p(T^{-\frac{1}{2}})\nonumber\\
    \ualpha_{S2}&-\frac{1}{T}\frac{T}{N}\Tr\left[\left(\frac{1}{p}\bH\bY\bY\T\bH\right)\pseudoinv\right]=O_p(T^{-\frac{1}{2}})\nonumber\\
    \ualpha_{S2}&-\frac{1}{\rho}\frac{1}{1-\frac{1}{\rho}}=O_p(T^{-\frac{1}{2}}).\nonumber
\end{align}
Thus, $\ualpha_{S2}\convergeInP\frac{1}{\rho-1}$.

\bigskip
\noindent\textbf{\underline{Limit of $\ualpha_{S3}$}}: Based on trace Lemma \ref{lemma:Trace_Lemma} we have
\begin{align}
    \ualpha_{S3}&-\frac{1}{T}\Tr\left[\frac{1}{T}\bH\bY\left[\calS\pseudoinv\right]^3\bY\T\bH\right]=O_p(T^{-\frac{1}{2}})\nonumber\\
    \ualpha_{S3}&-\frac{1}{T}\Tr\left[\left[\calS\pseudoinv\right]^2\right]=O_p(T^{-\frac{1}{2}})\nonumber.
\end{align}
Based on the same justification in $D3$, $\ualpha_{S3}\convergeInP\frac{\rho}{(\rho-1)^3}$.

\bigskip
\noindent\textbf{\underline{Limit of $\ualpha_{S4}$}}: Based on trace Lemma \ref{lemma:Trace_Lemma} we have
\begin{align}
    \ualpha_{S4}&-\frac{1}{T}\Tr\left[\frac{1}{T}\bH\bY\left[\calS\pseudoinv\right]^4\bY\T\bH\right]=O_p(T^{-\frac{1}{2}})\nonumber\\
    \ualpha_{S4}&-\frac{1}{T}\Tr\left[\left[\calS\pseudoinv\right]^3\right]=O_p(T^{-\frac{1}{2}})\nonumber.
\end{align}
Based on the same justification in $D3$, $\ualpha_{S4}\convergeInP\frac{\rho(\rho+1)}{(\rho-1)^5}$.

\bigskip
\noindent\textbf{\underline{Limit of $\ualpha_{IB}$}}:
\begin{align}
    \ualpha_{IB}&=\left(\frac{1}{T}\bY\T\bH\bz\right)\T\left(\bI_N-\bB\bB\pseudoinv\right)\left(\frac{1}{T}\bY\T\bH\bz\right)\nonumber\\
    &=\ualpha_{IS}-\frac{\sigma_f}{\sigmae}\ualpha_{S1}\ubeta_{0}-\frac{\sigma_f}{\sigmae}\uxi_1\uxi_4\ubeta_0-\uxi_1\uxi_5+\frac{\sigmae^2}{\tsigma_{\bB}}\uxi_3\uxi_5-\frac{\sigma_f\sigmae}{\tsigma_{\bB}}\uxi_3\uxi_8\ubeta_0\nonumber\\
    &\convergeInP\frac{\sigma_f^2}{\sigmae^2}\frac{1}{\tsigma_{\bB}}\tilde{b}^2\frac{\rho-1}{\rho}.\nonumber
\end{align}
Thus, $\ualpha_{IB}\convergeInP\frac{\sigma_f^2}{\sigmae^2}\frac{1}{\tsigma_{\bB}}\tilde{b}^2\frac{\rho-1}{\rho}$.

\bigskip
\noindent\textbf{\underline{Limit of $\ugamma_{IS}$}}:
\begin{align}
    \gamma_{IS}&=\bb\T\left(\bI_N-\calS\pseudoinv\calS\right)\bb\nonumber\\
    &=\lim_{\lambda\rightarrow 0^{+}}\bb\T\left[\left(\calS+\lambda\bI_N\right)\inv\left(\calS+\lambda\bI_N\right)-\left(\calS+\lambda\bI_N\right)\inv\calS\right]\bb\nonumber\\
    &=\lim_{\lambda\rightarrow 0^{+}}\bb\T\left[\lambda\left(\calS+\lambda\bI_N\right)\inv\right]\bb\nonumber\\
    &=\lim_{\lambda\rightarrow 0^{+}}\lambda\left\Vert\bb\right\Vert^2 m(-\lambda)\nonumber\\
    &=\left\Vert\bb\right\Vert^2\frac{\rho-1}{\rho}.\nonumber
\end{align}
Thus, $\ugamma_{IS}\convergeInP\tilde{b}^2\frac{\rho-1}{\rho}$.

\bigskip
\noindent\textbf{\underline{Limit of $\ugamma_{IB}$}}:
\begin{align}
    \ugamma_{IB}&=\bb\T\left(\bI_N-\bB\bB\pseudoinv\right)\bb\nonumber\\
    &=\ugamma_{IS}-\uxi_4\uxi_1+\frac{\sigmae^2}{\tsigma_{\bB}}\uxi_6\uxi_3-\frac{\sigma_f}{\sigmae}\ubeta_{0}\ugamma_{S1}-\frac{\sigma_f}{\sigmae}\ubeta_{0}\ugamma_{IS}\uxi_1+\frac{\sigma_f\sigmae}{\tsigma_{\bB}}\ubeta_{0}\uxi_9\uxi_3\nonumber\\
    &\Rightarrow\ugamma_{IS}.\nonumber
\end{align}
Thus, $\ugamma_{IB}\convergeInP\tilde{b}^2\frac{\rho-1}{\rho}$.

\bigskip
\noindent\textbf{\underline{Limit of $\ugamma_{B1}$}}:
\begin{align}
    \ugamma_{B1}&=\bb\T\bB\pseudoinv\bb\nonumber\\
    &=\frac{1}{\sigmae^2}\ugamma_{S1}+\frac{1}{\sigmae^2}\uxi_7\uxi_1-\frac{1}{\tsigma_{\bB}}\uxi_9\uxi_3\nonumber\\
    &\convergeInP\frac{1}{\tsigma_{\bB}}\frac{1}{\sigmae^2}\frac{\tilde{b}^2}{\rho(\rho-1)}.\nonumber
\end{align}
Thus, $\ugamma_{B1}\convergeInP\frac{1}{\tsigma_{\bB}}\frac{1}{\sigmae^2}\frac{\tilde{b}^2}{\rho(\rho-1)}$.

\bigskip

Now we are able to derive asymptotics for $E1$-$E6$. Similar to case of $\rho<1$, we organize the following into three parts: First, in \ref{B322}, we present the asymptotic results
for $E1$-$E6$. Second, in \ref{B324}, we list the asymptotic results for the intermediate terms that appear in \ref{B323}. Third, in \ref{B325}, we provide the proofs for the asymptotic results of the intermediate terms.

Recall that $\tphi=\frac{\sigma_f^2\tilde{b}^2}{\sigmae^2}$, thus $\frac{\mu_f^2}{\sigma_f^2}=\frac{\tphi+1}{\tphi}\ttheta$, $\tsigma_{\bB}=\frac{\tphi}{\rho}+1$.

\paragraph{\texorpdfstring{Asymptotic results of $E1$-$E6$}{Asymptotic results of E1-E6}}\label{B323}

\noindent\textbf{\underline{Limit of $E1$}}: We have 
\begin{align}
E1&=\bmu\T\hatbSig\pseudoinv\bmu\nonumber\\
&=\mu_f^2\bb\T\hatbSig\pseudoinv\bb\nonumber\\
&=\mu_f^2\ugamma_{\hatSig1}\nonumber\\
&\convergeInP \frac{\mu_f^2}{\sigma_f^2}\frac{(\tsigma_{\bB}-1)(\rho\tsigma_{\bB}-\rho+1)}{\tsigma_{\bB}^2(\rho-1)}=\frac{(\tphi+1)^2\rho\ttheta}{(\tphi+\rho)^2(\rho-1)}.\nonumber
\end{align}

\bigskip

\noindent\textbf{\underline{Limit of $E2$}}: We have 
\begin{align}
E2&=\hatbsm\T\hatbSig\pseudoinv\hatbsm\nonumber\\
&=\left(\frac{\sigma_f\bb\bz\T\be}{T}+\frac{\sigmae\bY\T\be}{T}\right)\T\hatbSig\pseudoinv\left(\frac{\sigma_f\bb\bz\T\be}{T}+\frac{\sigmae\bY\T\be}{T}\right)\nonumber\\
&=E21+E22+2E23,\nonumber
\end{align}
where 
\begin{align}
E21&=\sigma_f^2\left(\frac{\bz\T\be}{\sqrt{T}}\right)^2\frac{1}{T}\bb\T\hatbSig\pseudoinv\bb=\sigma_f^2\left(\frac{\bz\T\be}{\sqrt{T}}\right)^2\frac{1}{T}\ugamma_{\hatSig1}=O_p(T^{-1}),\nonumber\\
E22&=\sigmae^2\uttalpha_{\hatSig1}\convergeInP \frac{1}{\rho-1},\nonumber\\
E23&=\frac{1}{\sqrt{T}}\sigma_f\sigmae\left(\frac{\bz\T\be}{\sqrt{T}}\right)\utbeta_{\hatSig1}=O_p(T^{-1}).\nonumber
\end{align}
Thus, $E2\convergeInP \frac{1}{\rho-1}$.

\bigskip

\noindent\textbf{\underline{Limit of $E3$}}: We have 
\begin{align}
E3&=\bmu\T\hatbSig\pseudoinv\hatbsm\nonumber\\
&=\mu_f\bb\T\hatbSig\pseudoinv\left(\frac{\sigma_f\bb\bz\T\be}{T}+\frac{\sigmae\bY\T\be}{T}\right)\nonumber\\
&=E31+E32,\nonumber
\end{align}
where 
\begin{align}
E31&=\frac{1}{\sqrt{T}}\mu_f\sigma_f\left(\frac{\bz\T\be}{\sqrt{T}}\right)\bb\T\hatbSig\pseudoinv\bb=\frac{1}{\sqrt{T}}\mu_f\sigma_f\left(\frac{\bz\T\be}{\sqrt{T}}\right)\ugamma_{\hatSig1}=O_p(T^{-\frac{1}{2}}),\nonumber\\
E32&=\mu_f\sigmae\utbeta_{\hatSig1}=O_p(T^{-\frac{1}{2}}).\nonumber
\end{align}
Thus, $E3=O_p(T^{-\frac{1}{2}})$.

\bigskip

\noindent\textbf{\underline{Limit of $E4$}}: We have 
\begin{align}
E4&=\bmu\T\hatbSig\pseudoinv\bSig\hatbSig\pseudoinv\bmu\nonumber\\
&=\mu_f^2\bb\T\hatbSig\pseudoinv\left(\sigma_f^2\bb\bb\T+\sigmae^2\bI_N\right)\hatbSig\pseudoinv\bb\nonumber\\
&=\mu_f^2\sigma_f^2\left(\bb\T\hatbSig\pseudoinv\bb\right)^2+\mu_f^2\sigmae^2\left(\bb\T\hatbSig\pseudoinv\hatbSig\pseudoinv\bb\right)\nonumber\\
&=\mu_f^2\sigma_f^2\ugamma_{\hatSig1}^2+\mu_f^2\sigmae^2\ugamma_{\hatSig2}\nonumber\\
&\convergeInP \frac{\mu_f^2}{\sigma_f^2}\frac{(\tsigma_{\bB}-1)^2}{\tsigma_{\bB}^4}\frac{1}{(\rho-1)^2}(\rho\tsigma_{\bB}-\rho+1)\left[(\tsigma_{\bB}-1)(\rho-1)+\frac{\rho\tsigma_{\bB}^2}{(\tsigma_{\bB}-1)(\rho-1)}\right]\nonumber\\
&=\frac{\rho\ttheta\tphi^2(\tphi+1)^2}{(\rho-1)(\tphi+\rho)^4} + \frac{\rho^2\ttheta(\tphi+1)^2}{(\rho-1)^3(\tphi+\rho)^2}.\nonumber
\end{align}

\bigskip

\noindent\textbf{\underline{Limit of $E5$}}: We have 
\begin{align}
E5&=\hatbsm\T\hatbSig\pseudoinv\bSig\hatbSig\pseudoinv\hatbsm\nonumber\\
&=\left(\frac{\sigma_f\bb\bz\T\be}{T}+\frac{\sigmae\bY\T\be}{T}\right)\T\hatbSig\pseudoinv\left(\sigma_f^2\bb\bb\T+\sigmae^2\bI_N\right)\hatbSig\pseudoinv\left(\frac{\sigma_f\bb\bz\T\be}{T}+\frac{\sigmae\bY\T\be}{T}\right)\nonumber\\
&=E51+E52+2E53+E54+E55+2E56,\nonumber
\end{align}
where 
\begin{align}
E51&=\sigma_f^4\left(\frac{\bz\T\be}{\sqrt{T}}\right)^2\frac{1}{T}\left(\bb\T\hatbSig\pseudoinv\bb\right)^2=\sigma_f^4\left(\frac{\bz\T\be}{\sqrt{T}}\right)^2\frac{1}{T}\ugamma_{\hatSig1}^2=O_p(T^{-1}),\nonumber\\
E52&=\sigma_f^2\sigmae^2\utbeta_{\hatSig1}^2=O_p(T^{-1}),\nonumber\\
E53&=\sigma_f^3\sigmae\left(\frac{\bz\T\be}{\sqrt{T}}\right)\frac{1}{\sqrt{T}}\ugamma_{\hatSig1}\utbeta_{\hatSig1}=O_p(T^{-1}),\nonumber\\
E54&=\sigma_f^2\sigmae^2\left(\frac{\bz\T\be}{\sqrt{T}}\right)^2\frac{1}{T}\ugamma_{\hatSig2}=O_p(T^{-1})\nonumber\\
E55&=\sigmae^4\uttalpha_{\hatSig2}\convergeInP \frac{\rho}{(\rho-1)^3},\nonumber\\
E56 &=\sigma_f\sigmae^3\left(\frac{\bz\T\be}{\sqrt{T}}\right)\frac{1}{\sqrt{T}}\utbeta_{\hatSig2}=O_p(T^{-1})
\end{align}
Thus, $E5\convergeInP \frac{\rho}{(\rho-1)^3}$.

\bigskip

\noindent\textbf{\underline{Limit of $E6$}}: We have 
\begin{align}
E6&=\bmu\T\hatbSig\pseudoinv\bSig\hatbSig\pseudoinv\hatbsm\nonumber\\
&=\mu_f\bb\T\hatbSig\pseudoinv\left(\sigma_f^2\bb\bb\T+\sigmae^2\bI_N\right)\hatbSig\pseudoinv\left(\frac{\sigma_f\bb\bz\T\be}{T}+\frac{\sigmae\bY\T\be}{T}\right)\nonumber\\
&=E61+E62+E63+E64,\nonumber
\end{align}
where 
\begin{align}
E61&=\mu_f\sigma_f^3\left(\frac{\bz\T\be}{\sqrt{T}}\right)\left(\bb\T\hatbSig\pseudoinv\bb\right)^2\frac{1}{\sqrt{T}}=\mu_f\sigma_f^3\left(\frac{\bz\T\be}{\sqrt{T}}\right)\ugamma_{\hatSig1}^2\frac{1}{\sqrt{T}}=O_p(T^{-\frac{1}{2}}),\nonumber\\
E62&=\mu_f\sigmae\sigma_f^2\ugamma_{\hatSig1}\utbeta_{\hatSig1}=O_p(T^{-\frac{1}{2}}),\nonumber\\
E63&=\mu_f\sigmae^2\sigma_f\ugamma_{\hatSig2}\left(\frac{\bz\T\be}{\sqrt{T}}\right)\frac{1}{\sqrt{T}}=O_p(T^{-\frac{1}{2}}),\nonumber\\
E64&=\mu_f\sigmae^3\utbeta_{\hatSig2}=O_p(T^{-\frac{1}{2}}).\nonumber
\end{align}
Thus, $E6=O_p(T^{-\frac{1}{2}})$.

\paragraph{\texorpdfstring{Asymptotic results of intermediate terms used in \ref{B323}}{Asymptotic results of intermediate terms used in \ref{B323}}}\label{B324}

\begin{align}
\utbeta_{0}&=\left(\frac{1}{T}\bY\T\be\right)\T\bb=O_p\left(T^{-\frac{1}{2}}\right)\nonumber\\
\utbeta_{S1}&=\left(\frac{1}{T}\bY\T\be\right)\T\calS\pseudoinv\bb=O_p\left(T^{-\frac{1}{2}}\right)\nonumber\\
\utbeta_{S11}&=\left(\frac{1}{T}\bY\T\be\right)\T\calS\pseudoinv\calS\bb=O_p\left(T^{-\frac{1}{2}}\right)\nonumber\\
\utbeta_{S2}&=\left(\frac{1}{T}\bY\T\be\right)\T\left[\calS\pseudoinv\right]^2\bb=O_p\left(T^{-\frac{1}{2}}\right)\nonumber\\
\utbeta_{B1}&=\left(\frac{1}{T}\bY\T\be\right)\T\bB\pseudoinv\bb=O_p\left(T^{-\frac{1}{2}}\right)\nonumber\\
\utbeta_{B2}&=\left(\frac{1}{T}\bY\T\be\right)\T\left[\bB\pseudoinv\right]^2\bb=O_p\left(T^{-\frac{1}{2}}\right)\nonumber\\
\utbeta_{B11}&=\bb\T\bB\bB\pseudoinv\left(\frac{1}{T}\bY\T\be\right)=O_p\left(T^{-\frac{1}{2}}\right)\nonumber\\
\utbeta_{IB}&=\bb\T\left(\bI_N-\bB\bB\pseudoinv\right)\left(\frac{1}{T}\bY\T\be\right)=O_p\left(T^{-\frac{1}{2}}\right)\nonumber\\
\utbeta_{A1}&=\left(\frac{1}{T}\bY\T\be\right)\T\bA\pseudoinv\bb=O_p\left(T^{-\frac{1}{2}}\right)\nonumber\\
\utbeta_{A2}&=\left(\frac{1}{T}\bY\T\be\right)\T\left[\bA\pseudoinv\right]^2\bb=O_p\left(T^{-\frac{1}{2}}\right)\nonumber\\
\utbeta_{A3}&=\left(\frac{1}{T}\bY\T\be\right)\T\left[\bA\pseudoinv\right]^3\bb=O_p\left(T^{-\frac{1}{2}}\right)\nonumber\\
\utbeta_{\hatSig1}&=\left(\frac{1}{T}\bY\T\be\right)\T\hatbSig\pseudoinv\bb=O_p\left(T^{-\frac{1}{2}}\right)\nonumber\\
\utbeta_{\hatSig2}&=\left(\frac{1}{T}\bY\T\be\right)\T\left[\hatbSig\pseudoinv\right]^2\bb=O_p\left(T^{-\frac{1}{2}}\right)\nonumber\\
\utalpha_{S1} &= \left(\frac{1}{T}\bY\T\be\right)\T\calS\pseudoinv\left(\frac{1}{T}\bY\T\bH\bz\right)=O_p\left(T^{-\frac{1}{2}}\right)\nonumber\\
\utalpha_{S2} &= \left(\frac{1}{T}\bY\T\be\right)\T\left[\calS\pseudoinv\right]^2\left(\frac{1}{T}\bY\T\bH\bz\right)=O_p\left(T^{-\frac{1}{2}}\right)\nonumber\\
\utalpha_{S11} &= \left(\frac{1}{T}\bY\T\be\right)\T\calS\pseudoinv\calS\left(\frac{1}{T}\bY\T\bH\bz\right)=O_p\left(T^{-\frac{1}{2}}\right)\nonumber\\
\utalpha_{S3} &= \left(\frac{1}{T}\bY\T\be\right)\T\left[\calS\pseudoinv\right]^3\left(\frac{1}{T}\bY\T\bH\bz\right)=O_p\left(T^{-\frac{1}{2}}\right)\nonumber\\
\uttalpha_{S1} &= \left(\frac{1}{T}\bY\T\be\right)\T\calS\pseudoinv\left(\frac{1}{T}\bY\T\be\right)\convergeInP \frac{1}{\rho-1}\nonumber\\
\uttalpha_{S11} &= \left(\frac{1}{T}\bY\T\be\right)\T\calS\pseudoinv\calS\left(\frac{1}{T}\bY\T\be\right)\convergeInP 1\nonumber\\
\uttalpha_{S2} &= \left(\frac{1}{T}\bY\T\be\right)\T\left[\calS\pseudoinv\right]^2\left(\frac{1}{T}\bY\T\be\right)\convergeInP \frac{\rho}{(\rho-1)^3}\nonumber\\
\utalpha_{B1} &= \left(\frac{1}{T}\bY\T\bH\bz\right)\T\bB\pseudoinv\left(\frac{1}{T}\bY\T\be\right)=O_p\left(T^{-\frac{1}{2}}\right)\nonumber\\
\utalpha_{B2} &= \left(\frac{1}{T}\bY\T\be\right)\T\left[\bB\pseudoinv\right]^2\left(\frac{1}{T}\bY\T\bH\bz\right)=O_p\left(T^{-\frac{1}{2}}\right)\nonumber\\
\utalpha_{B11} &= \left(\frac{1}{T}\bY\T\be\right)\T\bB\pseudoinv\bB\left(\frac{1}{T}\bY\T\bH\bz\right)=O_p\left(T^{-\frac{1}{2}}\right)\nonumber\\
\utalpha_{B21} &= \left(\frac{1}{T}\bY\T\be\right)\T\left[\bB\pseudoinv\right]^2\bB\left(\frac{1}{T}\bY\T\bH\bz\right)=O_p\left(T^{-\frac{1}{2}}\right)\nonumber\\
\utalpha_{B31} &= \left(\frac{1}{T}\bY\T\be\right)\T\left[\bB\pseudoinv\right]^3\bB\left(\frac{1}{T}\bY\T\bH\bz\right)=O_p\left(T^{-\frac{1}{2}}\right)\nonumber\\
\uttalpha_{B1} &= \left(\frac{1}{T}\bY\T\be\right)\T\bB\pseudoinv\left(\frac{1}{T}\bY\T\be\right)\convergeInP\frac{1}{\sigmae^2(\rho-1)}\nonumber\\
\uttalpha_{A1} &= \left(\frac{1}{T}\bY\T\be\right)\T\bA\pseudoinv\left(\frac{1}{T}\bY\T\be\right)\convergeInP\frac{1}{\sigmae^2(\rho-1)}\nonumber\\
\uttalpha_{A2} &= \left(\frac{1}{T}\bY\T\be\right)\T\left[\bA\pseudoinv\right]^2\left(\frac{1}{T}\bY\T\be\right)\convergeInP\frac{\rho}{\sigmae^4(\rho-1)^3}\nonumber\\
\uttalpha_{\hatSig1} &= \left(\frac{1}{T}\bY\T\be\right)\T\hatbSig\pseudoinv\left(\frac{1}{T}\bY\T\be\right)\convergeInP \frac{1}{\sigmae^2}\frac{1}{\rho-1}\nonumber\\
\uttalpha_{\hatSig2} &= \left(\frac{1}{T}\bY\T\be\right)\T\left[\hatbSig\pseudoinv\right]^2\left(\frac{1}{T}\bY\T\be\right)\convergeInP \frac{1}{\sigmae^4}\frac{\rho}{(\rho-1)^3}\nonumber\\
\ugamma_{\hatSig1} &=\bb\T\hatbSig\pseudoinv\bb\convergeInP \frac{1}{\sigma_f^2}\frac{(\tsigma_{\bB}-1)(\rho\tsigma_{\bB}-\rho+1)}{\tsigma_{\bB}^2(\rho-1)}\nonumber\\
\ugamma_{\hatSig2} &=\bb\T\left[\hatbSig\pseudoinv\right]^2\bb\convergeInP \frac{1}{\sigma_f^4}\frac{1}{v_{\bA}^2}\left(\frac{\sigma_f^2}{\tsigma_{\bB}^2}\frac{1}{\rho-1}\right)^2\left(\tsigma_{\bB}-1\right)^2\left(\tsigma_{\bB}+\rho-1\right)\left(\rho\tsigma_{\bB}-\rho+1\right)\nonumber\\
\utzeta_{1}&=\left(\frac{1}{T}\bY\T\be\right)\T\bv_{\bB}\T=O_p(T^{-\frac{1}{2}})\nonumber\\
\utzeta_{2}&=\left(\frac{1}{T}\bY\T\be\right)\T\bk_{\bB}=O_p(T^{-\frac{1}{2}})\nonumber\\
\utzeta_{3}&=\left(\frac{1}{T}\bY\T\be\right)\T\bp_{\bB}=O_p(T^{-\frac{1}{2}})\nonumber\\
\utzeta_{4}&=\bk_{\bB}\T\calS\pseudoinv\left(\frac{1}{T}\bY\T\be\right)=O_p(T^{-\frac{1}{2}})\nonumber\\
\utzeta_{5}&=\bk_{\bB}\T\calS\pseudoinv\calS\left(\frac{1}{T}\bY\T\be\right)=O_p(T^{-\frac{1}{2}})\nonumber\\
\utzeta_{6}&=\bq_{\bB}\T\left(\frac{1}{T}\bY\T\be\right)=O_p(T^{-\frac{1}{2}})\nonumber\\
\utzeta_{7}&=\bq_{\bB}\T\calS\left(\frac{1}{T}\bY\T\be\right)=O_p(T^{-\frac{1}{2}})\nonumber\\
\utzeta_{8}&=\bu_{\bA}\pseudoinv\left(\frac{1}{T}\bY\T\be\right)=O_p(T^{-\frac{1}{2}})\nonumber\\
\utzeta_{9}&=\left(\frac{1}{T}\bY\T\be\right)\T\bv_{\bA}\pseudoinv=O_p(T^{-\frac{1}{2}})\nonumber\\
\utzeta_{10}&=\bh_{\bA}\left(\frac{1}{T}\bY\T\be\right)=O_p(T^{-\frac{1}{2}})\nonumber\\
\utzeta_{11}&=\left(\frac{1}{T}\bY\T\be\right)\T\bk_{\bA}=O_p(T^{-\frac{1}{2}})\nonumber\\
\utzeta_{12}&=\left(\frac{1}{T}\bY\T\be\right)\T\calS\pseudoinv\bp_{\bB}=O_p(T^{-\frac{1}{2}})\nonumber\\
\utzeta_{13}&=\bk_{\bB}\T\calS\pseudoinv\calS\pseudoinv\left(\frac{1}{T}\bY\T\be\right)=O_p(T^{-\frac{1}{2}})\nonumber\\
\utzeta_{14}&=\bq_{\bB}\T\calS\pseudoinv\left(\frac{1}{T}\bY\T\be\right)=O_p(T^{-\frac{1}{2}})\nonumber\\
\utzeta_{15}&=\bh_{\bB}\calS\pseudoinv\left(\frac{1}{T}\bY\T\be\right)=O_p(T^{-\frac{1}{2}})\nonumber\\
\utzeta_{16}&=\left(\frac{1}{T}\bY\T\be\right)\T\calS\pseudoinv\calS\pseudoinv\bp_{\bB}=O_p(T^{-\frac{1}{2}})\nonumber\\
\utzeta_{17}&=\left(\frac{1}{T}\bY\T\be\right)\T\bB\pseudoinv\bv_{\bA}\pseudoinv=O_p(T^{-\frac{1}{2}})\nonumber
\end{align}

Here we present some useful results:

\begin{enumerate}
\setcounter{enumi}{9}
\item $\left(\frac{1}{T}\bY\T\be\right)\T\bB\pseudoinv\bB\pseudoinv \Rightarrow\frac{1}{\sigmae^4}\left(\frac{1}{T}\bY\T\be\right)\T\calS\pseudoinv\calS\pseudoinv$
\item $\left(\frac{1}{T}\bY\T\be\right)\T\bA\pseudoinv\Rightarrow\left(\frac{1}{T}\bY\T\be\right)\T\bB\pseudoinv$
\item $\bA\pseudoinv\left(\frac{1}{T}\bY\T\be\right)\Rightarrow\bB\pseudoinv\left(\frac{1}{T}\bY\T\be\right)$
\end{enumerate}

\paragraph{\texorpdfstring{Proofs for \ref{B324}}{Proofs for \ref{B324}}}\label{B325}

\noindent\textbf{\underline{Limit of $\utbeta_{0}$}}: Consider the square of $\utbeta_{0}$, we have
\begin{equation}
    \utbeta_{0}^2=\left(\frac{1}{T}\bY\T\be\right)\T\bb\bb\T\left(\frac{1}{T}\bY\T\be\right)\nonumber.
\end{equation}
Based on trace Lemma \ref{lemma:Trace_Lemma} and same justification in $D3$, we have 
\begin{align}
\utbeta_{0}^2&-\frac{1}{T}\bb\T\bb=O_p(T^{-\frac{1}{2}}).\nonumber
\end{align}
Thus, $\utbeta_{0}=O_p(T^{-\frac{1}{2}})$.

\bigskip
\noindent\textbf{\underline{Limit of $\utbeta_{S1}$}}: By the same justification in $D3$, we have
\begin{align}
\utbeta_{S1}=\left(\frac{1}{T}\bY\T\be\right)\T\calS\pseudoinv\bb=O_p(T^{-\frac{1}{2}}).\nonumber
\end{align}

\bigskip
\noindent\textbf{\underline{Limit of $\uttalpha_{S1}$}}: By the same justification in $D2$, we have
\begin{equation}
\uttalpha_{S1}=\left(\frac{1}{T}\bY\T\be\right)\T\calS\pseudoinv\left(\frac{1}{T}\bY\T\be\right)\convergeInP \frac{1}{\rho-1}.\nonumber
\end{equation}

\bigskip
\noindent\textbf{\underline{Limit of $\utalpha_{S1}$}}: Consider the square of $\utalpha_{S1}$, we have
\begin{equation}
    T\utalpha_{S1}^2=\frac{1}{T}\bz\T\bH\bY\frac{1}{T}\calS\pseudoinv\bY\T\be\be\T\bY\calS\pseudoinv\frac{1}{T}\bY\T\bH\bz\nonumber.
\end{equation}
Based on trace Lemma \ref{lemma:Trace_Lemma}, we have 
\begin{align}
T\utalpha_{S1}^2&-\frac{1}{T}\be\T\bY\calS\pseudoinv\frac{1}{T}\bY\T\bH\bY\frac{1}{T}\calS\pseudoinv\bY\T\be=O_p(T^{-\frac{1}{2}}),\nonumber\\
T\utalpha_{S1}^2&-\uttalpha_{S1}=O_p(T^{-\frac{1}{2}}).\nonumber
\end{align}
Thus, $\utalpha_{S1}=O_p(T^{-\frac{1}{2}})$.

\bigskip
\noindent\textbf{\underline{Limit of $\utbeta_{S11}$}}: Consider the square of $\utbeta_{S11}$, we have
\begin{equation}
    \utbeta_{S11}^2=\left(\frac{1}{T}\bY\T\be\right)\T\calS\pseudoinv\calS\bb\bb\T\calS\calS\pseudoinv\left(\frac{1}{T}\bY\T\be\right)\nonumber.
\end{equation}
Based on trace Lemma \ref{lemma:Trace_Lemma}, we have 
\begin{align}
\utbeta_{S11}^2&-\frac{1}{T}\bb\T\calS\pseudoinv\calS\bb=O_p(T^{-\frac{1}{2}}).\nonumber
\end{align}
Based on $\ugamma_{S1}$, we have $\bb\T\calS\pseudoinv\calS\bb\convergeInP\tilde{b}^2/\rho$. Thus, $\utbeta_{S11}=O_p(T^{-\frac{1}{2}})$.

\bigskip
\noindent\textbf{\underline{Limit of $\uttalpha_{S11}$}}:
\begin{align}
\uttalpha_{S11}&=\left(\frac{1}{T}\bY\T\be\right)\T\calS\pseudoinv\calS\left(\frac{1}{T}\bY\T\be\right)\nonumber\\
&=\left(\frac{1}{T}\bY\T\be\right)\T\frac{1}{T}\bY\T\bH\left(\frac{1}{T}\bH\bY\bY\T\bH\right)^{-2}\bH\bY\frac{1}{T}\bY\T\bH\bY\left(\frac{1}{T}\bY\T\be\right)\nonumber\\
&=\left(\frac{1}{T}\bY\T\be\right)\T\frac{1}{T}\bY\T\bH\left(\frac{1}{T}\bH\bY\bY\T\bH\right)^{-1}\bH\bY\left(\frac{1}{T}\bY\T\be\right).\nonumber
\end{align}
Thus, $\uttalpha_{S11}-\rho\frac{1}{p}\Tr(I_{T})=O_p(T^{-\frac{1}{2}})$, $\uttalpha_{S11}\convergeInP 1$.

\bigskip
\noindent\textbf{\underline{Limit of $\utzeta_{1}$}}: 
\begin{align}
\utzeta_{1}&=\left(\frac{1}{T}\bY\T\be\right)\T\bv_{\bB}\T\nonumber\\
&=\left(\frac{1}{T}\bY\T\be\right)\T\left(\bI_N-\calS\pseudoinv\calS\right)\bb\nonumber\\
&=\utbeta_{0}-\utbeta_{S11}=O_p(T^{-\frac{1}{2}}).\nonumber
\end{align}

\bigskip
\noindent\textbf{\underline{Limit of $\utzeta_{2}$}}: 
\begin{align}
\utzeta_{2}&=\left(\frac{1}{T}\bY\T\be\right)\T\bk_{\bB}=\frac{\sigma_f}{\sigmae}\utalpha_{S1}=O_p(T^{-\frac{1}{2}}).\nonumber
\end{align}

\bigskip
\noindent\textbf{\underline{Limit of $\utzeta_{3}$}}: 
\begin{align}
\utzeta_{3}&=-\left\Vert\bk_{\bB}\right\Vert^2\utzeta_{1}-\utzeta_{2}=O_p(T^{-\frac{1}{2}}).\nonumber
\end{align}

\bigskip
\noindent\textbf{\underline{Limit of $\utalpha_{S2}$}}: Consider the square of $\utalpha_{S2}$, we have
\begin{equation}
    T\utalpha_{S2}^2=\frac{1}{T}\bz\T\bH\bY\calS\pseudoinv\calS\pseudoinv\frac{1}{T}\bY\T\be\frac{1}{T}\be\T\bY\calS\pseudoinv\calS\pseudoinv\bY\T\bH\bz\nonumber.
\end{equation}
Based on trace Lemma \ref{lemma:Trace_Lemma}, we have 
\begin{align}
T\utalpha_{S2}^2&-\frac{1}{T}\left(\frac{1}{T}\bY\T\be\right)\T\calS\pseudoinv\calS\pseudoinv\bY\T\bH\bH\bY\calS\pseudoinv\calS\pseudoinv\left(\frac{1}{T}\bY\T\be\right)=O_p(T^{-\frac{1}{2}}),\nonumber\\
T\utalpha_{S2}^2&-\left(\frac{1}{T}\bY\T\be\right)\T\left[\calS\pseudoinv\right]^3\left(\frac{1}{T}\bY\T\be\right)=O_p(T^{-\frac{1}{2}}),\nonumber\\
T\utalpha_{S2}^2&\convergeInP\frac{1}{\rho^3}\frac{1+1/\rho}{(1-1/\rho)^5}=\frac{\rho^2+\rho}{(\rho-1)^5}.\nonumber
\end{align}
Thus, $\utalpha_{S2}=O_p(T^{-\frac{1}{2}})$.

\bigskip
\noindent\textbf{\underline{Limit of $\utzeta_{4}$}}: $
\utzeta_{4}=\frac{\sigma_f}{\sigmae}\utalpha_{S2}=O_p(T^{-\frac{1}{2}}).$

\bigskip
\noindent\textbf{\underline{Limit of $\utzeta_{5}$}}: $
\utzeta_{5}=\frac{\sigma_f}{\sigmae}\utalpha_{S1}=O_p(T^{-\frac{1}{2}}).$

\bigskip
\noindent\textbf{\underline{Limit of $\utzeta_{6}$}}: $
\utzeta_{6}=-\frac{v_{\bB}^2}{\sigmae^2}\utzeta_{4}-\frac{1}{\sigmae^2}\utbeta_{S1}=O_p(T^{-\frac{1}{2}}).$

\bigskip
\noindent\textbf{\underline{Limit of $\utzeta_{7}$}}: $
\utzeta_{7}=-\frac{v_{\bB}^2}{\sigmae^2}\utzeta_{5}-\frac{1}{\sigmae^2}\utbeta_{S11}=O_p(T^{-\frac{1}{2}}).$

\bigskip
\noindent\textbf{\underline{Limit of $\utbeta_{B1}$}}: 
\begin{align}
    \utbeta_{B1}&=\left(\frac{1}{T}\bY\T\be\right)\bB\pseudoinv\bb\nonumber\\
    &=\frac{1}{\sigmae^2}\utbeta_{S1}+\frac{1}{\sigmae^2}\utzeta_{1}\uxi_5-\frac{1}{\tsigma_{\bB}}\utzeta_3\uxi_3\nonumber\\
    &=O_p(T^{-\frac{1}{2}}).\nonumber
\end{align}

\bigskip
\noindent\textbf{\underline{Limit of $\utbeta_{B11}$}}: 
\begin{align}
    \utbeta_{B11}&=\bb\T\bB\bB\pseudoinv\left(\frac{1}{T}\bY\T\be\right)\nonumber\\
    &\Rightarrow\bb\T\calS\calS\pseudoinv\left(\frac{1}{T}\bY\T\be\right)\nonumber\\
    &=\utbeta_{S11}=O_p(T^{-\frac{1}{2}}).\nonumber
\end{align}

\bigskip
\noindent\textbf{\underline{Limit of $\utbeta_{IB}$}}: $
\utbeta_{IB}=\utbeta_{0}-\utbeta_{B11}=O_p(T^{-\frac{1}{2}}).$

\bigskip
\noindent\textbf{\underline{Limit of $\utzeta_{8}$}}: $
\utzeta_{8}=\frac{1}{\left\Vert\bu_{\bA}\right\Vert^2}\utbeta_{IB}=O_p(T^{-\frac{1}{2}}).$

\bigskip
\noindent\textbf{\underline{Limit of $\utzeta_{9}$}}: $
\utzeta_{9}=\frac{\sigma_f\sigmae}{\left\Vert\bv_{\bA}\right\Vert^2}\utalpha_{S11}-\utalpha_{B11}=O_p(T^{-\frac{1}{2}}).$

\bigskip
\noindent\textbf{\underline{Limit of $\utalpha_{S11}$}}: Consider the square of $\utalpha_{S11}$, we have
\begin{equation}
    T\utalpha_{S11}^2=\frac{1}{T}\bz\T\bH\bY\calS\calS\pseudoinv\bY\T\be\frac{1}{T}\be\T\bY\calS\pseudoinv\calS\bY\T\bH\bz\nonumber.
\end{equation}
Based on trace Lemma \ref{lemma:Trace_Lemma}, we have 
\begin{align}
T\utalpha_{S11}^2&-\frac{1}{T}\left(\frac{1}{T}\bY\T\be\right)\T\calS\pseudoinv\calS\bY\T\bH\bH\bY\calS\calS\pseudoinv\left(\frac{1}{T}\bY\T\be\right)=O_p(T^{-\frac{1}{2}}),\nonumber\\
T\utalpha_{S11}^2&-\left(\frac{1}{T}\bY\T\be\right)\T\calS\left(\frac{1}{T}\bY\T\be\right)=O_p(T^{-\frac{1}{2}}),\nonumber\\
T\utalpha_{S11}^2&\convergeInP\rho.\nonumber
\end{align}
Thus, $\utalpha_{S11}=O_p(T^{-\frac{1}{2}})$.

\bigskip
\noindent\textbf{\underline{Limit of $\utalpha_{B11}$}}: Based on useful result 4 in \ref{B321}, we have 
\begin{align}
\utalpha_{B11} &= \left(\frac{1}{T}\bY\T\be\right)\T\bB\pseudoinv\bB\left(\frac{1}{T}\bY\T\bH\bz\right)\nonumber\\
&\Rightarrow\utalpha_{S11}+\utzeta_1\uzeta_5-\frac{\sigmae^2}{\tsigma_{\bB}}\utzeta_3\uzeta_6\nonumber\\
&=O_p(T^{-\frac{1}{2}}).\nonumber
\end{align}

\bigskip
\noindent\textbf{\underline{Limit of $\utalpha_{B1}$}}: 
\begin{align}
\utalpha_{B1} &= \left(\frac{1}{T}\bY\T\bH\bz\right)\T\bB\pseudoinv\left(\frac{1}{T}\bY\T\be\right)\nonumber\\
&\Rightarrow\frac{1}{\sigmae^2}\utalpha_{S1}+\frac{1}{\sigmae^2}\uzeta_1\utzeta_4-\frac{1}{\tsigma_{\bB}}\uzeta_3\utzeta_6\nonumber\\
&=O_p(T^{-\frac{1}{2}}).\nonumber
\end{align}

\bigskip
\noindent\textbf{\underline{Limit of $\utzeta_{10}$}}: $
\utzeta_{10}=\sigma_f\sigmae\utalpha_{B1}=O_p(T^{-\frac{1}{2}}).$

\bigskip
\noindent\textbf{\underline{Limit of $\utzeta_{11}$}}: $
\utzeta_{11}=\utbeta_{B1}=O_p(T^{-\frac{1}{2}}).$

\bigskip
\noindent\textbf{\underline{Limit of $\uttalpha_{B1}$}}:
\begin{align}
\uttalpha_{B1} &= \left(\frac{1}{T}\bY\T\be\right)\T\bB\pseudoinv\left(\frac{1}{T}\bY\T\be\right)\nonumber\\
&\Rightarrow\frac{1}{\sigmae^2}\uttalpha_{S1}+\frac{1}{\sigmae^2}\utzeta_1\utzeta_4-\frac{1}{\tsigma_{\bB}}\utzeta_3\utzeta_6\nonumber\\
&\convergeInP\frac{1}{\sigmae^2}\frac{1}{\rho-1}.\nonumber
\end{align}

\bigskip
\noindent\textbf{\underline{Limit of $\uttalpha_{A1}$}}:
\begin{align}
\uttalpha_{A1} &= \left(\frac{1}{T}\bY\T\be\right)\T\bA\pseudoinv\left(\frac{1}{T}\bY\T\be\right)\nonumber\\
&=\uttalpha_{B1}-\utzeta_{11}\utzeta_8-\utzeta_9\utzeta_{10}+\beta_{\bA}\utzeta_9\utzeta_8\nonumber\\
&\convergeInP\frac{1}{\sigmae^2}\frac{1}{\rho-1}.\nonumber
\end{align}

\bigskip
\noindent\textbf{\underline{Limit of $\utbeta_{A1}$}}: 
\begin{align}
\utbeta_{A1} &= \left(\frac{1}{T}\bY\T\be\right)\T\bA\pseudoinv\bb\nonumber\\
&=\utbeta_{B1}-\utzeta_{11}\uxi_{10}-\utzeta_9\uxi_{11}+\beta_{\bA}\utzeta_9\uxi_{10}\nonumber\\
&=O_p(T^{-\frac{1}{2}}).\nonumber
\end{align}

\bigskip
\noindent\textbf{\underline{Limit of $\utbeta_{A2}$}}: Based on useful result 7 in\ref{B321}, we have 
\begin{align}
\utbeta_{A2} &= \left(\frac{1}{T}\bY\T\be\right)\T\bA\pseudoinv\bA\pseudoinv\bb\nonumber\\
&=-\frac{1}{v_{\bA}^2}\utzeta_{10}+\tbeta_{\bA}\frac{1}{v_{\bA}^2}\utzeta_{8}\nonumber\\
&=O_p(T^{-\frac{1}{2}}).\nonumber
\end{align}

\bigskip
\noindent\textbf{\underline{Limit of $\uttalpha_{A2}$}}:  
\begin{align}
\uttalpha_{A2} &= \left(\frac{1}{T}\bY\T\be\right)\T\bA\pseudoinv\bA\pseudoinv\left(\frac{1}{T}\bY\T\be\right)\nonumber\\
&\Rightarrow\uttalpha_{B2}\nonumber\\
&\convergeInP\frac{1}{\sigmae^4}\frac{\rho}{(\rho-1)^3}.\nonumber
\end{align}

\bigskip
\noindent\textbf{\underline{Limit of $\uttalpha_{S2}$}}: By the same justification in $D5$, we have
\begin{equation}
\uttalpha_{S2}=\left(\frac{1}{T}\bY\T\be\right)\T\calS\pseudoinv\calS\pseudoinv\left(\frac{1}{T}\bY\T\be\right)\convergeInP \frac{\rho}{(\rho-1)^3}.\nonumber
\end{equation}

\bigskip
\noindent\textbf{\underline{Limit of $\utalpha_{S3}$}}: Consider the square of $\utalpha_{S3}$, we have
\begin{equation}
    T\utalpha_{S3}^2=\frac{1}{T}\bz\T\bH\bY\left[\calS\pseudoinv\right]^3\frac{1}{T}\bY\T\be\frac{1}{T}\be\T\bY\left[\calS\pseudoinv\right]^3\bY\T\bH\bz\nonumber.
\end{equation}
Based on trace Lemma \ref{lemma:Trace_Lemma}, we have 
\begin{align}
T\utalpha_{S3}^2&-\frac{1}{T}\left(\frac{1}{T}\bY\T\be\right)\T\left[\calS\pseudoinv\right]^3\bY\T\bH\bH\bY\left[\calS\pseudoinv\right]^3\left(\frac{1}{T}\bY\T\be\right)=O_p(T^{-\frac{1}{2}}),\nonumber\\
T\utalpha_{S3}^2&-\left(\frac{1}{T}\bY\T\be\right)\T\left[\calS\pseudoinv\right]^5\left(\frac{1}{T}\bY\T\be\right)=O_p(T^{-\frac{1}{2}}),\nonumber\\
T\utalpha_{S3}^2&=O_p(1).\nonumber
\end{align}
Thus, $\utalpha_{S2}=O_p(T^{-\frac{1}{2}})$.

\bigskip
\noindent\textbf{\underline{Limit of $\utzeta_{12}$}}: $
\utzeta_{12}=-\frac{\sigma_f}{\sigmae}\utalpha_{S2}=O_p(T^{-\frac{1}{2}}).$

\bigskip
\noindent\textbf{\underline{Limit of $\utzeta_{13}$}}: $
\utzeta_{13}=\frac{\sigma_f}{\sigmae}\utalpha_{S3}=O_p(T^{-\frac{1}{2}}).$

\bigskip
\noindent\textbf{\underline{Limit of $\utzeta_{14}$}}: $
\utzeta_{14}=-\frac{\left\Vert\bv_{\bB}\right\Vert^2}{\sigmae^2}\utzeta_{13}-\utzeta_{15}=O_p(T^{-\frac{1}{2}}).$

\bigskip
\noindent\textbf{\underline{Limit of $\utzeta_{15}$}}: $
\utzeta_{15}=\frac{1}{\sigmae^2}\utbeta_{S2}=O_p(T^{-\frac{1}{2}}).$

\bigskip
\noindent\textbf{\underline{Limit of $\utbeta_{S2}$}}: By the same justification in $D6$, we have
\begin{align}
\utbeta_{S2}=\left(\frac{1}{T}\bY\T\be\right)\T\calS\pseudoinv\calS\pseudoinv\bb=O_p(T^{-\frac{1}{2}}).\nonumber
\end{align}

\bigskip
\noindent\textbf{\underline{Limit of $\utbeta_{B2}$}}:  
\begin{align}
\utbeta_{B2} &= \left(\frac{1}{T}\bY\T\be\right)\T\bB\pseudoinv\bB\pseudoinv\bb\nonumber\\
&\Rightarrow\frac{1}{\sigmae^4}\utbeta_{S2}\nonumber\\
&=O_p(T^{-\frac{1}{2}}).\nonumber
\end{align}

\bigskip
\noindent\textbf{\underline{Limit of $\utalpha_{B2}$}}:  
\begin{align}
\utalpha_{B2} &= \left(\frac{1}{T}\bY\T\be\right)\T\bB\pseudoinv\bB\pseudoinv\left(\frac{1}{T}\bY\T\bH\bz\right)\nonumber\\
&\Rightarrow\frac{1}{\sigmae^4}\utalpha_{S2}\nonumber\\
&=O_p(T^{-\frac{1}{2}}).\nonumber
\end{align}

\bigskip
\noindent\textbf{\underline{Limit of $\utzeta_{16}$}}:
\begin{align}
\utzeta_{16}&=\left(\frac{1}{T}\bY\T\be\right)\T\calS\pseudoinv\calS\pseudoinv\bp_{\bB}\nonumber\\
&=-\left\Vert\bk_{\bB}\right\Vert^2\left(\frac{1}{T}\bY\T\be\right)\T\calS\pseudoinv\calS\pseudoinv\bv_{\bB}\T-\left(\frac{1}{T}\bY\T\be\right)\T\calS\pseudoinv\calS\pseudoinv\bk_{\bB}\nonumber\\
&=-\frac{\sigma_f}{\sigmae}\utalpha_{S3}=O_p(T^{-\frac{1}{2}})\nonumber
\end{align}

\bigskip
\noindent\textbf{\underline{Limit of $\utalpha_{B31}$}}:  
\begin{align}
\utalpha_{B31} &= \left(\frac{1}{T}\bY\T\be\right)\T\left[\bB\pseudoinv\right]^3\bB\left(\frac{1}{T}\bY\T\bH\bz\right)\nonumber\\
&\Rightarrow\frac{1}{\sigmae^4}\left(\frac{1}{T}\bY\T\be\right)\T\calS\pseudoinv\calS\pseudoinv\left[\calS\pseudoinv\calS\left(\frac{1}{T}\bY\T\bH\bz\right)+\bv_{\bB}\T\uzeta_5-\frac{\sigmae^2}{\tsigma_{\bB}}\bp_{\bB}\uzeta_6\right]\nonumber\\
&=\frac{1}{\sigmae^4}\left[\utalpha_{S2}+\frac{\sigmae^2}{\tsigma_{\bB}}\utzeta_{16}\uzeta_6\right]\nonumber\\
&=O_p(T^{-\frac{1}{2}}).\nonumber
\end{align}

\bigskip
\noindent\textbf{\underline{Limit of $\utalpha_{B21}$}}:  
\begin{align}
\utalpha_{B21} &= \left(\frac{1}{T}\bY\T\be\right)\T\left[\bB\pseudoinv\right]^2\bB\left(\frac{1}{T}\bY\T\bH\bz\right)\nonumber\\
&\Rightarrow\frac{1}{\sigmae^4}\left(\frac{1}{T}\bY\T\be\right)\T\calS\pseudoinv\calS\pseudoinv\left[\sigmae^2\calS
+\left(\sigma_f\sigmae\frac{1}{T}\bY\T\bH\bz\right)\bb\T\right]\left(\frac{1}{T}\bY\T\bH\bz\right)\nonumber\\
&=\frac{1}{\sigmae^2}\utalpha_{S1}+\frac{\sigma_f}{\sigmae^3}\utalpha_{S2}\ubeta_0\nonumber\\
&=O_p(T^{-\frac{1}{2}}).\nonumber
\end{align}

\bigskip
\noindent\textbf{\underline{Limit of $\utzeta_{17}$}}:  
\begin{align}
\utzeta_{17} &= \left(\frac{1}{T}\bY\T\be\right)\T\bB\pseudoinv\bv_{\bA}\pseudoinv\nonumber\\
&=\frac{\sigma_f\sigmae}{\left\Vert\bv_{\bA}\right\Vert^2}\left[\utalpha_{B1}-\utalpha_{B21}\right]\nonumber\\
&=O_p(T^{-\frac{1}{2}}).\nonumber
\end{align}

\bigskip
\noindent\textbf{\underline{Limit of $\utbeta_{A3}$}}: Based on results 8 and 11:
\begin{align}
\utbeta_{A3} &= \left(\frac{1}{T}\bY\T\be\right)\T\left[\bA\pseudoinv\right]^3\bb\nonumber\\
&\Rightarrow\left(\frac{1}{T}\bY\T\be\right)\T\bB\pseudoinv\left[\bB\pseudoinv\bv_{\bA}\pseudoinv-\bB\pseudoinv\bb\uxi_{21}-\bv_{\bA}\pseudoinv\uxi_{22}+\tbeta_{\bA}\bv_{\bA}\pseudoinv\uxi_{21}\right]\nonumber\\
&=\frac{1}{v_{\bA}^2}\sigma_f\sigmae\left(\utalpha_{B2}-\utalpha_{B31}\right)-\utbeta_{B2}\uxi_{21}-\utzeta_{17}\uxi_{22}+\tbeta_{\bA}\utzeta_{17}\uxi_{21}\nonumber\\
&=O_p(T^{-\frac{1}{2}}).\nonumber
\end{align}

\bigskip
\noindent\textbf{\underline{Limit of $\utbeta_{\hatSig1}$}}:
\begin{align}
\utbeta_{\hatSig1} &= \left(\frac{1}{T}\bY\T\be\right)\T\hatbSig\pseudoinv\bb\nonumber\\
&=\utbeta_{A1}-\frac{\utbeta_{A1}\ugamma_{A2}}{\ugamma_{A2}}-\frac{\utbeta_{A2}\ugamma_{A1}}{\ugamma_{A2}}+\frac{\ugamma_{A3}\utbeta_{A1}\ugamma_{A1}}{\ugamma_{A2}^2}\nonumber\\
&=O_p(T^{-\frac{1}{2}}).\nonumber
\end{align}

\bigskip
\noindent\textbf{\underline{Limit of $\uttalpha_{\hatSig1}$}}:
\begin{align}
\uttalpha_{\hatSig1} &= \left(\frac{1}{T}\bY\T\be\right)\T\hatbSig\pseudoinv\left(\frac{1}{T}\bY\T\be\right)\nonumber\\
&=\uttalpha_{A1}-\frac{\utbeta_{A1}\utbeta_{A2}}{\ugamma_{A2}}-\frac{\utbeta_{A2}\utbeta_{A1}}{\ugamma_{A2}}+\frac{\ugamma_{A3}\utbeta_{A1}\utbeta_{A1}}{\ugamma_{A2}^2}\nonumber\\
&\convergeInP\frac{1}{\sigmae^2}\frac{1}{\rho-1}.\nonumber
\end{align}

\bigskip
\noindent\textbf{\underline{Limit of $\uttalpha_{\hatSig2}$}}:
\begin{align}
\uttalpha_{\hatSig2} &= \left(\frac{1}{T}\bY\T\be\right)\T\hatbSig\pseudoinv\hatbSig\pseudoinv\left(\frac{1}{T}\bY\T\be\right)\nonumber\\
&\Rightarrow\uttalpha_{A2}\nonumber\\
&\convergeInP\frac{1}{\sigmae^2}\frac{\rho}{(\rho-1)^3}.\nonumber
\end{align}

\bigskip
\noindent\textbf{\underline{Limit of $\utbeta_{\hatSig2}$}}:
\begin{align}
\utbeta_{\hatSig2} &= \left(\frac{1}{T}\bY\T\be\right)\T\hatbSig\pseudoinv\hatbSig\pseudoinv\bb\nonumber\\
&=O_p(T^{-\frac{1}{2}}).\nonumber
\end{align}

\bigskip
\noindent\textbf{\underline{Limit of $\ugamma_{\hatSig1}$}}:
\begin{align}
\ugamma_{\hatSig1} &= \bb\T\hatbSig\pseudoinv\bb\nonumber\\
&=\ugamma_{A1}-\ugamma_{A1}-\frac{\ugamma_{A2}\ugamma_{A1}}{\ugamma_{A2}}+\frac{\ugamma_{A3}\ugamma_{A1}^2}{\ugamma_{A2}^2}\nonumber\\
&\convergeInP\frac{1}{\sigma_f^2}\frac{(\tsigma_{\bB}-1)(\rho\tsigma_{\bB}-\rho+1)}{\tsigma_{\bB}^2(\rho-1)}\nonumber
\end{align}

\bigskip
\noindent\textbf{\underline{Limit of $\ugamma_{\hatSig2}$}}:
\begin{align}
\ugamma_{\hatSig2} &= \bb\T\hatbSig\pseudoinv\hatbSig\pseudoinv\bb\nonumber\\
&=\ugamma_{A2}-\frac{1}{\ugamma_{A2}}\left(\ugamma_{A2}^2+\ugamma_{A3}\ugamma_{A1}+\ugamma_{A1}\ugamma_{A3}\right)\nonumber\\
&+\frac{\ugamma_{A3}}{\ugamma_{A2}^2}\left(\ugamma_{A2}\ugamma_{A1}+\ugamma_{A1}\ugamma_{A2}\right)+\frac{\ugamma_{A4}}{\ugamma_{A2}^2}\ugamma_{A1}^2-\frac{\ugamma_{A3}^2}{\ugamma_{A2}^3}\ugamma_{A1}^2\nonumber\\
&\convergeInP\frac{1}{\sigma_f^4}\frac{1}{v_{\bA}^2}\left(\frac{\sigma_f^2}{\tsigma_{\bB}^2}\frac{1}{\rho-1}\right)^2\left(\tsigma_{\bB}-1\right)^2\left(\tsigma_{\bB}+\rho-1\right)\left(\rho\tsigma_{\bB}-\rho+1\right)\nonumber
\end{align}
\end{proof}

\bigskip

\section{Asymptotic Results for Ridge Regularized Estimator}
In this section, our examination focus on the asymptotic performance of the ridge regularized estimator. 

\subsection{Ridge Regularized Estimator}

\begin{definition}[Regularized Estimator]
Given the sample mean vector $\hatbmu$ and the sample covariance matrix $\hatbSig$ (defined in main body Section \ref{Sec:Problem}) from observed data, along with a specified risk constraint $\sigma$, we define the following estimator:
\begin{align}\label{Def:regularized-estimator}
\hatbomeastlam = \dfrac{\sigma}{\sqrt{\hatbmu\T\hatbSiglam\inv\hatbmu}}\hatbSiglam\inv\hatbmu,
\end{align}
where $\hatbSiglam\inv =\left(\hatbSig+\lambda\bm{I}_N\right)\inv $ and $\lambda>0$ the regularized parameter.
\end{definition}

\subsection{Assumptions}

\begin{assumptionD}[High Dimensionality]\label{ASSU_D1:HD}
We consider asymptotic setup where $T, N\rightarrow\infty$, and $\rho_T\coloneq N/T\rightarrow\rho\in(0,1)\cup (1,\infty)$. 
\end{assumptionD}

\begin{assumptionD}[DGP]\label{ASSU_D2:DGP} The excess returns vector at time $t$ is generated as $\br_t=\bmu+\bSig^{\frac{1}{2}}\by_t$, where $\by_t$ is independently and identically distributed following $\NormDis\left(\bm{0},\bI_N\right)$ for each time period $t$.
\end{assumptionD}
Note that normality assumption is common in portfolio optimization. It is more strict than Assumption \ref{ASSU2:DGP}, which assumes an elliptical distribution for $\by_t$ to allow for more tail flexibility. The main reason for assuming normality is for technical proof: it ensures the independence of the sample mean and variance - a property unique to the normal distribution - therefore facilitating certain theoretical proofs.

\begin{assumptionD}[Population Covariance Matrix]\label{ASSU_D3:SIGMA}
\noindent
\begin{itemize}
    \item[(a)] The population covariance matrix $\bSig$ is a deterministic $N\times N$ positive semidefinite matrix.
     \item[(b)] Denote by $\bSig=\sum_{i=1}^{N}\tau_i\bupsilon_i\bupsilon_i\T$ the eigenvalue decomposition of $\bSig$ with $\tau_1\geq\tau_2\geq\dots\geq\tau_N\geq0$. The empirical spectral distribution (ESD) of population covariance matrix is defined as: $\hatH_{N}(\tau)=N^{-1}\sum_{i=1}^{N}\mathbb{I}_{\tau_i\leq \tau}$. It is assumed that $\hatH_{N}$ converges to a limiting law $H$, which is referred to as the limiting spectral distribution (LSD).
     \item[(c)] The support of $H$, $Supp(H)$, is included in a compact interval $[M_1,M_2]$, with $0<M_1\leq M_2<\infty$.
     \item[(d)] $\limsup_{N\rightarrow\infty}\tau_1< M$, $\liminf_{N\rightarrow\infty}\tau_N>0$, $\int \tau^{-1} \,d\hatH_{N}(\tau)<M$.
\end{itemize}
\end{assumptionD}
The assumption of a limiting population spectral distribution is common in large-dimensional asymptotics. Specifically, $(b)$ describes the cross-sectional distribution of population eigenvalues, denoted as $\hatH_N$. The subscript $N$ in $\hatH_N$ is to emphasize its dependence on a particular $\bSig$ with specific dimensionality. Note that the condition $(c)$ does not imply the fulfillment of $(d)$. This distinction arises because a minor fraction [expressed formally as $o(N)$] of the eigenvalues of $\bSig$ may either escalate indefinitely or diminish to zero, even though $(c)$ is satisfied. The last condition in $(c)$ requires the eigenvalues of $\bSig$ not to accumulate near 0.

\begin{assumptionD}[Reweighted Spectral Distribution]\label{ASSU_D4:mu_G} The empirical distribution of the expected returns $\bmu$, projected on the basis of eigenvectors $\left(\langle\bmu,\bupsilon_1\rangle,...,\langle\bmu,\bupsilon_N\rangle\right)$ is defined as: 
\[\hatG_{N}(\tau)=\dfrac{1}{\left\Vert\bmu\right\Vert_2^2}\sum_{i=1}^{N}\langle\bmu,\bupsilon_i\rangle^2 \mathbb{I}_{\tau_i\leq \tau}.
\] It is assumed that $\hatG_{N}$ converges to a limiting law $G$.
\end{assumptionD}
The distribution $\hatG_N$ represents the empirical distribution, reweighted through the projection of $\bmu$ onto the eigenvectors $\bupsilon_i$ of the covariance matrix $\bSig$. This reweighting process emphasizes the influence of $\bmu$'s orientation relative to the principal components defined by $\bSig$'s eigenvectors. In finance, this distribution allows investors to gauge whether expected returns are aligned with or diverge from the principal risk directions of the market.

\begin{assumptionD}[Sparse Mean Vector]\label{ASSU_D5:Sparse_Mean}
Assume that the mean vector $\bmu$ is sparse such that as $N\rightarrow\infty$, $\left\Vert\bmu(N)\right\Vert_2^2\convergeInP \tilde{\xi}^2$.
\end{assumptionD}

\begin{assumptionD}[Maximum Theoretical Sharpe Ratio]\label{ASSU_D6:MSR}
Assume that the square of theoretical Sharpe ratio, ${\theta}$ (equivalently, ${\bmu\T\bSig\inv\bmu}$), is a ``function" of $N$, represented as ${\theta(N)}$. As $N$ increases, ${\theta(N)}\convergeInP {\ttheta}$, where ${\ttheta}$ is bounded away from 0 and infinity.
\end{assumptionD}

\bigskip

\subsection{Main results}
First, we introduce some notations used in the following Theorems. Let $m_{F_{\hatbSig}}(z)$ represent the Stieltjes transform of the empirical spectral distribution (ESD) of sample covariance matrix $\hatbSig$. The ESD is defined as $F_{\hatbSig}(d)=N^{-1}\sum_{i=1}^{N}\mathbb{I}_{d_i\leq d}$, where $d_1\geq\dots\geq d_N$ are eigenvalues of $\hatbSig$. Accordingly, the Stieltjes transform is given by:
\[
m_{F_{\hatbSig}}(z) = \dfrac{1}{N}\sum_{i=1}^{N}\dfrac{1}{d_i-z}=\dfrac{1}{N}\Tr\left[\left(\hatbSig-z\bI_N\right)^{-1}\right].
\]
Also, we denote by $\textcolor{black}{m(z)}$ the limit that $m_{F_{\hatbSig}}(z)$ converges to almost surely. The limit $\textcolor{black}{m(z)}$ is the solution of the following equation (detailed in \cite{marchenko1967distribution}):
\begin{equation}\label{eqn:m(z)}
m(z) = \int \dfrac{1}{\tau\left(1-\rho-\rho z m(z)\right)} dH(\tau).
\end{equation}
Further define $m_{1}(z)$ via
\begin{align}\label{eq:m1def}
m_{1}(z) := \frac{\int \frac{\tau^2[1-\rho-\rho zm(z)]}{[\tau[1-\rho-\rho zm(z)] -z]^2}\, dH(\tau)}{1-\rho \int \frac{z\tau}{[\tau[1-\rho-\rho zm(z)] -z]^2}\, dH(\tau)}.\nonumber
\end{align}

To facilitate a deeper understanding of our analysis, under Assumption (\ref{ASSU_D1:HD}- \ref{ASSU_D5:Sparse_Mean}), we introduce and define several key intermediate terms:
\begin{align}
\Theta_1(\lambda,\rho)&=\dfrac{1-\lambda m(-\lambda)}{1-\rho\left(1-\lambda m(-\lambda)\right)},\nonumber\\
\Theta_2(\lambda,\rho)&=\dfrac{1-\lambda m(-\lambda)}{\left[1-\rho+\rho\lambda m(-\lambda)\right]^3}-\lambda\dfrac{m(-\lambda)-\lambda m'(-\lambda)}{\left[1-\rho+\rho\lambda m(-\lambda)\right]^4},\nonumber\\
\Phi_1(\lambda,\rho,H,G) &=\tilde{\xi}^2 \int \frac{1}{\lambda\left[1+ \left(\rho m(-\lambda)+\frac{\rho-1}{\lambda}\right)\tau\right]} dG(\tau),\nonumber\\
\Phi_2(\lambda,\rho,H,G) &=\tilde{\xi}^2 \left(1+\rho m_{1}(-\lambda)\right) \int \frac{\tau}{[\lambda +(1-\rho+\rho\lambda m(-\lambda))\tau]^2} dG(\tau).\nonumber
\end{align}

\begin{theorem}[Sharpe ratio]
\label{Thm:sr_ridge}
Under Assumptions (\ref{ASSU_D1:HD}-\ref{ASSU_D5:Sparse_Mean}), it holds {in probability} that
\begin{equation}
SR\left[\hatbomeastlam\right] \rightarrow \dfrac{\Phi_1}{\sqrt{\Phi_2+\rho\Theta_2}}.\nonumber
\end{equation}
\end{theorem}

\begin{theorem}[Out-of-sample prediction loss]
\label{Thm:pl_ridge}
Under Assumptions (\ref{ASSU_D1:HD}-\ref{ASSU_D6:MSR}), it holds {in probability} that
\begin{equation}
L_{\bR}\left[\hatbomeastlam;{\bomeast}\right] \rightarrow \sigma^2\left[\dfrac{\Phi_1^2+\Phi_2+\rho\Theta_2}{\Phi_1+\rho\Theta_1}-2\dfrac{\Phi_1\left(1+\ttheta\right)}{\sqrt{\Phi_1+\rho\Theta_1}\sqrt{\ttheta}}+1+\ttheta\right].\nonumber
\end{equation}
\end{theorem}

\subsection{Proofs for Theorem \ref{Thm:sr_ridge}, \ref{Thm:pl_ridge}}
\begin{proof}[\textbf{Proof:}]
Let us first recall that the ridge regularized estimator is defined by
\begin{align}
\hatbomeastlam = \dfrac{\sigma}{\sqrt{\hatbmu\T\hatbSiglam\inv\hatbmu}}\hatbSiglam\inv\hatbmu,\nonumber
\end{align}
then the prediction loss of this estimator can be expanded using some key intermediate terms 
\begin{align}
L_{\bR}\left[\hatbomeastlam;{\bomeast}\right]&={B_{\bR}\left(\hatbomeastlam;{\bomeast}\right)}
\nonumber\\
&=\left(\hatbomeastlam-{\bomeast}\right)\T\left(\bSig+\bmu\bmu\T\right)\left(\hatbomeastlam-{\bomeast}\right)
\nonumber\\
&=\sigma^2\left[\dfrac{\hatbmu\T\hatbSiglam\inv\bSig\hatbSiglam\inv\hatbmu}{\hatbmu\T\hatbSiglam\inv\hatbmu} - 2\dfrac{\hatbmu\T\hatbSiglam\inv{\bmu}}{\sqrt{\hatbmu\T\hatbSiglam\inv\hatbmu}\sqrt{{\bmu}\T{\bSig}\inv{\bmu}}}+1\right.
\nonumber\\
&\quad \text{     }+\left.\dfrac{\left(\hatbmu\T\hatbSiglam\inv\bmu\right)^2}{\hatbmu\T\hatbSiglam\inv\hatbmu} - 2\dfrac{\hatbmu\T\hatbSiglam\inv\bmu\sqrt{\bmu\T\bSig\inv{\bmu}}}{\sqrt{\hatbmu\T\hatbSiglam\inv\hatbmu}}+{{\bmu}\T{\bSig}\inv\bmu}\right].\nonumber
\end{align}

The Sharpe ratio can be expanded as 
\begin{align}
SR\left[\hatbomeastlam\right]=\dfrac{\hatbmu\T\hatbSiglam\inv\bmu}{\sqrt{\hatbmu\T\hatbSiglam\inv\bSig\hatbSiglam\inv\hatbmu}}.\nonumber
\end{align}

We denote
$F1=\bmu\T\hatbSiglam\inv\bmu$, $F2=\hatbsm\T\hatbSiglam\inv\hatbsm$, $F3=\bmu\T\hatbSiglam\inv\hatbsm$, $F4=\bmu\T\hatbSiglam\inv\bSig\hatbSiglam\inv\bmu$, $F5=\hatbsm\T\hatbSiglam\inv\bSig\hatbSiglam\inv\hatbsm$, $F6=\bmu\T\hatbSiglam\inv\bSig\hatbSiglam\inv\hatbsm$, with $\hatbsm=\hatbmu-\bmu$.

\bigskip

\noindent\textbf{\underline{Limit of $F1$}}:
By using the anisotropic local law of Theorem 3.16 in \cite{knowles2017anisotropic}, we obatin that for any $D>0, \varepsilon>0$, where exists $C=C(D,\varepsilon)$ such that, with probability at least $1-CT^{-D}$,
\begin{equation}
    \left\vert \bmu\T\left(\hatbSig+\lambda\bm{I}_N\right)\inv\bmu-\dfrac{1}{\lambda}\bmu\T\left[\bm{I}_N+r(-\lambda)\bSig\right]\inv\bmu \right\vert\leq\sqrt{\dfrac{\Im\left(r(-\lambda)\right)}{\Im\left(-\lambda\right)T^{1-\varepsilon}}}\nonumber
\end{equation}
uniformly over $\Re(\lambda)>T^{-2/3+(1/\textcolor{black}{M})}, \Im(-\lambda)>0$, where $\Re$ and $\Im$ are real and imaginar part of a complex number. Here $r(z)$ is defined as the unique solution of 
\begin{equation}
    \dfrac{1}{r(z)}=-z+\rho\int\dfrac{\tau}{1+\tau r(z)}dH(\tau). \nonumber
\end{equation}
Existence and uniqueness of the solution $r(z)$ is given, for instance, in Lemma 2.2 of \cite{knowles2017anisotropic}. This lemma also states that $r(z)$ is the Stieltjes transorm of a probability measure with support in $[0,C(\textcolor{black}{M})]$. As a consequence, for $\lambda=\lambda_1+i\lambda_2,\lambda_1>0$
\begin{align}
\Im(r(-\lambda))&=\Im\int\frac{1}{x+\lambda}d\pi(x)=\int\frac{-\lambda_2}{(x+\lambda_1)^2+\lambda_2^2}d\pi(x)\nonumber\\
&\Rightarrow\left\vert \Im(r(-\lambda)) \right\vert\leq\dfrac{\left\vert \Im(\lambda) \right\vert}{\Re(\lambda)^2}.\nonumber
\end{align}
Therefore, 
\begin{equation}
    \left\vert \bmu\T\left(\hatbSig+\lambda\bm{I}_N\right)\inv\bmu-\dfrac{1}{\lambda}\bmu\T\left[\bm{I}_N+r(-\lambda)\bSig\right]\inv\bmu \right\vert\leq\dfrac{1}{\left\vert\Re(\lambda)\right\vert T^{\frac{1-\varepsilon}{2}}}.\nonumber
\end{equation}
Also,
\begin{align}
    &\dfrac{1}{\lambda}\bmu\T\left[\bm{I}_N+r(-\lambda)\bSig\right]\inv\bmu\nonumber\\
    =&\dfrac{1}{\lambda}\bmu\T\left[\bm{I}_N+\left(\rho m(-\lambda)+\frac{\rho-1}{\lambda}\right)\bSig\right]\inv\bmu\nonumber\\
    =&{\|\bmu\|^2} \int \frac{1}{\lambda\left[1+ \left(\rho m(-\lambda)+\frac{\rho-1}{\lambda}\right)\tau\right]} dG(\tau),\nonumber
\end{align}
where the first equality follows by the definition
\begin{equation}
    m(z) = \dfrac{1-\rho+ z r(z)}{\rho z}\nonumber.
\end{equation}

\bigskip

\noindent\textbf{\underline{Limit of $F2$}}:
Recall that \( \hatbsm = \bSig^{\frac{1}{2}}\bY\T\be/T \), thus we have 
\begin{align}
\hatbsm\T\hatbSiglam\inv\hatbsm
=& \left(\bY\T\be/T\right)\T\bSig^{\frac{1}{2}}\left(\hatbSig+\lambda\bI_N\right)\inv\bSig^{\frac{1}{2}}\left(\bY\T\be/T\right)
\nonumber\\
=& 
\dfrac{1}{T}\left(\sqrt{T}\barbs\right)\T\bSig^{\frac{1}{2}}\left(\hatbSig+\lambda\bI_N\right)\inv\bSig^{\frac{1}{2}}\left(\sqrt{T}\barbs\right),
\nonumber
\end{align}
where $\barbs=\bY\T\be/T$. Note that $\sqrt{T}\barbs$ is considered a vector of $N$ random variables, each following a normal distribution $\NormDis(0,1)$. In accordance with Assumption \ref{ASSU_D2:DGP} (Normality Assumption), the sample mean and sample covariance matrix are independent. This independence allows us to invoke the trace Lemma (Lemma \ref{lemma:Trace_Lemma}), yielding:
\[
\dfrac{T}{N}\hatbsm\T\hatbSiglam\inv\hatbsm - \dfrac{1}{N}\Tr\left[\bSig^{\frac{1}{2}}\left(\hatbSig+\lambda\bI_N\right)\inv\bSig^{\frac{1}{2}}\right]\rightarrow 0,\quad\text{in probability}.
\]
By applying the technique of using the Stieltjes transform of the empirical spectral distribution of a matrix proposed in \cite{ledoit2011eigenvectors}, we have
\begin{equation}\label{eqn:Tr(hSig_Sig)}
\dfrac{1}{N}\Tr\left[\bSig^{\frac{1}{2}}\left(\hatbSig+\lambda\bI_N\right)\inv\bSig^{\frac{1}{2}}\right] \convergeInP \Theta_1(\lambda,\rho),
\end{equation}
where $\Theta_1(\lambda,\rho)=\dfrac{1-\lambda m(-\lambda)}{1-\rho\left(1-\lambda m(-\lambda)\right)}$ and $m(-\lambda)$ is defined in (\ref{eqn:m(z)}). Therefore,
\begin{equation}
\hatbsm\T\hatbSiglam\inv\hatbsm \convergeInP \rho\Theta_1(\lambda,\rho).\nonumber
\end{equation}
Here is the detailed proof of (\ref{eqn:Tr(hSig_Sig)}): Denote the resolvent of $\hatbSig$ by $\rsvt(z)=\left(\hatbSig-z\bm{I}_N\right)^{-1}$. Note that the sample covariance matrix $\hatbSig=\frac{1}{T}\sum_{t=1}^{T}\bSig^{\frac{1}{2}}\by_t\by_t\T\bSig^{\frac{1}{2}}$, by denoting $\bty_t=\frac{1}{\sqrt{T}}\by_t$, we can rewrite $\hatbSig=\sum_{t=1}^{T}\bSig^{\frac{1}{2}}\bty_t\bty_t\T\bSig^{\frac{1}{2}}$. Therefore, for any $1\leq t\leq T$,
\begin{align}
    \rsvt(z)&=\left[\left(\hatbSig-\bSig^{\frac{1}{2}}\bty_t\bty_t\T\bSig^{\frac{1}{2}}\right)-z\bm{I}_N+\bSig^{\frac{1}{2}}\bty_t\bty_t\T\bSig^{\frac{1}{2}}\right]^{-1}\nonumber\\
    &=\rsvt^{t}(z)-\dfrac{\rsvt^{t}(z)\bSig^{\frac{1}{2}}\bty_t\bty_t\T\bSig^{\frac{1}{2}}\rsvt^{t}(z)}{1+\bty_t\T\bSig^{\frac{1}{2}}\rsvt^{t}(z)\bSig^{\frac{1}{2}}\bty_t},\label{eqn:27}
\end{align}
where $\rsvt^{t}(z)=\left[\left(\hatbSig-\bSig^{\frac{1}{2}}\bty_t\bty_t\T\bSig^{\frac{1}{2}}\right)-z\bm{I}_N\right]^{-1}$. Note that $\bty_t$ and $\rsvt^{t}(z)$ are independent for all $z\in\mathbb{C}$.

Multiplying $\rsvt(z)$ to both sides of the identity $\left(\hatbSig-z\bm{I}_N\right) + z\bm{I}_N = \hatbSig$, we have 
\begin{equation}\label{eqn:hSig_R(z)}
\bm{I}_N+z\rsvt(z)=\hatbSig\rsvt(z)=\sum_{t=1}^{T}\bSig^{\frac{1}{2}}\bty_t\bty_t\T\bSig^{\frac{1}{2}}\rsvt(z).
\end{equation}

Now let us set $z=-\lambda$. Taking trace on both sides of (\ref{eqn:hSig_R(z)}), dividing both sides by $N$, and using (\ref{eqn:27}), we have
\begin{align}
    1-\dfrac{\lambda}{N}\Tr\left[\rsvt(-\lambda)\right] &= \dfrac{1}{N}\sum_{t=1}^{T}\left[\bty_t\T\bSig^{\frac{1}{2}}\rsvt^{t}(-\lambda)\bSig^{\frac{1}{2}}\bty_t - \dfrac{\left(\bty_t\T\bSig^{\frac{1}{2}}\rsvt^{t}(-\lambda)\bSig^{\frac{1}{2}}\bty_t \right)^2}{1+\bty_t\T\bSig^{\frac{1}{2}}\rsvt^{t}(-\lambda)\bSig^{\frac{1}{2}}\bty_t }\right]\nonumber\\
    &=\dfrac{1}{N}\sum_{t=1}^{T}\dfrac{\bty_t\T\bSig^{\frac{1}{2}}\rsvt^{t}(-\lambda)\bSig^{\frac{1}{2}}\bty_t}{1+\bty_t\T\bSig^{\frac{1}{2}}\rsvt^{t}(-\lambda)\bSig^{\frac{1}{2}}\bty_t}\nonumber\\
    &=\dfrac{T}{N} - \dfrac{1}{N}\sum_{t=1}^{T}\dfrac{1}{1+\bty_t\T\bSig^{\frac{1}{2}}\rsvt^{t}(-\lambda)\bSig^{\frac{1}{2}}\bty_t}\nonumber\\
    &=\dfrac{T}{N} - \dfrac{1}{N}\sum_{t=1}^{T}\frac{1}{1+\dfrac{1}{T}\Tr\left[\bSig^{\frac{1}{2}}\rsvt^{t}(-\lambda)\bSig^{\frac{1}{2}}\right]}+\delta_1\nonumber\\
    &=\dfrac{T}{N} - \dfrac{T}{N}\frac{1}{1+\dfrac{p}{T}\dfrac{1}{N}\Tr\left[\rsvt(-\lambda)\bSig\right]}+\delta_1+\delta_2\label{eqn:29}
\end{align}
with residual terms $\delta_1$ and $\delta_2$. The detailed proof of $\max\left(\delta_1,\delta_2\right)=o_p\left(T^{-{1}/{2}}\right)$ can be found in the supplementary material of \cite{chen2011regularized}. Also, we know that
\begin{equation}\label{eqn:30}
\dfrac{1}{N}\Tr\left[\rsvt(-\lambda)\right]\convergeInP m(-\lambda).
\end{equation}
Thus based on (\ref{eqn:29}) and (\ref{eqn:30}), it is easy to get (\ref{eqn:Tr(hSig_Sig)}).

\bigskip

\noindent\textbf{\underline{Limit of $F3$}}: Similar to the proof for $D3$, we examine the square of $F3$, $\hatbsm\T\hatbSiglam\inv\bmu\bmu\T\hatbSiglam\inv\hatbsm$, where the vectors on both the left and right sides of the quadratic form are identical. We have 
\begin{align}
\hatbsm\T\hatbSiglam\inv\bmu\bmu\T\hatbSiglam\inv\hatbsm
=& \left(\dfrac{\bY\T\be}{T}\right)\T\bSig^{\frac{1}{2}}\left(\hatbSig+\lambda\bm{I}_N\right)\inv\bmu\bmu\T\left(\hatbSig+\lambda\bm{I}_N\right)\inv\bSig^{\frac{1}{2}}\left(\dfrac{\bY\T\be}{T}\right)
\nonumber\\
=& 
\dfrac{1}{T}\left(\sqrt{T}\barbs\right)\T\bSig^{\frac{1}{2}}\left(\hatbSig+\lambda\bm{I}_N\right)\inv\bmu\bmu\T\left(\hatbSig+\lambda\bm{I}_N\right)\inv\bSig^{\frac{1}{2}}\left(\sqrt{T}\barbs\right).
\nonumber
\end{align}
By adopting this approach, we can apply the same reasoning used in $F2$, where $\sqrt{T}\barbs$ is considered a vector of $N$ random variables, each following a normal distribution $\NormDis(0,1)$. This allows us to invoke the trace Lemma (Lemma \ref{lemma:Trace_Lemma}), yielding:
\[
\dfrac{T}{N}\hatbsm\T\hatbSiglam\inv\bmu\bmu\T\hatbSiglam\inv\hatbsm - \dfrac{1}{N}\Tr\left(\bSig^{\frac{1}{2}}\left(\hatbSig+\lambda\bm{I}_N\right)\inv\bmu\bmu\T\left(\hatbSig+\lambda\bm{I}_N\right)\inv\bSig^{\frac{1}{2}}\right)\rightarrow 0,\quad\text{in probability}.
\]
Subsequently, $\Tr\left(\bSig^{\frac{1}{2}}\left(\hatbSig+\lambda\bm{I}_N\right)\inv\bmu\bmu\T\left(\hatbSig+\lambda\bm{I}_N\right)\inv\bSig^{\frac{1}{2}}\right)=\bmu\T\left(\hatbSig+\lambda\bm{I}_N\right)\inv\bSig\left(\hatbSig+\lambda\bm{I}_N\right)\inv\bmu$, is easily identifiable as exactly matching $F4$.

To sum up,
\begin{equation}
\left(\hatbsm\T\hatbSiglam\inv\bmu\right)^2-\frac{1}{T}\cdot\Phi_2(\lambda,\rho,H,G)\convergeInP 0.\nonumber
\end{equation}
Given Assumption \ref{ASSU_D1:HD} which assumes $T\rightarrow\infty$, it follows that $\bmu\T\hatbSiglam\inv\hatbsm$ converges to 0.

\bigskip

\noindent\textbf{\underline{Limit of $F4$}}:
$F4$ is the same as the expression for bias of ridge regression in the paper \cite{hastie2022surprises}. Thus the proof can be found in its Supplementary document for proving for (A.12). What we have for $F4$ is that
\begin{equation}
\bmu\T\hatbSiglam\inv\bSig\hatbSiglam\inv\bmu \convergeInP \Phi_2(\lambda,\rho,H,G).
\end{equation}

\bigskip

\noindent\textbf{\underline{Limit of $F5$}}: Similar to the proof in $F2$, we can rewrite $F5$ as
\begin{align}
\hatbsm\T\hatbSiglam\inv\bSig\hatbSiglam\inv\hatbsm
=& \left(\bY\T\be/T\right)\T\bSig^{\frac{1}{2}}\left(\hatbSig+\lambda\bI_N\right)\inv\bSig\left(\hatbSig+\lambda\bI_N\right)\inv\bSig^{\frac{1}{2}}\left(\bY\T\be/T\right)
\nonumber\\
=& 
\dfrac{1}{T}\left(\sqrt{T}\barbs\right)\T\bSig^{\frac{1}{2}}\left(\hatbSig+\lambda\bI_N\right)\inv\bSig\left(\hatbSig+\lambda\bI_N\right)\inv\bSig^{\frac{1}{2}}\left(\sqrt{T}\barbs\right),
\nonumber
\end{align}
where $\barbs=\bY\T\be/T$. By the same argument in $F2$, we have
\[
\dfrac{T}{N}\hatbsm\T\hatbSiglam\inv\bSig\hatbSiglam\inv\hatbsm - \dfrac{1}{N}\Tr\left[\left(\hatbSig+\lambda\bI_N\right)\inv\bSig\left(\hatbSig+\lambda\bI_N\right)\inv\bSig\right]\rightarrow 0,\quad\text{in probability}.
\]
Also, we have
\begin{equation}\label{eqn:Tr(hSig_Sig_hSig_Sig)}
\dfrac{1}{N}\Tr\left[\left(\hatbSig+\lambda\bI_N\right)\inv\bSig\left(\hatbSig+\lambda\bI_N\right)\inv\bSig\right] \convergeInP \Theta_2(\lambda,\rho),
\end{equation}
where $\Theta_2(\lambda,\rho)=\dfrac{1-\lambda m(-\lambda)}{\left[1-\rho+\rho\lambda m(-\lambda)\right]^3}-\lambda\dfrac{m(-\lambda)-\lambda m'(-\lambda)}{\left[1-\rho+\rho\lambda m(-\lambda)\right]^4}$. Therefore,
\begin{equation}
\hatbsm\T\hatbSiglam\inv\bSig\hatbSiglam\inv\hatbsm\convergeInP \rho\Theta_2(\lambda,\rho).\nonumber
\end{equation}
The following is the detailed proof of (\ref{eqn:Tr(hSig_Sig_hSig_Sig)}), it employs the technique following \cite{ledoit2011eigenvectors} based on the Stieltjes transform of the ESD of a random matrix. Multiplying $\bSig\rsvt(-\lambda)$ to both sides of the identity (\ref{eqn:hSig_R(z)}) (with $z=-\lambda$), taking trace, dividing by $N$, and using (\ref{eqn:27}),  we have 
\begin{align}
&\dfrac{1}{N}\Tr\left[ \rsvt(-\lambda)\bSig \right]-\lambda\dfrac{1}{N}\Tr\left[ \left(\rsvt(-\lambda)\right)^2 \bSig\right]\nonumber\\
=&\dfrac{1}{N}\sum_{t=1}^{T}\bty_t\T\bSig^{\frac{1}{2}}\rsvt(-\lambda)\bSig\rsvt(-\lambda)\bSig^{\frac{1}{2}}\bty_t \nonumber\\
=&\dfrac{1}{N}\sum_{t=1}^{T}\left[\bty_t\T\bSig^{\frac{1}{2}}\rsvt^t(-\lambda)\bSig\rsvt(-\lambda)\bSig^{\frac{1}{2}}\bty_t \right.\nonumber\\
&\left.- \dfrac{\left(\bty_t\T\bSig^{\frac{1}{2}}\rsvt^{t}(-\lambda)\bSig^{\frac{1}{2}}\bty_t \right)\left(\bty_t\T\bSig^{\frac{1}{2}}\rsvt^t(-\lambda)\bSig\rsvt(-\lambda)\bSig^{\frac{1}{2}}\bty_t \right)}{1+\bty_t\T\bSig^{\frac{1}{2}}\rsvt^{t}(-\lambda)\bSig^{\frac{1}{2}}\bty_t} \right]\nonumber\\
=&\dfrac{1}{N}\sum_{t=1}^{T} \dfrac{\bty_t\T\bSig^{\frac{1}{2}}\rsvt^t(-\lambda)\bSig\rsvt(-\lambda)\bSig^{\frac{1}{2}}\bty_t}{1+\bty_t\T\bSig^{\frac{1}{2}}\rsvt^{t}(-\lambda)\bSig^{\frac{1}{2}}\bty_t}\nonumber\\
=&\dfrac{1}{N}\sum_{t=1}^{T} \dfrac{\bty_t\T\bSig^{\frac{1}{2}}\rsvt^t(-\lambda)\bSig\rsvt^t(-\lambda)\bSig^{\frac{1}{2}}\bty_t}{\left(1+\bty_t\T\bSig^{\frac{1}{2}}\rsvt^{t}(-\lambda)\bSig^{\frac{1}{2}}\bty_t\right)^2}\nonumber\\
=&\dfrac{1}{N}\sum_{t=1}^{T} \dfrac{(1/T)\Tr\left[\bSig^{\frac{1}{2}}\rsvt^t(-\lambda)\bSig\rsvt^t(-\lambda)\bSig^{\frac{1}{2}}\right]}{\left(1+(1/T)\Tr\left[\bSig^{\frac{1}{2}}\rsvt^{t}(-\lambda)\bSig^{\frac{1}{2}}\right]\right)^2}+\delta_3\nonumber\\
=&\dfrac{(1/p)\Tr\left[\rsvt(-\lambda)\bSig\rsvt(-\lambda)\bSig\right]}{\left(1+(N/T)(1/p)\Tr\left[\rsvt(-\lambda)\bSig\right]\right)^2}+\delta_3+\delta_4\label{eqn:32}
\end{align}
with residuals $\delta_3$ and $\delta_4$. Also we know that 
\[
\dfrac{d}{d\lambda}\left[\dfrac{1}{N}\Tr\left[\rsvt(-\lambda)\bSig\right]\right]=-\dfrac{1}{N}\Tr\left[(\rsvt(-\lambda))^2\bSig\right],
\]
and observe that
\[
\dfrac{1}{N}\Tr\left[\rsvt(-\lambda)\bSig\right] \convergeInP \Theta_1(\lambda,\rho).
\]
From these premises, we define:
\[\delta_5=-\dfrac{1}{N}\Tr\left[(\rsvt(-\lambda))^2\bSig\right]-\dfrac{\partial}{\partial\lambda}\Theta_1(\lambda,\rho).
\]
To substantiate the convergence of $\max\left(\delta_3,\delta_4,\delta_5\right)=o_p\left(1\right)$, readers are referred to the supplementary material provided by \cite{chen2011regularized}. By reevaluating and rearranging Equation (\ref{eqn:32}), we can deduce the identity specified in Equation (\ref{eqn:Tr(hSig_Sig_hSig_Sig)}).

\bigskip

\noindent\textbf{\underline{Limit of $F6$}}: Mirroring the approach for $F3$, $F6$ also presents an imbalance in its quadratic form. To address this, we analyze the square of $F6$, given by $\hatbsm\T\hatbSiglam\inv\bSig\hatbSiglam\inv\bmu\bmu\T\hatbSiglam\inv\bSig\hatbSiglam\inv\hatbsm$. This formulation simplifies to:
\begin{align}
\left(\hatbsm\T\hatbSiglam\inv\bSig\hatbSiglam\inv\bmu\right)^2
=& \left(\dfrac{\bY\T\be}{T}\right)\T\bSig^{\frac{1}{2}}\hatbSiglam\inv\bSig\hatbSiglam\inv\bmu\bmu\T\hatbSiglam\inv\bSig\hatbSiglam\inv\bSig^{\frac{1}{2}}\left(\dfrac{\bY\T\be}{T}\right)
\nonumber\\
=& 
\dfrac{1}{T}\left(\sqrt{T}\barbs\right)\T\bSig^{\frac{1}{2}}\hatbSiglam\inv\bSig\hatbSiglam\inv\bmu\bmu\T\hatbSiglam\inv\bSig\hatbSiglam\inv\bSig^{\frac{1}{2}}\left(\sqrt{T}\barbs\right).
\nonumber
\end{align}
Using the same rationale as in $F5$ and invoking the Trace Lemma \ref{lemma:Trace_Lemma}, we deduce that
\[
\dfrac{T}{N}\left(\hatbsm\T\hatbSiglam\inv\bSig\hatbSiglam\inv\bmu\right)^2 - \dfrac{1}{N}\Tr\left(\bSig^{\frac{1}{2}}\hatbSiglam\inv\bSig\hatbSiglam\inv\bmu\bmu\T\hatbSiglam\inv\bSig\hatbSiglam\inv\bSig^{\frac{1}{2}}\right)\rightarrow 0,\quad\text{in probability}.
\]
Furthermore, $\Tr\left(\bSig^{\frac{1}{2}}\hatbSiglam\inv\bSig\hatbSiglam\inv\bmu\bmu\T\hatbSiglam\inv\bSig\hatbSiglam\inv\bSig^{\frac{1}{2}}\right) =\bmu\T\hatbSiglam\inv\bSig\hatbSiglam\inv\bSig\hatbSiglam\inv\bSig\hatbSiglam\inv\bmu$.

To sum up,
\begin{equation}
\left(\hatbsm\T\hatbSiglam\inv\bSig\hatbSiglam\inv\bmu\right)^2-\frac{1}{T}\cdot\bmu\T\hatbSiglam\inv\bSig\hatbSiglam\inv\bSig\hatbSiglam\inv\bSig\hatbSiglam\inv\bmu\convergeInP
0.\nonumber
\end{equation}
Given Assumption \ref{ASSU_D1:HD} which assumes $T\rightarrow\infty$, it follows that $\hatbsm\T\hatbSiglam\inv\bSig\hatbSiglam\inv\bmu$ converges to 0.
\end{proof}


\end{document}